\documentclass[
a4paper, 
titlepage, 
bibliography=totoc, 
11pt, 
BCOR17mm, 
DIV12, 
headinclude, 
footinclude=false]
{scrbook}

\usepackage{graphicx}				
\usepackage{pdfpages}
\usepackage{here}
\usepackage{epstopdf}

\usepackage[english]{babel}
\usepackage[utf8]{inputenc}
\usepackage[T1]{fontenc}

\usepackage{fancyhdr}
\usepackage{setspace}
\usepackage{xcolor}

\usepackage{amsmath}
\usepackage{amsfonts}
\usepackage{amssymb}
\usepackage{hyperref}
\hypersetup{
	colorlinks,
	citecolor=blue,
	filecolor=black,
	linkcolor=black,
	urlcolor=black
}
\usepackage[super]{natbib}
\usepackage{comment}
\usepackage{slashed}
\usepackage{ytableau}
\usepackage{wasysym}
\usepackage{geometry}
\usepackage[hang,flushmargin]{footmisc}
\usepackage{setspace}
\ytableausetup{smalltableaux}
\usepackage{graphicx}
\usepackage{amsmath}
\usepackage{mathrsfs}
\usepackage{dsfont}
\usepackage{amssymb}
\usepackage{subfigure}
\usepackage{color}
\usepackage{blkarray}
\usepackage{hyperref}
\usepackage{verbatim}
\usepackage{amsthm}
\usepackage{enumerate}
\usepackage{bbm}
\usepackage{hyperref}
\usepackage{braket}
\usepackage{booktabs}
\usepackage{mathtools}
\usepackage{array,booktabs}
\usepackage{multirow}

\newcommand*{\hbb}{\mathbb{H}}
\newcommand*{\Tr}{\mathrm{Tr}}

\newcommand*{\bnd}{\partial\gamma}
\newcommand*{\core}{\rho_{\text{core}}}
\newcommand*{\avg}[1]{\langle #1 \rangle}
\newcommand*{\sw}{\mathcal{S}}

\newcommand{\ind}[1]{\vec{\textbf{#1}}}

\newcommand*{\coloneqq}{\mathrel{\vcenter{\baselineskip0.5ex \lineskiplimit0pt \hbox{\scriptsize.}\hbox{\scriptsize.}}} =}
\newcommand*{\coloneqqrev}{=\mathrel{\vcenter{\baselineskip0.5ex \lineskiplimit0pt \hbox{\scriptsize.}\hbox{\scriptsize.}}} }

\addto\captionsngerman{}

\definecolor{indigoblue}{cmyk}{1, .77, .34, .21}

\newcommand{\LMUTitle}[9]{
	\thispagestyle{empty}				
	\vspace*{\stretch{1}}				
	{\parindent0cm\rule{\linewidth}{.7ex}}  			
	\vspace*{\stretch{1}}	
	{\parindent0cm\rule{\linewidth}{.3ex}}  
	
	\begin{flushright}
		\vspace*{\stretch{1}}
		\bfseries\scshape\Huge #1\\
		\vspace*{\stretch{1}}
		\sffamily\bfseries\large #2\\
		\vspace*{\stretch{1}}
	\end{flushright}
	
	{\parindent0cm\rule{\linewidth}{.3ex}}
	\vspace*{\stretch{1}} 
	{\parindent0cm\rule{\linewidth}{.7ex}}
	\vspace*{\stretch{5}}
	
	\begin{center}
		\includegraphics[width=10cm]{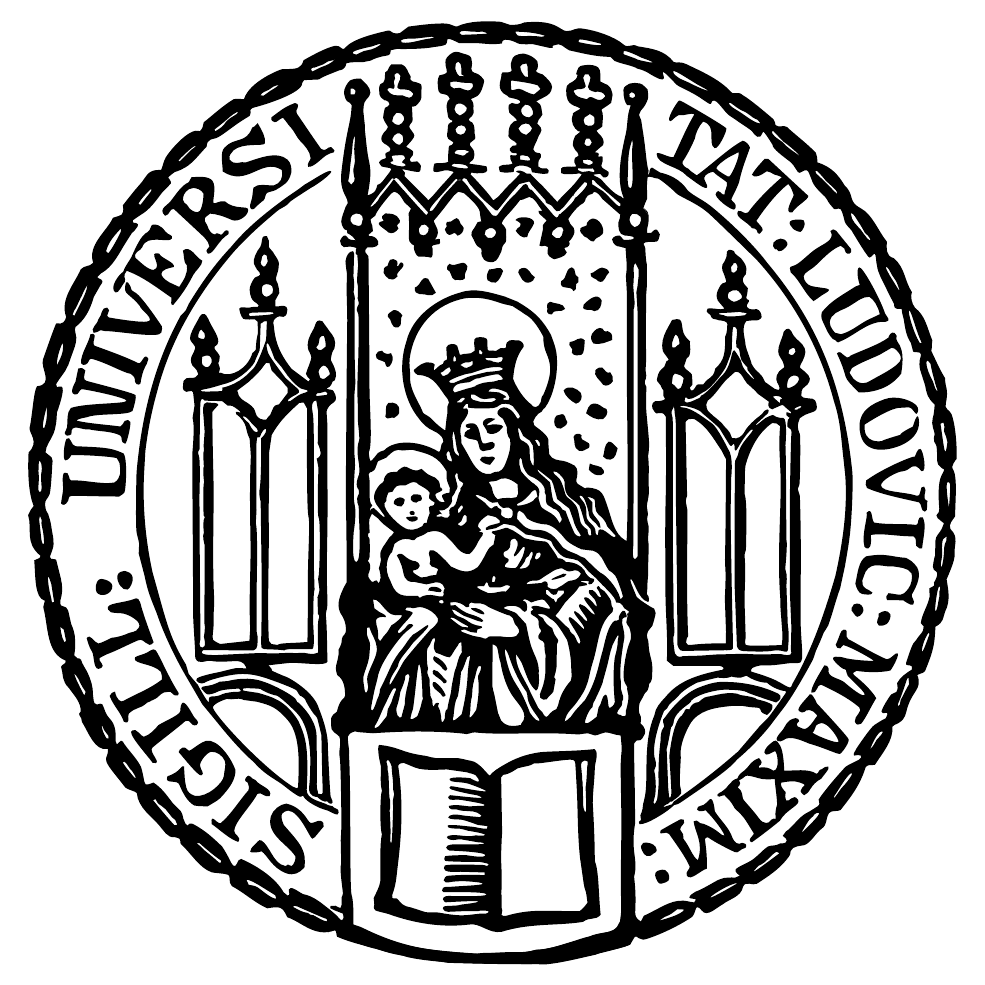}
	\end{center}
	
	\vspace*{\stretch{1}}
	\begin{center}\sffamily\LARGE{#5}
	\end{center}
	
	\newpage								
	\thispagestyle{empty}
	
	\cleardoublepage
	\thispagestyle{empty}
	
	\vspace*{\stretch{1}}
	{\parindent0cm
		\rule{\linewidth}{.7ex}}
	\begin{center}
		\vspace*{\stretch{1}}
		\bfseries\Huge #1\\
		\vspace*{\stretch{1}}
	\end{center}
	\rule{\linewidth}{.7ex}
	
	\vspace*{\stretch{4}}
	\begin{center}
		\Large\textbf{Master's thesis}\\
		\Large Faculty of physics\\
		\Large Ludwig--Maximilians--University\\
		\Large Munich\\
		\vspace*{\stretch{1}}
		\Large submitted by\\
		\Large\textbf{#2}\\
		\Large from Munich\\
		\vspace*{\stretch{2}}
		\Large Munich, #6
	\end{center}

	\newpage
	\thispagestyle{empty}
	\textbf{Acknowledgements}\vspace*{0.5cm}
	I wish to express my gratitude to
	\begin{itemize}
		\item My supervisors Dr. Daniele Oriti and Eugenia Colafranceschi for their continuous support, feedback and direction during this project, as well as providing me with a PhD student position.
		\item Professor Ulrich Schollwöck for his interest in our work and being the second referee for this thesis' defense.
		\item My many university colleagues from the physics and TMP program as well as the geometry groups for many helpful discussions and moral support.
		\item My family and particularly my parents for their support throughout the years it took me to get here.
	\end{itemize}
	
	\vspace*{\stretch{2}}
	
	\begin{flushleft}
		\large Supervisor: #7 \\[1mm]
	\end{flushleft}
	
	\cleardoublepage
}

\begin{document}

\LMUTitle
      {Superposed Random Spin Tensor Networks and their Holographic Properties}	
      {Simon Langenscheidt}		
      {Munich}					
      {Faculty of physics}     
      {Munich 2022}            
      {\today}             		
      {Daniele Oriti}  

  \cleardoublepage
  \markboth{Abstract}{Abstract}
  We study criteria for and properties of boundary-to-boundary holography in a class of spin network states defined by analogy to projected entangled pair states (PEPS). In particular, we consider superpositions of states corresponding to well-defined, discrete geometries on a graph. By applying random tensor averaging techniques, we map entropy calculations to a random Ising model on the same graph, with distribution of couplings determined by the relative sizes of the involved geometries. The superposition of tensor network states with variable bond dimension used here presents a picture of a genuine quantum sum over geometric backgrounds. We find that, whenever each individual geometry produces an isometric mapping of a fixed boundary region $C$ to its complement, then their superposition does so iff the relative weight going into each geometry is inversely proportional to its size. Additionally, we calculate average and variance of the area of the given boundary region and find that the average is bounded from below and above by the mean and sum of the individual areas, respectively. Finally, we give an outlook on possible extensions to our program and highlight conceptual limitations to implementing these.

  \renewcommand\contentsname{Table of contents}
  \tableofcontents
  \markboth{Table of contents}{Table of contents}

\mainmatter\setcounter{page}{1}


\chapter{Introduction}
	Within approaches to providing quantised UV completions to general relativity (GR), there are several distinct directions. Some prefer to stay in the continuum, such as Asymptotic Safety, some discretise degrees of freedom with the intent of recovering a continuum regime, as in Loop Quantum Gravity(LQG), while yet others recover low-energy phenomenology in more indirect ways, like through consistency conditions and effective actions in String Theory. However, certain features of gravitational theories are shared among them. From general thermodynamical and information theoretical arguments, one expects black holes, quantum or not, to fulfil a Bekenstein-Hawking bound\cite{bekensteinBlackHolesEntropy1973} 
	\begin{equation}\label{BHBHBound}
		S \leq \frac{A}{4G_N}
	\end{equation}
	for an appropriately chosen entropy. This states, through microstate counting interpretations of entropies, that the number of degrees of freedom of a black hole should not scale with \textit{volume}, as it would in most systems, but rather with the \textit{area bounding said volume}. This implies a form of so-called (informational) \textit{holography}: Information on the degrees of freedom of the system can be or is encoded on the boundary of a region, from where it may be recovered. This is termed after optical holography, where volumetric information (phase differences) can be encoded on a photoplate boundary, from where it may be reconstructed using the right techniques, displaying the original 3-dimensional image from a 2-dimensional data carrier.\\
	Other forms of more concrete holography have been developed which feature a more direct relation to gravity: Starting from Maldacena\cite{maldacenaLargeLimitSuperconformal1999}, String theorists were able to construct several examples of gravitational systems in Anti-deSitter spaces whose degrees of freedom, when suitably mapped to the conformal boundary, give rise to a set of operators generating a conformal field theory. This AdS/CFT correspondence has also been extended to 3D euclidean gravity or asymptotically flat, 4D spacetimes\cite{goellerQuasiLocal3DQuantum2020,freidelWeylBMSGroup2021a}, where the asymptotic symmetry is given by variants of the BMS algebra, extending conformal symmetry. What all of these examples have in common is that they describe a relation between a \textit{bulk} gravitational theory and a seperate set of degrees of freedom on an \textit{asymptotic boundary}, describing the same physics. \\
	However, a different type of holography for finite regions of space/spacetime appears to exist in a variety of contexts, in particular through entropy bounds. For example, recent work in classical gravity suggests that corner charges of GR provide an encoding of bulk information\cite{donnellyGravitationalEdgeModes2021,freidelEdgeModesGravity2020a,freidelExtendedCornerSymmetry2021}, which applies to any finite region of space with boundary. In condensed matter physics, the ground states of local Hamiltonians on lattice systems are often found to be \textit{short-range entangled}\cite{ciracRenormalizationTensorProduct2009b}: The entropy of any finite region scales at most as the size of its boundary. States with such properties are highly desirable: They provide a more manageable subset of the space of states to begin searches for ground states in, but also have properties like exponentially decaying correlations between regions mimicking a local lightcone structure through Lieb-Robinson bounds\cite{nachtergaeleLiebRobinsonBoundsQuantum2010,bravyiLiebRobinsonBoundsGeneration2006}.\\
	As such, given the many insights from local and global holography in quantum gravity (QG) contexts, it is natural to characterise classes of quantum geometries which feature this behaviour on a small scale. A natural starting point of this investigation are \textit{spin network states}, which form both a conceptual and concrete basis of spatial quantum geometry. These states were originally envisioned by Penrose\cite{penroseANGULARMOMENTUMAPPROACH,ApplicationsNegativeDimensional} and later recovered in the LQG canonical quantisation programme\cite{rovelliSpinNetworksQuantum1995}. They also feature in any \textit{spin foam }approach to constructing QG path integrals\cite{perezSpinFoamApproach2013a} as well as their completions in Group Field Theories (GFT)\cite{krajewskiGroupFieldTheories2012,oritiGroupFieldTheory2014,oritiMicroscopicDynamicsQuantum2011} as basis states of discretised geometries with definite area and volume values.\\
	The goal of this work is to elaborate on criteria for holographic mappings between patches of the boundary of a finite spatial region to exist. This region is modeled by a superposition of spin network states. So far, only spin network states with a fixed graph structure and definite, fixed spin values have been considered, where areas take definite values. This thesis extends existing results to a superposition of different spin values, covering a much larger class of states. In this way, the work presented here provides a first step into studying generic local holography in multiple QG approaches, with possible applications to semiclassical states where a large number of different spin network states are superposed.\\
	The general structure of this work is as follows: After a brief introduction of spin networks, their structure as glued vertex states and tensor networks, we review notions of holography and the work done in the 'no-superposition' case. We then perform a similar analysis on states which are superpositions of individual tetrahedra with different areas, but fixed given glueing arrangement. Importantly, we make a typicity\footnote{By a typicity statement we mean propositions of the form "The typical object from class X has property P", as given by a probability distribution. Such a distribution, for simplicity, might be approximated by a Gaussian, in which case the main parameters are the mean and variance. Then, a typicity statement is better the lower the variance is, as the probability of finding an object in class X with the given property deviating from the mean is very low. } statement about holographic properties of what we will call \textit{Random Spin Tensor Networks} (RSTN). We find that it is convenient to write the purity of boundary states as a probability average in a \textit{distribution of geometries}, determined by relative weights of the individual geometries in the state. We illustrate a few examples and determine a condition on said distribution to ensure holography. A short calculation of the area value of a holographic boundary region reveals a simple formula for these special superpositions. Finally, we discuss some natural extensions of the superposition of spins setting and highlight limitations in pursueing these.

\section{Spin network states as models of quantum geometries}
To start, we review the notion of spin networks as abstract combinatorial objects, as well as their use in labeling states in various fields of physics, particularly in relation to lattice gauge theory and quantum gravity. In doing so, we introduce the key concept of a spin network state. \\
For a given compact Lie group $ G $, typically chosen to be $ SU(2) $, a standard definition of a spin network can be given as follows: It consists of
\begin{enumerate}
	\item A graph $ \gamma  $ with vertex and edge sets $ (V,E) $.
	\item A colouring of its edges by representations $ \rho $ of $ G $.
	\item A colouring of its vertices by \textit{intertwining maps} $ \iota $ of the representations on the adjacent edges. Equivalently, these are invariant tensors $ \text{Inv}_{G_{Diag}}(\bigotimes_{ \alpha} V_{\rho_\alpha}) $ in the tensor product of representation spaces associated to the edges at the vertex.
\end{enumerate}
\begin{figure}
	\centering
	\includegraphics[width=0.7\linewidth]{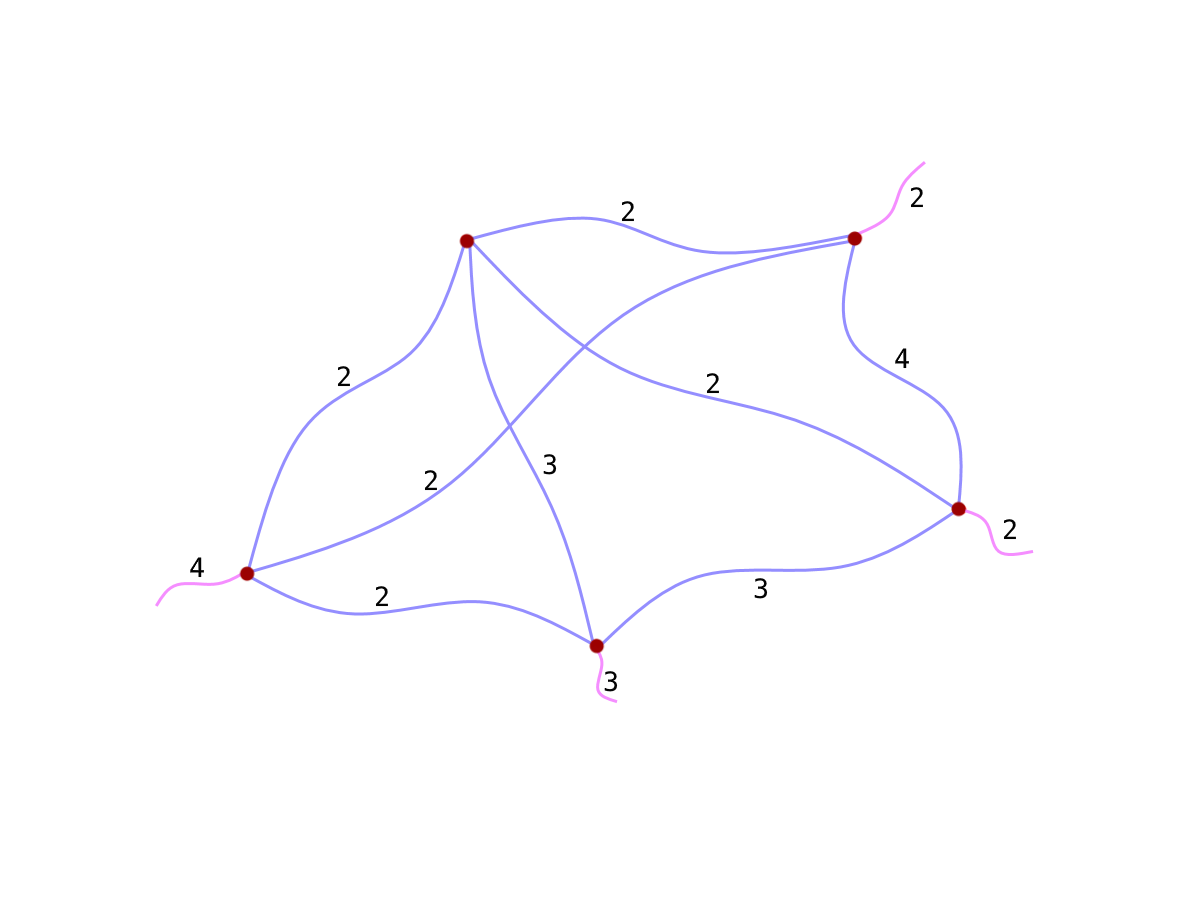}
	\caption{Example of a simple spin network labeling a state. Links are labeled by representations of $ SU(2) $, here in terms of their dimension $ d_j $. Intertwiners are situated at vertices, but not specified graphically.}
	\label{fig:snexample1}
\end{figure}
It is worth emphasizing the question of boundary edges. To allow a spin network to have a boundary, one either allows the edge set to contain semiedges/semilinks starting from any vertex, or introduces a colouring of the graph by adding a second vertex set of \textit{boundary vertices}. In the former, the semiedges do not end on a vertex, meaning we do not work with graphs in the usual sense. The latter amends this by attaching a 1-valent vertex to every boundary semilink. Then, a boundary edge/link is identified by its end being a boundary vertex. Practically, these might be distinguished by a coloring of the vertices into grey and red. Since these two variants are translateable into each other, we will at times switch between them when convenient.\\
Additionally, boundary semiedges should be labeled by relevant vectors from the representations associated to these edges. For the example of $ SU(2) $, this will typically be an azimuthal quantum number $ n \in[-j:j] $ labeling the standard basis of the spin-$ j $ representation. 
\begin{figure}
	\centering
	\includegraphics[width=0.6\linewidth]{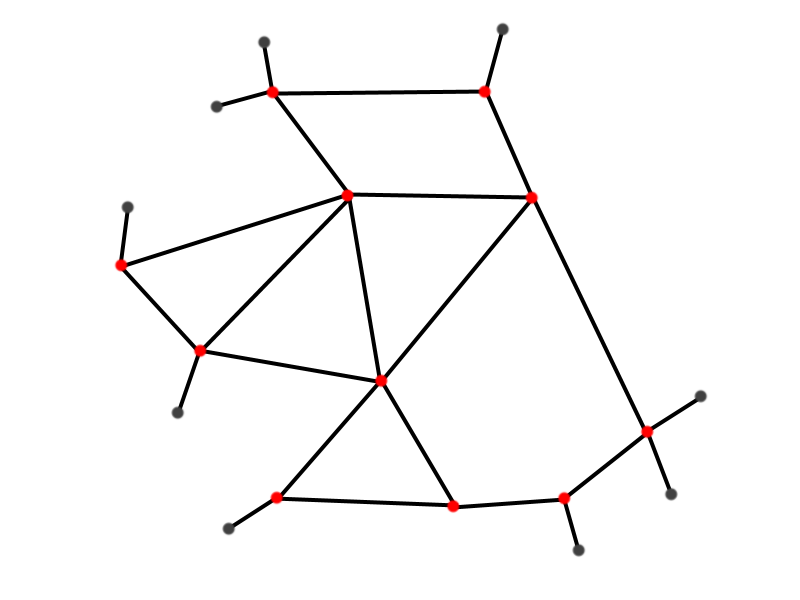}
	\caption{An example graph with boundary/bulk colouring in grey/red. Removing the grey vertices results in an open graph with semiedges.}
	\label{fig:graphexample1}
\end{figure}

This definition can be generalised and restricted in many different ways\cite{smolinFutureSpinNetworks1997,baezSpinNetworksNonperturbative1998}. In particular, one might restrict the set of graphs to be of a fixed valency or to be bipartite or oriented. On the other hand, one may replace the compact Lie group $ G $ by other objects with a similar representation theory, such as quantum groups or even Hopf algebras. Furthermore, one may furnish the graph with some imbedding into a manifold or additionally with a framing. 
Historically, the notion of spin networks was invented by Penrose\cite{penroseANGULARMOMENTUMAPPROACH,ApplicationsNegativeDimensional} as a suggestion that continuum structures such as Euclidean space and continuous probabilities might be limiting cases of discrete substructures. In particular, a set of 3-valent $ SU(2) $-spin networks with a large number of boundary links were shown to admit limits in which they describe continuum angles in 3-dimensional Euclidean space.
However, they have since found use in a variety of contexts independent of these ideas. Before presenting these contexts, we will present an independent motivation for them as representatives of quantum (twisted) simplicial geometry by starting from the quantisation of a single tetrahedron. We believe that this is the most direct way to reach spin network states without appealing to specific problems in other approaches.
\subsection{Quantisation of a single tetrahedron}
The goal of this section is to quantise the classical degrees of freedom of a 3-simplex and its embedding into a larger simplicial complex by analogy to the standard quantisation of a point particle.\cite{barbieriQuantumTetrahedraSimplicial1998a} We will in this way produce a first idea of quantum geometry which naturally admits spin network states as a basis.
\begin{figure}
	\centering
	\includegraphics[width=0.7\linewidth]{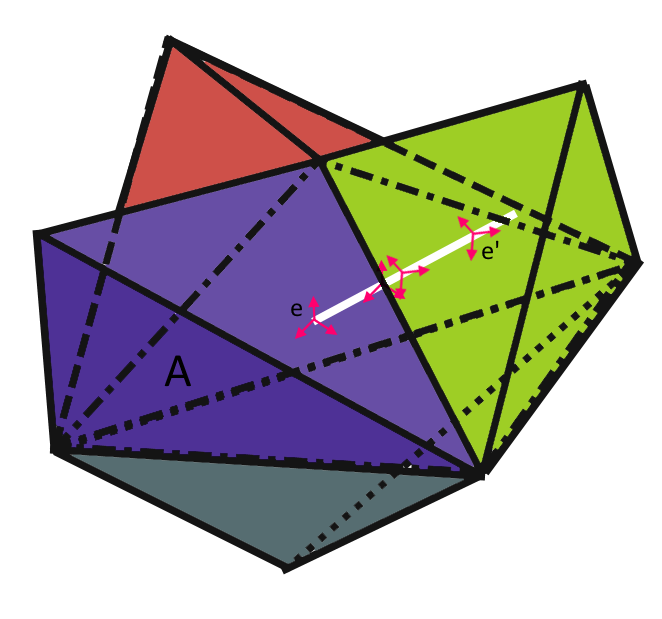}
	\caption{A simple example of the type of classical geometry we wish to quantise. 5 abstract simplices are glued along their faces. The geometric properties of the full space are given by the areas of the faces (here labeled by A) and a connection for transporting local frames around. Such a transport is exemplified along the white line crossing the face between the central simplex and a peripheral one: The connection is flat inside the simplices, meaning that transport is trivial before and after crossing the face. At the face, the initial frame $ e $ is rotated into $ e' $ using a group element from the rotation group. Flatness on interiors implies that these types of connections are specified completely through a group element on each face. These connections together with the areas specify a metric on the entirety of the simplices and thus the whole space. \\
	Importantly, the space is not smooth, but still piecewise linear.}
	\label{fig:simpcomp}
\end{figure}
In the classical realm, we consider spaces with some discrete geometric features: It shall be a space made by attaching tetrahedra to each other at their boundaries, giving the faces area values and providing a connection or parallel transport of local frames in the space. Such geometries are sometimes referred to as \textit{twisted}\cite{freidelTwistedGeometriesGeometric2010}.
Each tetrahedral cell will be supposed flat in its interior, so the parallel transport is trivial, while transport across faces will induce nontrivial rotations of frames.\\ Additionally, each face of the cells will have an area value, which, together with the flatness of the interior, will allow one to construct a unique metric and geometry on the whole cell. Providing areas for the faces induces a metric on each cell through embeddings into Euclidean space.
An embedding of a classical tetrahedron into Euclidean space may be described easily by its 4 normal vectors $ J_\alpha \in \mathbb{R}^3 $ ($ \alpha \in [1:4] $), each with length given by the area of its associated equilateral face. These 4 normals are not independent, for they satisfy 
\begin{equation}\label{ClosureConstraint}
	\sum_\alpha \vec{J}_\alpha = 0
\end{equation}
which is conventionally referred to as the geometric \textit{closure constraint}. Notably, we do not impose the nondegeneracy condition that the normal vectors should not be collinear or coplanar. In some states, the tetrahedron might rather look like a triangle or even line.\\
This embedding defines a metric on the tetrahedron. 
There may eventually be nontrivial glueings with other tetrahedra across faces. If we consider a local coordinate vector frame $ \{\vec{e_i}\} $ at the center of it, we may describe the nontrivial embedding into $ \mathbb{R}^3 $ by how this frame is transported to the faces - a flat geometry will have trivial transport. A frame such as this may only be rotated into another frame by this operation, so when transporting a frame from the interior of a tetrahedron to another's, we will have an associated $ \overrightarrow{e'_i} = \overrightarrow{R_f e_i} $ with rotations $ R_f \in SO(3) $. \\
So, we consider the state of a classical tetrahedral cell in our complex to be given by the 4 parallel transports across its faces, as well as the normals to them. This encodes how the simplices of different sizes and orientations, which are intrinsically all identical, are glued together to form a larger space. In a way, we are prescribing the geometric data attached to an immersion 
\begin{equation}\label{Imbedding}
	\Delta_3 \hookrightarrow \Sigma
\end{equation}
of a 3-simplex into the 3-dimensional space $ \Sigma $ we construct, as well as an embedding of each simplex into Euclidean space. By providing said data, we can then actually build the space $ \Sigma $ just from the abstract simplices alone. Quantising the data then leads to a quantum version of the same space.
The quantisation procedure proceeds in analogy to that of a point particle: \\
The classical phase space of a particle is given by $ P= T^\ast Q $, the cotangent bundle of the configuration space $ Q $. The algebra of quantities of interest is given by functions on $ P $, equipped with the  Poisson bracket induced by the canonical symplectic structure on $ P $. A particular focus lies on the subalgebra of polynomials in some coordinates on $ P $, such as the positions $ q $ and momenta $ p $. Quantisation realises/represents (a deformation of) this Poisson algebra as operators on a separable Hilbert space of states $ \hbb $. There are in principle inequivalent representations of this, which turn out to all be unitarily equivalent to one another. One typical choice is to let $ \hbb = L^2(Q) $ and represent polynomial functions $ f(x,p) $ by multiplication and derivative operators, respectively. Another, called the Segal-Bargmann representation, uses different starting coordinates on $ P $,  $ (z,q) $ where $ z= x + i p $ is a complex coordinate. The Hilbert space in question is then $ L^2_{hol}(P) $, where only holomorphic functions of $ z $ are used\footnote{Technically, there is an additional Gaussian weight $ e^{-|z|^2} $ in the inner product, which is irrelevant to our discussion}. We will quantise the tetrahedron by taking the normal vectors $ J $ as coordinates analogous to $ p $, while the role of $ q $ will be played by the parallel transports. These group elements (we will choose them rather from the rotation group's double cover Spin(3) = SU(2)) play the role of coordinates on $ Q = SU(2)^4$. The usefulness of this follows from the differential geometric fact that $ T^\ast G \cong G\times \mathfrak{g} $ for any Lie group, where $ \mathfrak{g} $ is the Lie algebra of the group. 
See each normal vector as an element of $ su(2) \cong \mathbb{R}^3$.  From this, we may identify the 4 normals as giving half of the phase space, corresponding to momenta. As soon as we glue together multiple tetrahedra, the parallel transports will provide curvature defects between faces, allowing for local geometry that is not globally flat.\\
Accepting $ P=T^\ast(SU(2)^4) $ as our classical phase space, we can easily quantise using $ \hbb = L^2(SU(2)^4) $. Upon quantisation, each of the normal vectors becomes a basis for its corresponding Lie algebra component. However, we still have the closure constraint to impose, which was neglected so far - as well as an $ SU(2)_{\text{Diagonal}} $-invariance of the parallel transports, $ (g_1,g_2,g_3,g_4)\sim (hg_1h^{-1},hg_2h^{-1},hg_3h^{-1},hg_4h^{-1}) $. It is easy to interpret this constraint when looking at an exponential element of $ SU(2)^4 $:
\begin{equation}\label{ExponentialElement}
	g(t_\alpha) = \exp(-i \sum^4_{\alpha=1} \vec{t}_\alpha \vec{J}_\alpha) \stackrel{\vec{t}_\alpha \equiv \vec{t}}{=} \exp(-i \vec{t}\sum^4_{\alpha=1} \vec{J}_\alpha)
\end{equation}
\begin{figure}
	\centering
	\includegraphics[width=0.7\linewidth]{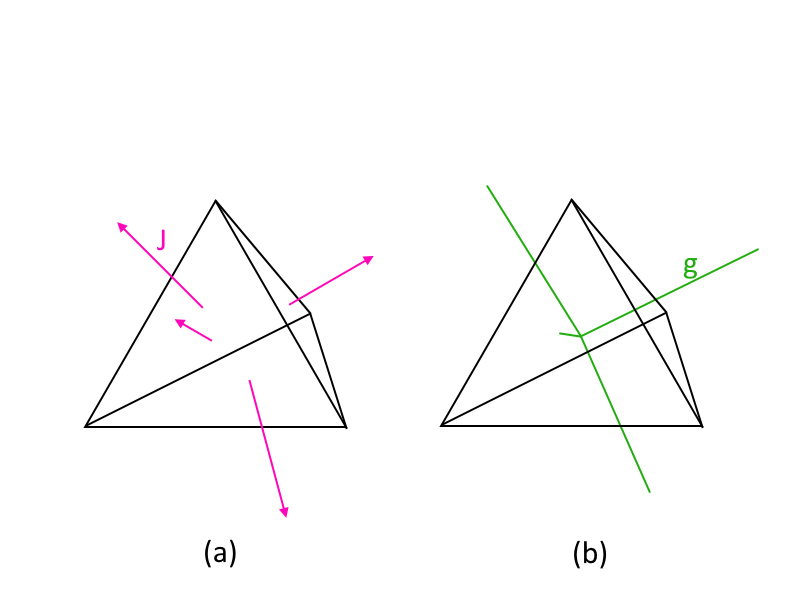}
	\caption{(a) Normal vectors to a tetrahedron, with length specifying the area of the faces. (b) Dual edges labeled by parallel transports.}
	\label{fig:tetra1}
\end{figure}

by taking $ \vec{t}_\alpha \equiv \vec{t} $, the closure constraint becomes the condition that the diagonal subgroup acts trivially, as required. In other words, we rather consider the configuration space and, therefore, the Hilbert space, to be
\begin{equation}\label{ConfSpaceQuant}
	Q= \frac{SU(2)^4}{SU(2)_{Diag}} \qquad \hbb = L^2(\frac{SU(2)^4}{SU(2)_{Diag}})
\end{equation}
where we choose the normalised \textit{Haar measure} of $ SU(2) \cong S_3 $ on the homogeneous space to provide us with the means of integral norms on $ Q $.\\
So, the state of a quantum tetrahedron or 3-simplex can be specified as a square integrable function on 4 $ SU(2) $-variables which is invariant under diagonal conjugation. By the Peter-Weyl decomposition of $ L^2(G) $\footnote{An analogue of the Fourier decomposition of square integrable functions on the circle, which, for compact Lie groups, splits a function into a sum over representations of the group.} for compact groups such as the one we are considering\cite{hallLieGroupsLie2015}, we may write a function of 4 group elements as
\begin{equation}\label{PeterWeylDecomp1}
	f(g_1,g_2,g_3,g_4) = \sum_{j_1,j_2,j_3,j_4\in \mathbb{Z}^+} \sum^{j_\alpha}_{m_\alpha,n_\alpha=-j_\alpha} f^{j_1,j_2,j_3,j_4}_{m_1,m_2,m_3,m_4;n_1,n_2,n_3,n_4} 
	D^{j_1}_{m_1;n_1}(g_1)D^{j_2}_{m_2;n_2}(g_2)D^{j_3}_{m_3;n_3}(g_3)D^{j_4}_{m_4;n_4}(g_4)
\end{equation}
which we abbreviate, due to its cumbersome length, by the following notation: Whenever 4-tuples associated to a vertex appear, they are written as a boldface representative, e.g. $ (j_1,j_2,j_3,j_4) = \mathbf{j} $. Once we work with $ N $ tetrahedra, we will often collect these into yet another vector, denoted by $ \ind{j} = (\mathbf{j}^1,\dots,\mathbf{j}^N) $.
In fact, the more common normalisation of coefficients is given as
\begin{equation}\label{PeterWeylDecomp2}
	f(\mathbf{g}) = \sum_{\mathbf{j}}\sum_{\mathbf{m},\mathbf{n}} f^{\mathbf{j}}_{\mathbf{m},\mathbf{n}} d_{\mathbf{j}} D^{\mathbf{j}}_{\mathbf{m},\mathbf{n}}(\mathbf{g})
\end{equation}
where the dimensions of the representations of $ SU(2) $, labeled by half-integers $ j_\alpha $, have been factored out of the coefficients. In both of these expressions, $ m,n $ label a basis of the representation space of the $ j $-representation, $ V^{j} $. The functions $ D^j_{m;n}(g) $ form such a basis in the $ L^2 $-functions on the group and are known as Wigner matrices. \\
If the closure constraint is satisfied, there exist relations between the coefficients $ f^j_{m;n} $. To find and resolve these, it is useful to look at the Hilbert space decomposition given by Peter-Weyl:
\begin{equation}\label{unconstrainedDecomp}
	\hbb \cong \bigoplus_{\mathbf{j}} V^{\mathbf{j}}\otimes\overline{V^{\mathbf{j}}}
\end{equation}
One may impose the closure constraint by replacing the second factor in each sector/direct summand by $ \mathcal{I}_\mathbf{j}= \text{Inv}(\overline{V^{\mathbf{j}}}) $ (dimension $ \mathcal{D}_{\mathbf{j}} $), the maximal subspace invariant under the diagonal group action. In terms of coefficients, this amounts to specifying a dependence
\begin{equation}\label{IntertwinerDecomp1}
	f^{\mathbf{j}}_{\mathbf{m},\mathbf{n}} = \sum_\iota f^{\mathbf{j}}_{\mathbf{m},\iota} C^{\mathbf{j},\iota}_{\mathbf{n}} \implies
	f(\mathbf{g}) = \sum_{\mathbf{j},\mathbf{m},\iota} f^{\mathbf{j}}_{\mathbf{m},\iota} 
	d_{\mathbf{j}}\sum_{\mathbf{n}}C^{\mathbf{j},\iota}_{\mathbf{n}}  D^{\mathbf{j}}_{\mathbf{m},\mathbf{n}}(\mathbf{g})
\end{equation}
with the $ 3j $-symbol of Wigner, $ C^{\mathbf{j},\iota}_{\mathbf{n}} $. The functions
\begin{equation}\label{IntertwinerDecomp2}
	\langle\mathbf{g}\ket{\mathbf{j} \mathbf{m} \iota} = \psi_{\mathbf{j}\mathbf{m}\iota}(\mathbf{g}) = \sum_{\mathbf{n}}C^{\mathbf{j},\iota}_{\mathbf{n}}  D^{\mathbf{j}}_{\mathbf{m},\mathbf{n}}(\mathbf{g})
\end{equation}
form a basis of the gauge invariant Hilbert space and will be called spin network vertex functions.
The space $ \mathcal{I}_\mathbf{j} $ consists of \textit{intertwiners} between the representations associated to the faces of the 3-simplex, and when a basis of it is expressed in terms of the Wigner matrices, the $ 3j $-symbols provide the coefficients.\\
When viewing the simplex as dual to a \textit{4-valent graph vertex} with its edges piercing the faces, one can associate both holonomies and representations with the 4 edges originating at the center. This dual vertex picture will be central in what follows - it allows us to prescribe a (classical) simplicial complex through a dual, coloured graph.
Let us also mention that the basis $ \ket{\mathbf{j} \mathbf{n} \iota} $ forms a set of eigenvectors of the \textit{area operators} $ A_\alpha = \sum_{\mathbf{j}} \sqrt{j_\alpha (j_\alpha +1)}\ket{\mathbf{j} \mathbf{n} \iota}\bra{\mathbf{j} \mathbf{n} \iota} $ acting on one of the edges each. There is also an analogue of a \textit{volume operator} given by a properly ordered quantisation of the triple product $ \vec{J}_1\cdot(\vec{J}_2\times \vec{J}_3) $ that turns out to be diagonal in the intertwiner basis. In this way, the tetrahedral quantisation and Peter-Weyl decomposition leads to the \textit{spin basis} in which areas of faces and volumes take sharp values. The same cannot be said about edge lengths or angles, however - not all quantities commute. Therefore, the geometry of glued quantum simplices will not be exactly that of a simplicial complex, but it will show definite values for areas and volumes in spin network vertex states. Angles, edge lengths and other quantities will not be sharp.\\
This is a picture of a quantum geometry, in the sense that measurements of properties of the system will not commute and present a Markov process. Measuring an edge length of the tetrahedron will in general produce random results. One can still, however, speak about general statistical properties of these measurements like averages, variances etc.

\subsection*{Glueing of tetrahedra}
Having established the single-tetrahedron Hilbert space, we now want to glue them together to form larger complexes, according to a minimal procedure in terms of matching their areas up by perfect correlation. For concreteness, we will focus on the case of $ N=2 $ tetrahedra. We wish to glue the tetrahedral states along the faces coloured $ \alpha = 0 $. In the spin basis, where the faces are labeled by areas $ j_\alpha $, this involves fixing the areas of the two faces to be the same, as well as making their orientation numbers $ m $ match oppositely.\footnote{Another alternative view is to require the normal vectors to be opposite to each other: $ \vec{J}_x+\vec{J}_y =0 $, which implies the gauge invariance shown below.} As such, in the spin decomposition of a product state of the two,
\begin{equation}\label{SpinDecomp2Tetras}
	\ket{\Psi_1}\otimes\ket{\Psi_2} = \sum_{\mathbf{j}^x \mathbf{m}^x \iota^x}\sum_{\mathbf{j}^y \mathbf{m}^y \iota^y}\Psi_{1,\mathbf{j}^x \mathbf{m}^x \iota^x}\Psi_{1,\mathbf{j}^y \mathbf{m}^y \iota^y} \ket{\mathbf{j}^x \mathbf{m}^x \iota^x}\ket{\mathbf{j}^y \mathbf{m}^y \iota^y}
\end{equation}
we insert a Kronecker delta $ \delta_{j^x_0,j^y_0}\delta_{m^x_0+m^y_0,0} $ forcing the two faces to be glued. The same procedure can be applied to non-product states as well.\\
\begin{figure}
	\centering
	\includegraphics[width=0.7\linewidth]{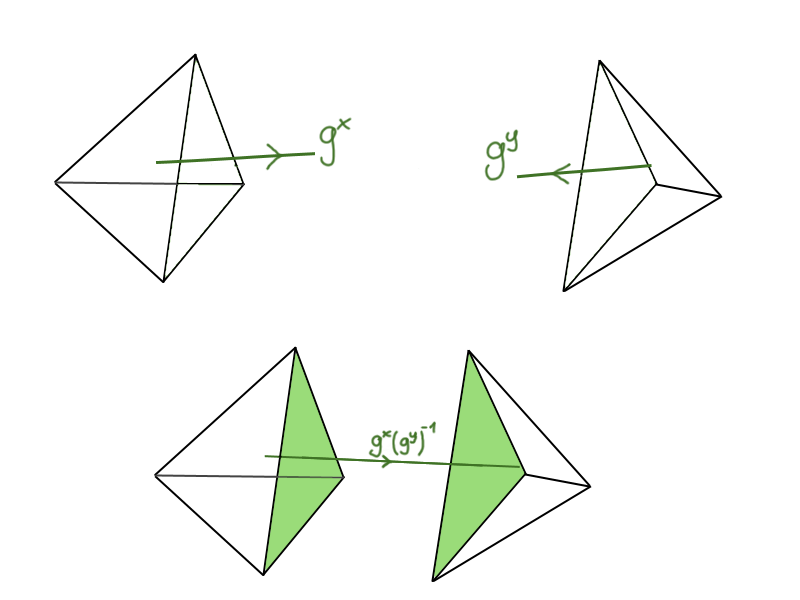}
	\caption{Even if the two tetrahedra have initially independent parallel transports, after glueing they only have a single group element between them.}
	\label{fig:tetra2}
\end{figure}

Alternatively, we may look at this as a projection process: Consider the group basis both before and after a glueing. While the two unglued links/faces of the tetrahedra both possess a seperate parallel transport element, once they are glued we expect there to be only a single link and thus only a single parallel transport. So, a glueing in the group basis amounts to removing the dependence on one of the two group elements in an indiscriminate fashion. In fact, if we see the parallel transport defined by an element $ g $ as outgoing, the glued link from vertex $ x $ to $ y $ should have parallel transport first by the element $ g^x $ and then the reverse transport of $ g^y $. The sought-for dependence on $ g^x (g^y)^{-1} $ is easily produced by a right average: Consider, for purposes of demonstration, only a function of these two elements, $ f(g^x,g^y) $, and perform a right group action average:
\begin{equation}\label{RightAverage1}
\begin{split}
	\tilde{f}(g^x,g^y) = \int_{SU(2)} d\mu(h) f(g^xh,g^yh) \\= 
	\int_{SU(2)} d\mu((g^y)^{-1}\tilde{h}) f(g^x (g^y)^{-1}\tilde{h},\tilde{h}) 
	= \int_{SU(2)} d\mu(\tilde{h}) f(g^x (g^y)^{-1}\tilde{h},\tilde{h})
\end{split}
\end{equation}
where in the last equality we used left invariance of the Haar measure. This averaged function only depends on the correct combination and has full right-invariance:
\begin{equation}\label{RightAverage1}
	\tilde{f}(g^xh,g^yh) = \tilde{f}(g^x,g^y) \, \forall h \in SU(2)
\end{equation}
We now have a projection operator that performs the glueing, expressed in the group basis. In the spin basis, this same operator is rather a projection onto a \textit{maximally entangled state} of the two links:
\begin{equation}\label{MaxEntState1}
	\ket{e_{j_e}} = \frac{1}{\sqrt{d_{j_e}}} \sum^{j_e}_{m_e = -j_e} (-1)^{j+m}\ket{j_e m_e}_x \otimes \ket{j_e -m_e}_y
\end{equation}
and the projection is given by $ \ket{e}\bra{e} $. In this state, the two subsystems share their information perfectly. It is called maximally entangled as for each fixed spin, the reduced density matrix 
\begin{equation}\label{MaxEntState2}
	\rho_x = \Tr_y[\ket{e_j}\bra{e_j}] = \mathbb{I}_j
\end{equation}
gives the maximally mixed state - the entanglement entropy reaches its maximal value
\begin{equation}\label{MaxEntState2}
	S(\mathbb{I}_j) = \log(d_{j})
\end{equation} 
given by the logarithm of the dimension.\\
With this in mind, we can glue further simplices together by chaining these projectors together. 
For a product of vertex basis states, after projection a state of the form
\begin{equation}\label{ProjOntoMaxEntState}
	\bigotimes_{e \in E} \bra{e} \bigotimes_x \ket{\mathbf{j}^x \mathbf{m}^x \iota^x}
\end{equation} 
emerges. Such a state is what we will call a \textit{spin network state}, as the 4-valent graph dual to the simplicial complex gives rise to a spin network with representation labels $ \ind{j} $, intertwiners $ \vec{\iota} $ and open edge labels $ \{m_e:e\in\bnd\} $. Expressed as a function of group elements, this will in fact be a diagonally invariant function of parallel transports along the edges of the graph, as the glueing procedure gives each link in the graph precisely one group element.

Since they apply to different links, the projectors for different links commute and do not interfere with each other. In this work, we will only perform \textit{coloured} glueings: Each factor in $ SU(2)^4 $ is assigned one of 4 colours, whose corresponding links may be glued together. This means that the class of graphs dual to our simplicial complexes are \textit{4-edge-coloured, 4(or 1 at boundary)-valent graphs}. The colouring, in a very direct way by the ordering it induces, produces an orientation of the simplices and thus of the simplicial complexes. In this way, we study states labeled by \textit{oriented, pure simplicial 3-complexes}. This class may be motivated by starting from a smooth oriented 3-manifold, which always admits a \textit{smooth oriented triangulation}. While not all simplicial complexes used as labels here are homeomorphic to manifolds, they are to \textit{oriented pseudomanifolds}. They therefore provide a controllable set of labels for states by spatial topologies that extends that of manifolds. However, due to the quantisation, it is a priori not so clear whether these quantum geometries will behave like classical ones. Already, even spin network states, once they are superposed, can not be expected to behave like nice geometries. \\
The glueing procedure can be understood in terms of measurements as follows: If measurements of two glued faces of tetrahedra are possible independently, their results will show perfect correlation. It is worth emphasizing that the question of \textit{proximity} of the two tetrahedra is not answered in this way. The only property matched between the tetrahedra is their shared area - no mention of lengths has been made, per se. This is, however, the notion of glueing that reproduces spin network states.
\subsection*{An aside on constraints and spacetime dynamics}
Before moving on to other elements of our investigation, it is worth to give a short remark about the relation to space\textit{times}. As a canonical Hamiltonian analysis of (pure) gravity on globally hyperbolic spacetimes $ M = \mathbb{R}\times \Sigma $ shows, the Hamiltonian of classical GR consists only of a number of classical constraints. Let $ q,p_q $ collect whichever other regular variables there may be, for example spatial metrics and extrinsic curvature. Then, the transition from the Lagrangian to the Hamiltonian formalism introduces canonical momenta $ p_q $ for each variable $ q $ through $ p_q = \frac{\partial L}{\partial \dot{q}} $. This may turn out to not be a function invertible for the velocities $ \dot{q} $, in which case one instead imposes a corresponding constraint on the classical phase space, $ C(q,p_q) = 0 $. A simple example may come from the point particle Lagrangian
\begin{equation}\label{ExampleLagrangina}
	L = \dot{q} e A(q)
\end{equation}
which can be interpreted as a particle of 0 mass coupled to an electromagnetic background $ A $. The constraint here is $ p = e A(q) $. Assuming that $ \dot{q} $ will be given by some function on the phase space, the Hamiltonian is.\\
\begin{equation}\label{key}
	 H(q,p) = \dot{q}p - L = \dot{q}(p- e A(q)) 
\end{equation}
In order for the constraint to hold at all times, one requires 
\begin{equation}\label{constraint}
	\{p-eA,H\} = (p-eA)\{p-eA,\dot{q}\} = 0
\end{equation}
which is automatically true on the constraint surface. In fact, one may achieve the same thing by introducing a new pair of conjugate variables $ N,P_N $ instead of seeing $ \dot{q} $ as a function on phase space. If $ H = N (p-eA(q)) $, then $ \dot{q} = \{q,H\} = N  $, while 
\begin{equation}\label{key}
	\dot{N} = \{N,H\} = 0, \dot{P_N} = \{P_N,H\} = -(p-eA(q)),
\end{equation}
which means that (1) the time evolution of $ q $ is actually arbitrary and (2) the constraint being imposed is equivalent to the conjugate momentum $ P_N $ being constant.\\
In general relativity without sources, one encounters the same type of situation.\cite{} There are a number of Lagrange multipliers $ N_\alpha $ and constraints $ C_\alpha $ and the Hamiltonian takes the form
\begin{equation}\label{GRConstraint}
	H(N_\alpha,P^\alpha_N; q,p_q) = \int_\Sigma vol_\Sigma(x) \sum_\alpha N_\alpha(x) \,  C_\alpha(q,p_q)(x)
\end{equation}
Being a Lagrange multiplier means that the conjugate momenta $ P_N $ must be constant.
Since they themselves do not appear, their equations of motion $ \frac{\partial H}{\partial P^\alpha_N} = \dot{N_\alpha} = 0$ show that the values of these multipliers are arbitrary. \\
This implies, through Poisson brackets, that $ C_\alpha (x) = 0 $ at all times and for all points in the spatial surface.\footnote{It is assumed that the constraints $ C_\alpha $ already form a closed Poisson subalgebra, so that no further constraints are generated.}\\
In fact, these constraints are dual (by Noether's second theorem) to the local symmetries of gravity: Particularly, spatial and timelike, foliation-preserving diffeomorphisms\footnote{These symmetries map the fields to their pushforwards under a diffeomorphism of the manifold.} are a symmetry of gravity. These local symmetries induce corresponding local constraints, analogous to how gauge invariance of classical electromagnetism gives rise to the electric Gauss law $ \nabla\cdot E = \rho_e = 0 $.\\
The constraints on gravity's naive phase space (often called kinematical phase space) are fittingly called the \textit{spatial and timelike Diffeomorphism (or Hamiltonian) constraints} as well as a \textit{Gauss constraint} in setups where spin geometry is used and bundle automorphisms or gauge transformations of the spin frame bundle provide another symmetry. \\
Given this information, we can see that the time evolution of a spatial geometry is, in simple words, \textit{fictitious}. Even after fixing some values of the Lagrange multipliers, once a spatial geometry is found to satisfy the constraints, the Hamiltonian will act trivially upon it. A way to reclaim the 'naive' equations of motion is to fix values of the multipliers and compute the equations of motion before using the constraints. In our earlier example, this is given by
\begin{equation}\label{key}
	\dot{q} = N = \text{const}, \qquad  \dot{p} = e\nabla_q A = \frac{e}{N} \frac{d}{dt}A(q) \implies \qquad p-\frac{e}{N}A(q) =  \text{const}
\end{equation}

The choice of a time evolution is still, however, an arbitrary one, as the Lagrangian evolution is trivial. The first and foremost task of quantum gravity is therefore to impose, in a meaningful, productive and controlled way, the constraints on the kinematical phase space of the starting variables and to reconstruct spatial geometries with the right data to produce a regime where smooth spaces appear as a reasonable approximation.\\
Let us now introduce a few contexts in which spin networks naturally appear to give a perspective on their usefulness.
\subsection{Lattice gauge theories and their loop representation}
The first practical context in which we would like to present spin networks is for quantised lattice gauge theories in the loop representation. For our purposes, a lattice gauge theory is a system with degrees of freedom given by a connection 1-form $ A_\mu $ analogous to Yang-Mills or Electromagnetism, living on an oriented lattice $ \Lambda $. Gauge invariance means that the only quantities of physical interest are those lattice objects which are somehow related to the curvature. The elementary variables in the classical system are given by \textit{parallel transports} of the gauge potential $ A $ along edges or longer paths $ e $ in the lattice:
\begin{equation}\label{ParTransport1}
	h_e(A) = \mathcal{P}\exp(\int_e A) \in G
\end{equation}
These gauge group variables describe how to compare local gauges, say for matter living on the lattice that is charged under the group $ G $. As can for example be seen most easily in the abelian case where 
\begin{equation}\label{ParTransport2}
	\int_e A \mapsto \int_e A + \int_e d\chi = \int_e A + \chi(y)-\chi(x) , 
\end{equation}
under a continuum gauge transformation the potential changes as  $ A(x) \mapsto S^{-1}(x) (A(x)+ d)S(x) , \, S(x)\in G $, only the endpoints experience a change: 
\begin{equation}\label{ParTransport3}
	h_e(A) \mapsto S(y)h_e(A)S(x)^{-1}
\end{equation} 
Clearly, if we choose the path to have the same start- and endpoint, the resulting group element will be just conjugated under gauge transformations. These group elements associated to loops are commonly known as \textit{holonomies}. In the loop representation of lattice gauge theories, their traces are chosen as elementary, gauge invariant variables for quantisation.\\
By starting from a Segal-Bargmann type representation of the variables, gauge invariance can be automatically imposed on the states upon quantisation. Choose as coordinates on the (infinite dimensional) classical phase space the traces of all the above holonomies as well as the canonically conjugate \textit{fluxes} $ E $, analogous to the electric field in canonical quantisation of Maxwell theory, satisfying $ \{A,E\}\sim 1 $ with the non-exponentiated connection.\\
These form an analogy to the $ (z,q) $ variables presented earlier.
Then, the space of physical quantum states will be given by gauge invariant, square integrable functionals $ F[A] $ of the connection $ A $. It can be shown that any such functional may be understood as some composition of various holonomy maps with different loops, in the following way: Let $ H: \Omega \Lambda \times \mathcal{A} \rightarrow G$ be the holonomy map from loops and connections to group elements. Then, any gauge invariant functional will be given by taking some combination of $ H(\gamma_i,-) $'s, turning them into numbers using a trace, and lastly evaluating them all simultaneously on $ A $.
We obtain a set of states spanning the Hilbert space:
\begin{equation}\label{Loop1}
	\ket{\{\gamma_i\}_i} \leftrightarrow (A\mapsto \Tr[\prod_i H_{\gamma_i}(A)E_i^{b_i}])
\end{equation}
which are the functionals associated to a set of loops $ \{\gamma_i\}_i $ and booleans $ b_i \in \{0,1\} $ indicate that fluxes may be inserted at arbitrary points along the loops. A great feature of constructing the Hilbert space from this set is that all states are automatically gauge invariant and thus physical. However, it is overcomplete: There are relations between traces of matrices called \textit{Mandelstam identities} which mean that some of the above states may be linearly related to one another\cite{rovelliSpinNetworksQuantum1995}. \\
Thus, to obtain a basis of the gauge invariant Hilbert space, one needs to superpose these loop states to obtain linearly independent states. By a definite graphical procedure, one can associate to each $ G $-spin network embedded in the lattice such a superposition of loop states. And, when collecting these states over all spin networks, one obtains a basis of the gauge invariant Hilbert space. These basis states are appropriately named $ G $-\textit{spin network states} and form a natural description of the quantum degrees of freedom of a discretised gauge theory. 
\subsection{Spin network states in canonical Loop Quantum Gravity}
An analogous story introduces spin networks in quantum gravity\cite{thiemannIntroductionModernCanonical2001,rovelliQuantumGravity2004}. The typical starting point of canonical quantisation of GR is to foliate a globally hyperbolic spacetime into spatial slices, which leads to a 3+1D split of the dynamical variables. The main variables are the spatial 3-metrics on the slices as well as extrinsic curvatures from the embedding. Using a sequence of canonical transformations\cite{henneauxDerivationAshtekarVariables1989}, one can transform these and the Hamiltonian into expressions dependent on connection variables, analogous to the Palatini formulation of GR, but clearly distinct: A convenient parametrisation turns out to be in terms of the complexified, self-dual Ashtekar-Barbero-Sen (ABS) connection $ A $ and a densitised spatial triad $ E $ \cite{ashtekarNewHamiltonianFormulation1987}. This is a complex linear combination of the 3D Levi-civita spin connection and the extrinsic curvatures, most easily understood in the time gauge, where representing the 4D spin connection $ \omega $, valued in $ \text{spin}(1,3) $, on left-handed Weyl spinors yields the ABS connection 
\begin{equation}\label{LHrep}
	A = (\rho_{LH})_\ast (\omega) = (\frac{1}{2}\omega^{ij}\epsilon_{ijk} + i \omega^0_k ) \sigma_k
\end{equation}
by seeing the Pauli matrices as a representation of $ su(2) $. So, it may be understood as the connection for left-handed spinors on the spacetime. On the other hand, the \textit{self-duality} is to be understood when looking at the complexified Lie algebra:
\begin{equation}\label{liealgsplit}
	\text{spin}(1,3)_\mathbb{C} \cong \text{spin}(3)_\mathbb{C}\oplus\text{spin}(3)_\mathbb{C} \cong \text{spin}(4)_\mathbb{C}
\end{equation}
in which the duality is in terms of a volume form on $ \text{spin}(4) $, in a basis described by the Levi-Civita symbol
\begin{equation}\label{internaldual}
	(\star\omega)_{ab} = \frac{1}{2}\epsilon_{abcd}\omega^{cd}.
\end{equation}
Using this connection variable, the constraints of the classical theory due to diffeomorphisms take a much more amenable form. To maximise its usefulness, one also uses as a conjugate variable the densitised spatial triad $ \tilde{E} $. Both of these variables are understood to take values in the complexified algebra $ su(2)_\mathbb{C} $ and will have to be restricted by so-called \textit{reality conditions}\cite{thiemannRealityConditionsInducing1996} to recover real Lorentzian GR.\footnote{In practice, this approach of complexification turns out to be quite involved as the reality conditions are difficult to implement. Instead, other variants of the Ashtekar connection are used which depend on the Immirzi parameter of the Holst action, but for the purpose of exposition we decide to stick to this more historical idea.}
With these variables, one can quantise in strong analogy to a lattice gauge theory for $ SU(2) $. Using again the Loop representation, one can construct gauge invariant states of the system as holomorphic functionals of $ A $, again with a basis in correspondence with spin networks.\cite{rovelliSpinNetworksQuantum1995} The key differences to \textit{lattice} gauge theories are the following:
\begin{itemize}
	\item Instead of being embedded in a lattice, the spin networks are equipped with an imbedding in the spatial slice $ \Sigma $ through which they can encode knotting information among other things.
	\item In LQG, there are additional constraints on the gauge invariant states to be considered physical, given by the Diffeomorphism and Hamiltonian, Scalar or Master constraints. The former set of these can be implemented directly on spin network states, leading to states labeled by s-knots, which are akin to isotopy classes of imbedded spin networks and depend on generalised link invariants of the spatial slice.
\end{itemize}
In particular, the action of the Hamiltonian and Diffeomorphism constraints is significantly simpler on this basis compared to the quantisation program in metric variables. In fact, several solutions to the Hamiltonian constraint may be found in this basis, providing physical states of quantum general relativity.\\
Additionally, the quantisation of area and volume operators defined by analogy from classical GR turns out to be diagonal in spin network states - representation labels correspond to areas and intertwiners to volumes. Therefore, these states of quantum GR have some definite geometric properties. However, they should not be confused for exact classical geometries, either, as lengths do not attain definite values. The situation is therefore entirely analogous to the states labeled by simplicial complexes we introduced earlier\cite{oritiGroupFieldTheory2014,colafranceschiQuantumGravityStates2021}.\\
The natural occurrence of spin network states in canonically quantised general relativity gives credence to the idea of them corresponding to states of quantum geometry. However, the attached imbedding remains a piece of information from the continuum, which means that even still, quantum geometry is partially dictated by a classical one. In response to problems connected to the continuum and, in part, due to hopes of recovering GR from purely discrete systems, it has been suggested that one should rather start from a microscopic model based on spin network states without imbedding data and recover a continuum space in some limit. There have also been ideas that such data may be incorporated in spin network states through a change of data from Lie groups to Hopf algebras, fully deconfining the idea of quantum from classical geometry.\cite{smolinFutureSpinNetworks1997}
\subsection{Spin foam models and Group Field Theories}
\begin{figure}
	\centering
	\includegraphics[width=1.1\linewidth]{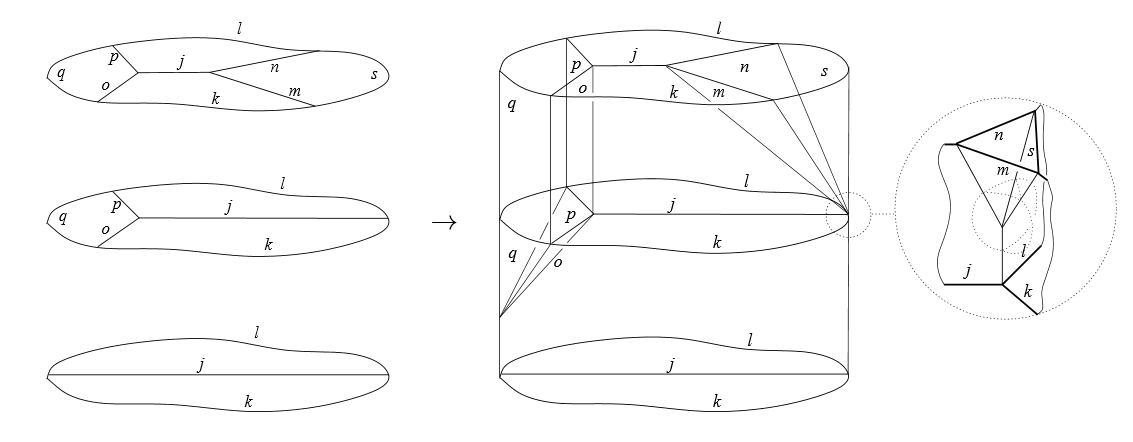}
	\caption{A sequence of 3 spin networks, connected by a spin foam with compatible labels. There are two \textit{spin foam vertices} in the 2-complex introducing new structures as the spin networks evolve discretely. A closeup of these vertices is given on the right.\\ Note: Reprinted from "The Spin Foam Approach to Quantum Gravity"\cite{perezSpinFoamApproach2013a}, p. 29, CC BY-4.0}
	\label{fig:spinfoamslice}
\end{figure}

\begin{figure}
	\centering
	\includegraphics[width=0.4\linewidth]{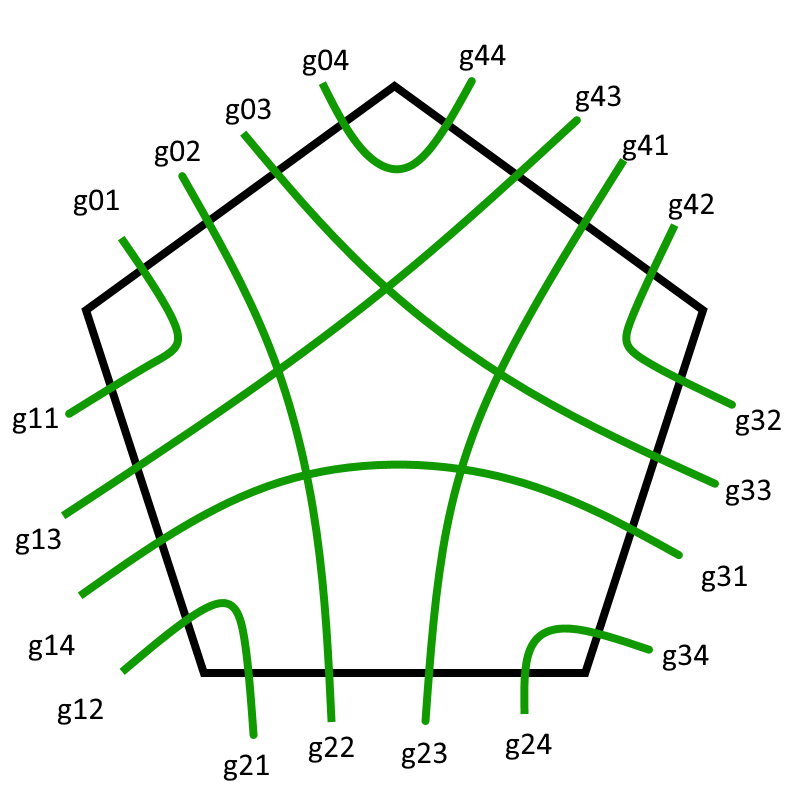}
	\caption{Schematic of how contractions happen in a 4D uncoloured simplicial interaction. The arguments of the 5 involved group fields, four each, are contracted with group Dirac delta functions $ \delta(g h^{-1}) $ according to the green lines connecting the arguments in this schematic. \\
	The purpose of this combinatorial scheme is that, interpreting the 4 arguments of each field as the parallel transports along faces of a tetrahedron, this mimicks the glueing of 3-simplices to form a 4-simplex.}
	\label{fig:pengaton}
\end{figure}

Whether the spin network states one wishes to consider are from LQG or abstract, the question of dynamics for these states remains. For example, the Hamiltonian constraint\footnote{Rather, one might impose instead the Master constraint $ M = \int d^3x \frac{H(x)^2}{\sqrt{|h|(x)}} $ which imposes all the Hamiltonian constraints simultaneously.} may be imposed by a standard oscillatory integral representing a projection operator: 
\begin{equation}\label{key}
	P = \int \mathcal{D}N \exp(i N H)
\end{equation}
Applied to a state, all but the kernel of $ H $ will be washed out by the integral over the 'Lapse function' $ N $. 
Then, the exponential operator may be turned into a path integral by insertions of resolutions of the identity through spin network states. The Feynman diagrams of such a path integral then correspond to discrete 2-complexes called \textit{spin foams}\cite{perezSpinFoamApproach2013a}. Independently from this argument, this same structure has been proposed by Baez \cite{baezIntroductionSpinFoam1999,baezSpinFoamModels1998} as a categorical arrow between spin network objects. In this general definition, a spin foam $ S $ mapping spin networks $ s_1 $ into $ s_2 $ is
\begin{enumerate}
	\item A directed 2-complex $ \Gamma $ with boundary given by the graphs of $ s_{1,2} $.
	\item A labeling of its 2-cells (corresponding to time evolutions of edges in a spin network) by representations compatibly with the bounding spin networks.
	\item A labeling of its 1-cells (corresponding to time evolutions of vertices in a spin network) labeled by intertwiners, also chosen compatibly.
\end{enumerate}
In this definition, compatibility is to be understood as choosing the dual representation for outgoing edges of a 2-face, etc. and one may make adjustments for spin networks with boundaries.\\
Given this definition, one can define a scalar product between spin network states through a \textit{spin foam model}:
\begin{equation}\label{key}
	\avg{s_1,s_2} := \sum_{S : \partial S = s_1 \cup \bar{s_2}} \mathcal{A}(S)
\end{equation}
So, an amplitude is given for each spin foam mapping between the two spin networks and a sum is performed over all such spin foams. This is entirely analogous to the sum over Feynman diagram amplitudes in a normal relativistic QFT. 
In this context, spin network states play the role of scattering states in the relativistic QFT, which are themselves geometrically given by the points of the point particles. Any 'time-slice' of the Feynman diagram will give another configuration of points. Similarly, any slice of a spin foam will yield a spin network, as in figure \ref{fig:spinfoamslice}.\\
 However, there is no scattering process happening - this should all be understood as in the boundary formulation of QFT, where transition amplitudes between boundary states are all that needs to be specified for theories to make sense\cite{oecklGeneralBoundaryQuantum2008}. Specifying the amplitude for any two spin network states will therefore give the full definition of a quantum system whose Hilbert space is spanned by these states.\\
Spin foam models can then also be taken as an independent, possibly covariant alternative quantum theory of geometry, in which the spin foam takes on the identity of a quantum pre-spacetime. In certain limits, some spin foam models have been shown to indeed produce the semiclassical behaviour expected of a lattice gravity path integral.\cite{perezSpinFoamApproach2013a} In this setting, one has only few restrictions on which types of boundary states one should use. 
Past the specification of an amplitude for each spin foam, the question of recovering a classical geometry remains. There are essentially two options here: The first is to consider limits of refinement of the spin foam and networks. The other is to construct a sum over spin foams that organises these in a controlled fashion. One systematic way of doing the latter is through \textit{Group Field Theories} (GFT)\cite{krajewskiGroupFieldTheories2012,oritiGroupFieldTheory2014,oritiMicroscopicDynamicsQuantum2011}:\\
Interpreting spin foams as Feynman diagrams allows one to consider the transition amplitudes given by a spin foam model in a productive way. One sees them as contributing to a correlation function of two \textit{spin network operators} in a quantum field theory. Said QFT needs to be defined on a number of copies of the Lie group $ G $. This can be most easily seen by another analogy to scattering diagrams: There, labels on the edges (analogies of the 2-faces in a spin foam) are given by momenta, which are representations of the translation group $ \mathbb{R}^4 $. The QFT's Hilbert space can be shown to be spanned by abstract spin network states, similar to the situation in LQG. In fact, when the QFT is taken to be unitarily equivalent to a free one, the Hilbert space is precisely the Fock space $ \mathbb{F}(L^2(\frac{G^4}{G_{Diag}})) $, built on the same Hilbert space of the single tetrahedron that we constructed in the beginning. The analogue of the diagonal group action in the relativistic case is given by the mass constraint $ p^2 = m^2 $, reducing $ L^2(\mathbb{R}^{1,3})  $ to $ L^2(\mathbb{R}^{1,3}/\mathbb{R}) \cong L^2(\mathbb{R}^{3}_{m^2}) $.\\
The dynamics of such a GFT are given, in general, by a path integral of some form, which specifies a partition function\cite{chircoStatisticalEquilibriumTetrahedra2019,livineEffectiveHamiltonianConstraint2011}
\begin{equation}\label{key}
	Z = \int \mathcal{D}\Phi\mathcal{D}\bar{\Phi} e^{-S(\bar{\Phi},\Phi)} 
\end{equation}
where the \textit{group fields} $ \Phi: G^D \rightarrow \mathbb{C} $ are to satisfy some gauge invariance condition. In the free case, these fields give rise to creation/annihilation operators associated to tetrahedra. There are many different ways these systems relate to other quantum gravity approaches, due to the many different representations that (wave)functions on a group can have.\cite{baratinGroupFieldTheory2012a} In particular, the Peter-Weyl decomposition allows extracting a spin foam model from any GFT's Feynman diagrams.\\
The specifics of the dynamics depend largely on which class of spin foams are to be generated in the perturbative expansion. One may include multiple fields and several interaction terms, which all should correspond to some building block of the spin foam. A particularly simple class of these interactions are the \textit{simplicial} ones, in which 5 tetrahedra are glued to form a 4-simplex. In group variables, it takes the schematic form
\begin{equation}\label{key}
	\mathcal{V} = \int_{G^{20}} d\mathbf{g_0}d\mathbf{g_1}d\mathbf{g_2}d\mathbf{g_3}d\mathbf{g_4} \Phi(\mathbf{g_0})\Phi(\mathbf{g_1})\Phi(\mathbf{g_2})\Phi(\mathbf{g_4})\Phi(\mathbf{g_5}) V(\mathbf{g_0},\mathbf{g_1},\mathbf{g_2},\mathbf{g_3},\mathbf{g_4})
\end{equation}
where the combinatorial factor $ V $ is visually encoded in a stranded diagram as in Figure \ref{fig:pengaton}.\\
Spin network states, as we have seen, can also be understood as glued or entangled tetrahedral states. In this way, they form an analogue of entangled many-body particle states of the GFT. This holds not just for GFT states, but in generality for all spin network states regardless of the approach.\cite{colafranceschiQuantumGravityStates2021} In the GFT formalism, though, the connections to entanglement are perhaps more transparent. However, even discarding the imbedding information in LQG and staying at the prefock (non-symmetrised) space level of the (free) GFT, the way these states are organised in the full Hilbert space is very different between approaches: While, for fixed number of vertices the scalar products agree in the two theories, GFT spin networks are orthogonal if their number of vertices is not identical\cite{oritiGroupFieldTheory2015}. This may be, in part, understood as a difference in dynamics: The free GFT does not have the same physical scalar product as LQG as it does not have the proper constraints imposed on its states. However, the dynamics of the GFT may effectively implement a different scalar product to make the relation\cite{livineEffectiveHamiltonianConstraint2011}.\\
To understand the function of spin networks as tensor networks, we first give a bit of background on tensor network states. Then, we review classes of holographic codes and their natural progression towards random tensor networks, which form the foundation of this project.
\subsection{Spin network states as tensor networks}
A tensor network\cite{orusPracticalIntroductionTensor2014}, for our purposes, is a multidimensional array 
\begin{equation}\label{key}
	T^{\mu_1,\dots\mu_n}_{\nu_1,\dots,\nu_n} \stackrel{ex.}{=} \sum_{\alpha_1,\dots,\alpha_n} \prod_i V^{\mu_i \alpha_i}_{\nu_i} \Pi_{\alpha_1,\dots,\alpha_n}
\end{equation}
composed from smaller arrays of various sizes by contraction according to rules specified by a graphical calculus. Above, we have given a very simple example where a rank (n,n)-tensor is decomposed into n tensors $ V $ of rank (2,1) and contracted using another tensor $ \Pi $.\\
A corresponding concept in physics is that of a \textit{tensor network state}: Starting from a factorised Hilbert space, the simplest types of states one can form from those of subsystems are tensor product states. Tensor network states form a natural, but controlled extension of this class where the subsystem states function as tensor blocks and are contracted or projected in a way that, when expressed in bases of the subsystems, will produce a tensor network.
A simple class of these states is given by so-called \textit{projected entangled pair states}(PEPS)\cite{ciracRenormalizationTensorProduct2009b,orusPracticalIntroductionTensor2014}: Elementary vertex states $ \ket{V_x} $ are projected onto a set of maximally entangled link states (also known as EPR pairs) $ \bigotimes_e \ket{e} $ that encode a connectivity pattern.\\
Consider, for motivation, a system of up/down spins $ S_x $ on a lattice (V,E), say a rectangular one. The total Hilbert space factorises into 
\begin{equation}\label{PEPSampleHS}
	\hbb = \bigotimes_{x\in V} \mathbb{C}^2_x
\end{equation}
but a typical Hamiltonian will respect the lattice structure of the system in some form, for example by only including nearest-neighbour interactions
\begin{equation}\label{key}
	\hat{H} = \sum_{e \in E} \lambda_e \frac{1-\hat{S_x}\hat{S_y}}{2}.
\end{equation}
This type of Hamiltonian is \textit{local} in the sense that any site has vanishing\footnote{More generally, a power-law or exponential decay with spatial distance suffices for qualitatively similar statements.} interaction with other sites outside of a finite region. Sites that are two edges apart will only couple indirectly.\\
In the above Hamiltonian, a ground state is fairly easy to find: if, say, all edge couplings $ \lambda_e>0 $, then the energy is minimised by having all spins equal. This is a simple product state
\begin{equation}\label{key}
	\ket{FM}= \bigotimes_{ x} \ket{\uparrow_x}.
\end{equation}
However, if the couplings are inhomogeneous and some are negative, the ground state is not known in general. An approach to approximate it is to add auxiliary Hilbert spaces $ V_{x,\alpha} $ to the sites, two for each, left and right. These may be intertwined with the existing site vertex Hilbert space $ \mathbb{C}^2_x $ by writing the full state in $ \hbb  $ as a tensor network state in tensor blocks coming from $ V_x  =\otimes_\alpha V_{x,\alpha} $. Then, the tensors have auxiliars indices ("legs") that can be contracted in accordance to the shape of the lattice or some sublattice. The auxiliary legs of two adjacent sites are projected onto a \textit{maximally entangled state} $ \ket{e} $, for example
\begin{equation}\label{key}
	\ket{e_{xy}} = \frac{1}{\sqrt{D}} \sum_m \ket{m}_x\ket{m}_y,
\end{equation}
which amounts to contracting the indices in the tensor network with Kronecker deltas. For a 1-dimensional lattice with $ N $ sites, placing vertex tensors $ T^{x,s^x}_{t^x_0,t^x_1} $ on each site with auxiliary legs $ t^x_0,t^x_1 $, this might look like
\begin{equation}\label{key}
	\ket{\Psi} = \sum_{s^1,\dots,s^N \in \{\pm 1\}}
	\Psi^{s^1,\dots,s^N} \ket{s^1,\dots,s^N}
	\qquad
	\Psi^{s^1,\dots,s^N} = \sum_{\{t^x_\alpha\}} \prod_{x} T^{x,s^x}_{t^x_0,t^x_1} \prod_{i<j} \delta_{t^i_1,t^j_0}
\end{equation}
where we have obtained this state as
\begin{equation}\label{key}
	\ket{\Psi} =  \bigoplus_{i<j} \bra{e}_{ij} \bigotimes_{x}\ket{T^x}.
\end{equation}
\begin{figure}
	\centering
	\includegraphics[width=0.7\linewidth]{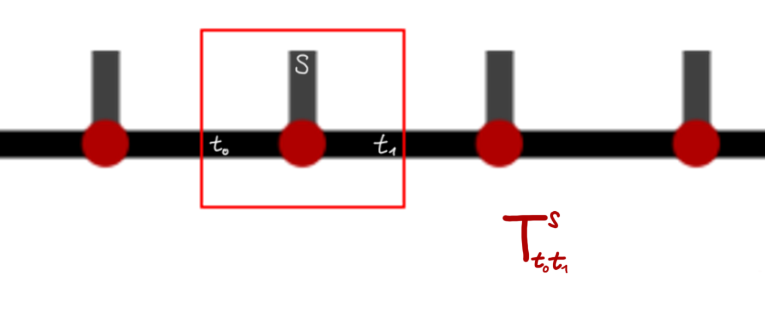}
	\caption{Tensor network contraction of tensors $ T^s_{t_0,t_1} $ into a 1D chain. Each left auxiliary leg, labeled by $ 0 $, is contracted with the right one of the site on its left, labeled by $ 1 $. Free legs give the physical indices.}
	\label{fig:tnexample1}
\end{figure}

The above general form of a product state projected onto a set of maximally entangled link states is a PEPS.\\
By itself, the product state $ \bigotimes_{x}\ket{T^x} $ has no entanglement between the site spaces $ \tilde{\hbb}_x \cong \mathbb{C}^2_x \otimes V_x $. The projection, however, introduces entanglement in a geometrically transparent, combinatorial way and can be controlled easily by adjusting the combinatorial pattern. Practically, one will prepare an Ansatz for a variational problem using these easily controlled elements of entanglement between sites and approximate the ground state through it.\\
At this point, it should be clear that the spin network basis states we introduced earlier are highly analogous to PEPS states. For fixed spins, the tensors $ \ket{\mathbf{j}^x \mathbf{m}^x \iota^x} $ correspond to $ \hbb_x $ in the above, and the auxiliary legs are now 'real' legs $ V^{\mathbf{j}^x} $. They are contracted by projecting on maximally entangled link states. However, a few key difference arise to standard PEPS:
\begin{itemize}
	\item The auxiliary leg spaces are of physical relevance in spin networks. Not all of them are contracted for this reason.
	\item The intertwiners function as the original 'site' legs, but depend in their dimension on the dimension of the surrounding auxiliary spaces.
	\item In the spin network case, these states only form a basis of the full Hilbert space. There is, however, always the option of \textit{superposing} these states in a certain form that preserves the combinatorial structure, as we will do later.
\end{itemize}

\section{Holography and (Random) tensor networks}
There have been many instances of concrete correspondences between boundary and bulk field theories, with or without gravity, since the initial proposal of by Maldacena\cite{maldacenaLargeLimitSuperconformal1999}. The earliest and perhaps most well known of these is the general $ AdS_{d+1}/CFT_d $ conjecture. In this line of ideas, one makes use of the asymptotic symmetry of the conformal boundary of Anti-deSitter spacetime (concretely realised as a projective lightcone) to establish an exact equality of partition functions of two field theories\cite{nilssonRoadEmergentSpacetime2019}. Typically, one considers low-energy effective actions of string theory in the bulk, but the duality is in principle not limited to this example. The crucial steps involve the following: First, a set of valid boundary conditions for the bulk theory must be found for which the solution to the classical equations of motion is uniquely determined by their boundary values. In this step, the causal compactness of the spacetime is crucial. Any two points on the asymptotic boundary are connected by a geodesic of finite proper time. This feature forces the bulk solution to heavily depend on the boundary values and in some cases determines it uniquely.

Second, one introduces Ansätze for field theories on both ends, for example as classical actions. The corresponding partition function then depend on the specified boundary values, as for example in the bulk case through
\begin{equation}\label{AdsCFTPFs}
	Z_{b}[\phi] = \int_{\Phi|_{\partial}= \phi}\mathcal{D}\Phi e^{-S_{b}[\Phi]},
\end{equation}
while the boundary CFT couples to the boundary value through some operator $ \mathcal{O} $. Requiring $ Z_b[\phi]=Z_{\partial}[\phi] $ is the statement of full duality and can be analysed through a semiclassical approximation.
On the level of actions, uniqueness of the bulk solution makes the correspondence unambiguous. The two theories can then be matched by comparing their correlators, typically starting with the 2-point function. 

To characterise the geometry of this correspondence better, Ryu and Takayanagi proposed a criterion\cite{ryuHolographicDerivationEntanglement2006}. When a general system is given a bipartite split into regions $ A $ and $ B $, for example a time slice of the boundary CFT, one can study its entanglement (von Neumann) entropy through path integral and other methods. When properly regulated, this entropy is, at leading order, proportional to the area of the surface dividing $ A $ and $ B $. In order to find a bulk pendent to this quantity, they turned to a known case of entropy scaling like a surface in gravity: That of a black hole's horizon, acting as a surface dividing interior from exterior. Their proposal is, in general, to associate to each boundary CFT region $ A $ a surface $ \gamma_A $ in the bulk, sharing the same boundary $ \partial A $. This surface is further required to be homologous to $ A $ and have minimal area. Then, the CFT entanglement area of $ A $ would be given by
\begin{equation}\label{RTformula}
	S_A = \frac{\text{Vol}(\gamma_A)}{4G_N}
\end{equation}
so precisely the Bekenstein-Hawking formula. In fact, this formula has been shown to hold in fair generality in the context of $  AdS/CFT $. The crucial idea of geometry emergent from holography stems from this; There is a duality between $ (d-1) $-boundary regions and $ (d-2) $-surfaces in the bulk, even with precise areas. This suggests that, with knowledge of all entanglement entropies in the CFT, one might be able to reconstruct the full geometry of the bulk. Of course, not every entanglement structure of the boundary will yield a reasonable bulk geometry. However, there are even clear cases where the minimal surfaces never enter a region of the bulk, such as a black hole's interior, conventionally called the \textit{entanglement shadow}. This region's geometry is inaccessible through the Ryu-Takayanagi (RT) formula. Without further corrections to the formula or additional data in the boundary theory, entanglement is then not enough to construct the interior completely. In fact, it has been suggested by work of Susskind and Maldacena\cite{maldacenaCoolHorizonsEntangled2013} on maximally extended black hole spacetimes that one needs additional data from the boundary to recover entanglement shadowed regions and correctly relate distances to entanglement. Rather one should see entanglement as providing ‘nontraversible wormholes’ and their length being given by some state-dependent quantity in the CFT dual. 

Additionally, there are certain tensor networks such as MERA whose application to critical systems in low dimensions have invited their usage in the same type of correspondence\cite{qiExactHolographicMapping2013}.\\
\begin{figure}
	\centering
	\includegraphics[width=0.7\linewidth]{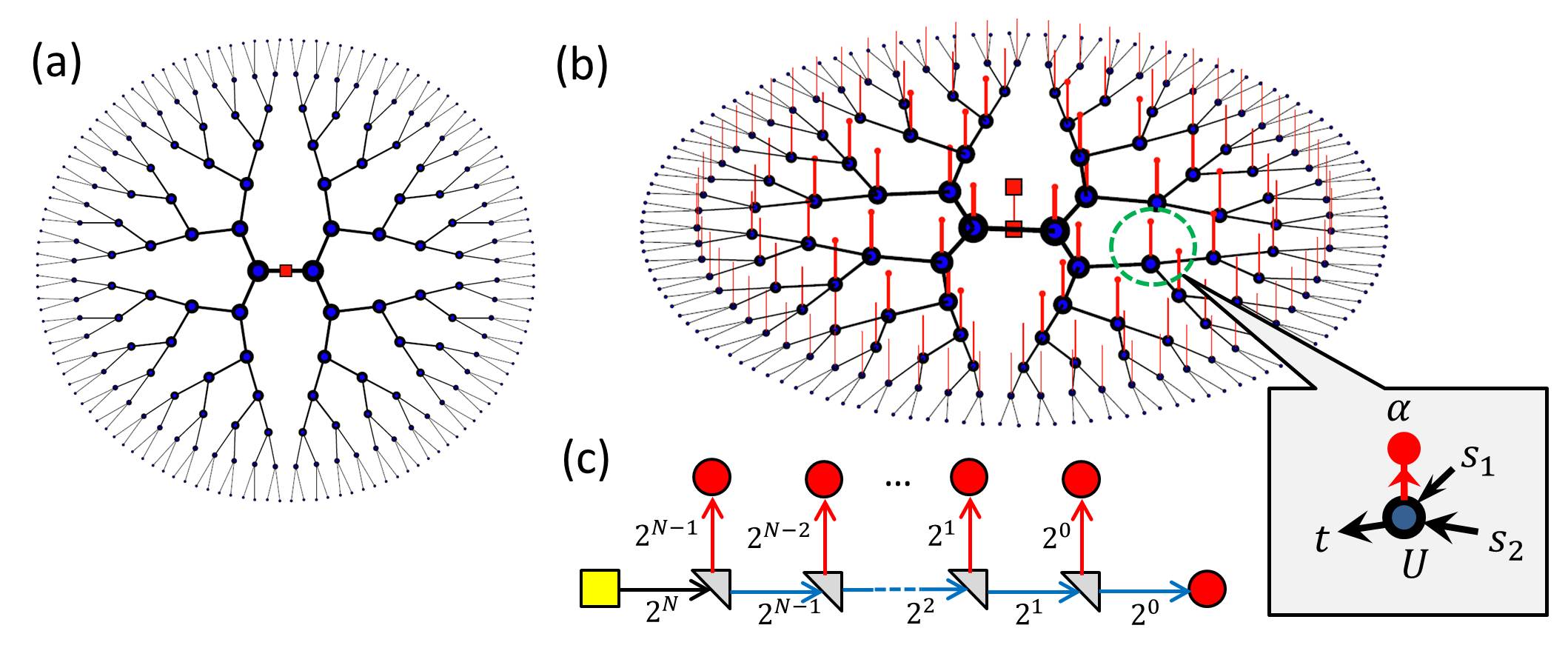}
	\caption{(a),(b): A visual representation of a MERA network.\\
	Note: Reproduced from "Exact holographic mapping and emergent space-time geometry"\cite{qiExactHolographicMapping2013}, p.2}
	\label{fig:mera}
\end{figure}
In one of their layers, correlation functions in these networks behave as in a CFT; they function as a boundary. Then, a state can be fixed in this layer, which induces a corresponding one in the bulk Hilbert space. The bulk can then be given an effective geometry through a distance function determined through the mutual information of two bulk sites:
\begin{equation}\label{MERADistance}
	d(x,y) := -\xi \log\frac{I(x,y)}{I_0},
\end{equation}
where the constant $ I_0 $ is chosen such that neighbouring maximally entangled sites are seen as having $ 0 $ distance between them. The correlation length $ \xi $, in turn, determines the overall scale of the metric. In a way, this only fixes the metric up to a conformal factor. The definition is motivated by the typical behaviour of correlations of operators $ A,B $ observed in ground states of gapped local Hamiltonians, where
\begin{equation}\label{Correlationbound}
	\frac{|\avg{A(x)B(y)}|}{||A|| \cdot ||B||} \leq C_0 e^{-\frac{d(x,y)}{\xi}} 
\end{equation}
provides a universal bound.\cite{nachtergaeleLiebRobinsonBoundsQuantum2010,bravyiLiebRobinsonBoundsGeneration2006}. It is worth noting that this exponential decay of correlations depends crucially on properties of the ground state, as selected by the dynamics of the system. While not all properties of $ AdS/CFT $ can be directly recovered from MERA alone, it has been suggested\cite{caochunjunTheoryQuantumGravity2018} that a MERA with highly entangled boundary state may show features such as an RT formula.
This example shows that in a certain sense, the correspondence between bulk geometries and boundary CFTs is universal and motivates looking for further examples. 

Such examples of holography are easily found in 3D and 4D asymptotically flat gravity, where, similar to the AdS case, symmetries can be used to make precise mappings. 
In 3D gravity the local triviality of the dynamics (every solution of Einstein's equations in vacuum is locally flat) allows a complete solution of Einstein's equations near asymptotic or even finite distance boundaries\cite{goellerQuasiLocal3DQuantum2020} with flat boundary conditions. Then, infinitesimal symmetries of the boundary, given by (extensions of) the Bondi-Metzner-Sachs (BMS) algebra may be used to study boundary quantities that correspond to bulk operators in a very similar fashion to the AdS case, where a part of the boundary symmetries is given by conformal transformations. 

In a perhaps more physically interesting setting like 4D spacetimes, we have other types of holographic correspondences. When self-dual boundary conditions are put on the horizon in a Schwarzschild spacetime\cite{smolinFutureSpinNetworks1997}, one can create a correspondence between the space of conformal blocks of a Chern-Simons theory on the horizon and a subspace of the \textit{physical} Hilbert space of the quantum geometry around it (in the sense that they satisfy the constraints of LQG). Additionally, these dimensions of these Hilbert spaces satisfy, in a certain limit, the Bekenstein-Hawking bound in the sense that
\begin{equation}\label{HorizonAreaLowIdk}
	\log(\dim(\hbb_{CB,S})) \sim A_{S},
\end{equation}
so, the logarithm of the dimension is given by the area in Planck units up to a factor. If indeed one assumes that all physical quantum gravity states must come from Hilbert spaces with such dimensionality, then there is a one-to-one correspondence between the Chern-Simons theory and the quantum geometry outside the horizon.\\
If instead asymptotically flat spacetimes are studied\cite{donnayBMSFluxAlgebra2021}, one finds the notion of \textit{celestial holography}: The asymptotic boundary is here replaced by the celestial sphere of the spacetime, so the dual theory lives on a 2-dimensional Euclidean sphere. Once again, the symmetries are found to be an extension of the BMS algebra. Particle scattering in the asymptotically flat spacetime may be expressed through correlation functions of equivalent operators on the celestial sphere and various connections to infrared properties of general relativity such as soft graviton theorems may be established.

What most of these examples have in common is that they are highly geometric in their approach; they depend on a specific spacetime or at least its asymptotic behaviour. While this is not a restriction in principle, it presents certain practical difficulties in gaining general insights on holographic correspondences, particularly in the non-asymptotic boundary case.
An alternative is to turn to a class of systems that are easier to control, such as tensor networks similar to MERA.
These are easier to control due to their combinatorial buildup, can to some extent be studied numerically, often have very simple Hilbert spaces with factorisation properties, their entanglement can be studied using elementary means, and seem to capture a lot of the same features of holography as full space(time)s. 

Two clear steps in this direction are given by the network proposed by Harlow, Pastawski, Preskill and Yoshida (known as the HaPPY code) as well as that by Yang, Qi and Hayden (YQH code in the following):
In the HaPPY code, the disk model of hyperbolic space is tiled with pentagons\cite{pastawskiHolographicQuantumErrorcorrecting2015}. \\
\begin{figure}
	\centering
	\includegraphics[width=0.5\linewidth]{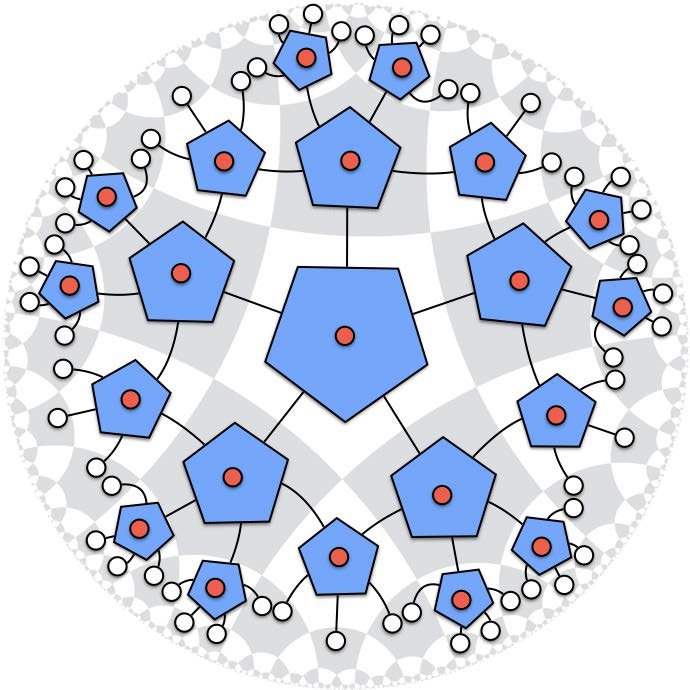}
	\caption{A graphical depiction of the pentagonal tensor network named introduced by Harlow et al. Five of the six legs of each tensor are contracted with surrounding legs, while the remaining leg, shown as a central dot on each tensor, is kept as an input. The network has a finite size beyond which it ends in a number of boundary legs.\\
	Note: Reprinted from "Holographic quantum error-correcting codes: toy models for the bulk/boundary correspondence"\cite{pastawskiHolographicQuantumErrorcorrecting2015}, p.7, CC BY-4.0 }
	\label{fig:happycode}
\end{figure}
Each tile is decorated with a 6-index \textit{perfect tensor}, of which 5 indices are contracted with the adjacent tiles' tensors. The additional index functions as an input or output. Perfectness here means that any bipartition of its indices yields an isometric map from any one set of indices to the rest. Everywhere in the network, the same bond dimension $ v $ is chosen. This network has a number of interesting properties: 
\begin{itemize}
	\item A lattice RT formula holds; Given a region of boundary legs $ A $, there is a minimal surface in the network near it $ \gamma_A $ such that $ S_A = ||\gamma_A|| \log(v) $, proportional to the number of links piercing the minimal surface. Importantly, this suggests an interpretation of the logarithm of the bond dimension as a quantity of area, if this is to be directly analogous to the usual RT formula.
	\item There is a \textit{causal wedge} $ C_A $ associated to each connected boundary region $ A $, from which any operator may be reconstructed as an operator supported in the region $ A $.
	\item The network is \textit{quantum error-correcting}: As the mapping of bulk operators to boundary ones is not single-valued, one may remove parts of the boundary and is still able to reconstruct said operators. Said differently, one needs only a subset of the boundary to cover the entire bulk by its causal wedges. 
\end{itemize}
Meanwhile, the YQH code uses no fixed contraction pattern, but even more special \textit{pluperfect} tensors with 4 contractible legs of bond dimension $ D $ as well as a physical one of dimension $ D^4 $\cite{yangBidirectionalHolographicCodes2016}. The physical Hilbert space of the bulk is taken as the image of the boundary under the YQH and is constrained by a type of $ SU(D) $ gauge invariance acting on each tensor. Said gauge invariance implies that local operators in the bulk are mapped to the trivial one on the boundary - there are no physical local observables, as in a diffeomorphism invariant theory. However, this can be remedied by first specifying a \textit{background bulk state} (perhaps to be thought of as a vacuum), on which a local operator can act as an excitation. The resulting state, now consisting of the background as well as a local excitation, can be mapped to the boundary as usual. In fact, the correlations of such excitations were found to behave like those of a Quantum field theory on a geometric background as long as there are not too many excitations present. In this way, the YQH code features an effective, perturbative locality of correlations.

While both of these examples are incredibly promising insights into general principles of holographic behaviour, their tensors are fairly constrained. However, as explained by YQH, pluperfect tensors may be understood as an idealised approximation of \textit{random tensors}\cite{pastawskiHolographicQuantumErrorcorrecting2015}: In the limit of large bond dimensions, random tensors behave approximately as pluperfect ones. It is this limit of large bond dimensions that we will see resurface over and over throughout this work in the form of a lower cutoff on spins.\\
Random tensor networks, importantly, themselves also feature holographic behaviour. As this case is more directly relevant to this work, it is worth elaborating on in more detail. Before that, we wish to mention that all properties mentioned for the other tensor networks hold, still: For connected boundary regions, there are entanglement wedges with error correction properties and RT formula. Local bulk operators are trivialised upon transport to the boundary. Given a background bulk state, one can determine local 'code subspaces' of the bulk Hilbert spaces which describe QFT-like excitations. However, in addition to these properties, there are two new insights: 
\begin{itemize}
	\item Boundary 2-point correlation functions between operators $ O_A,O_B $ with support resp. in $ A,B $ with disjoint entanglement wedges can be shown to have the same singular value decomposition as a corresponding bulk correlation function. In particular, as the regions on the boundary become small, the entanglement wedges approach the boundary. In this sense, boundary 2-point functions are in one-to-one correspondence with bulk ones. 
	\item By placing a random entangled bulk state in the interior of the network, one can create an entanglement shadow around the area of the bulk state, and thus create a kind of phase transition when the rest of the network is modeled after an Anti-deSitter space.
\end{itemize}
All of these properties will continue to hold in some capacity in the remainder of this work, as we will work with the same techniques.
Consequently, tensor network holography provides an exciting and very concrete setting to explore universal features of holographic correspondences that may even extend past toy models. On the other hand, the connection of these tensor networks to true geometry is less clear and often needs to be put in by hand, as in MERA through the mutual information metric.

\subsection{Random tensor networks}
In random tensor networks\cite{haydenHolographicDualityRandom2016}, individual tensors are randomly picked according to a probability distribution, under which one may then compute average quantities. In other words, with a different choice of setting and interpretation, one can obtain the same type of tensor network holography as for pluperfect tensors.\\
The setup is as follows: In a PEPS state, write each individual vertex state as a unitary operator applied to some reference state:
\begin{equation}\label{key}
	\ket{\Psi_x} = U_x \ket{\Psi_{ref}}
\end{equation}
The \textit{unitaries} are then picked randomly from a distribution $ \rho $ on the group $ \mathcal{U}(\hbb_x) $. Conventionally, this distribution is chosen to be just the Haar measure on the unitary group, with a constant function as density. This is the distribution of maximal entropy on the group when no constraints are imposed on it. However, as will be explored later in section \ref{AltDist}, there are, in principle, other choices for this distribution.\\
With these random pure states, one can construct a PEPS state with a fixed entanglement pattern. The key question of holography in this context is to establish an isometry between a set of \textit{bulk} Hilbert spaces and one of \textit{boundary} output spaces, or to calculate entropies in order to establish an RT formula. To specify the method, we first elaborate on a specific, strong notion of holography based on mappings of operators.

\subsection*{A notion of holography}
The projected state lives in a factorised Hilbert space. In this setting of a bi/tripartite system (bulk/boundary or a partition of the boundary space), we can study holographic properties of the state. For this, in general, consider a system described by a Hilbert space with tripartition $ \hbb \cong \hbb_A \otimes \hbb_{B} \otimes \hbb_C $, where we interpret A as a subsystem from which information is read (output), C as one where information is inserted (input) and B as the entire rest of the system, acting as an environment, bath or background.\\
\begin{figure}
	\centering
	\includegraphics[width=1.0\linewidth]{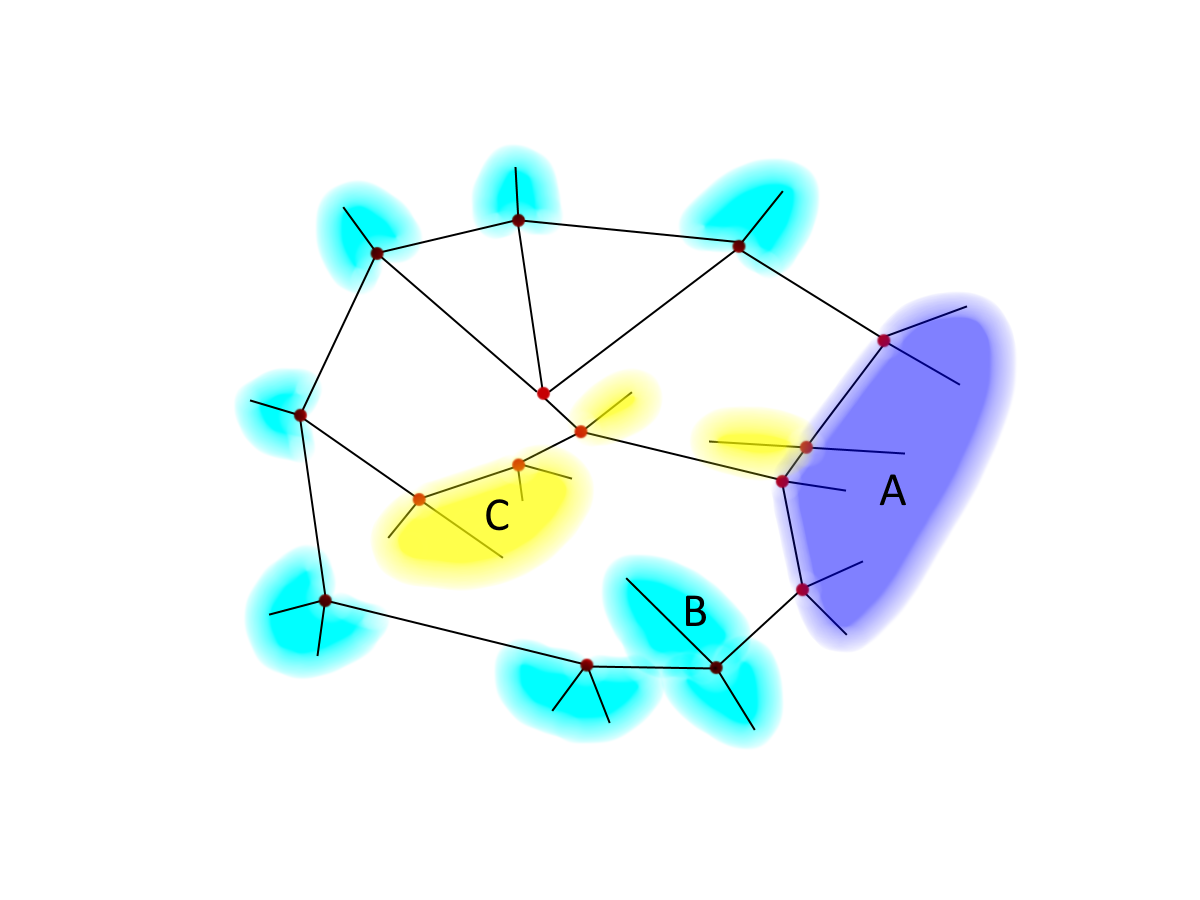}
	\caption{An example setup for holography particularly relevant to this work. The Hilbert space is given by the completely factorised set of boundary link spaces $ \bigotimes_e \hbb_e $.\\
	The tripartition in question is labeled by colours. If the full state induces an isometric transport map $ \mathcal{T}_{C,A} $, we may take any operator acting on the boundary links in C and turn it into one acting on those of A instead without losing information.}
	\label{fig:holographysetup}
\end{figure}
In this setting, consider a pure state $ \ket{\Phi} $ of the system using the natural self-duality of Hilbert spaces $ \hbb_i \cong (\hbb_i)^\ast $. Since $ \ket{\phi} \in \hbb_A \otimes \hbb_{B} \otimes \hbb_C \cong \hbb_A \otimes \hbb_{B} \otimes (\hbb_C)^\ast $, we may see it (by turning kets into bras) as a map from subsystem C to subsystems A and B. Schematically, if we consider a factorised state $ \ket{\phi}=\ket{\phi}_{AB}\otimes\ket{\phi}_C $,
\begin{equation}\label{flipping}
	\ket{\phi_{AB}}\otimes\ket{\phi_C} \hat{=} \ket{\phi_{AB}}\otimes\bra{\phi_C} = \ket{\phi_{AB}}\bra{\phi_C} \in \hom(\hbb_C,\hbb_A \otimes \hbb_{B}).
\end{equation}
This may straightforwardly extended into an (anti)linear map $ \hbb_{AB}\otimes \hbb_C  \rightarrow \hbb_{AB}\otimes (\hbb_C)^\ast \cong\hom(\hbb_C,\hbb_A \otimes \hbb_{B})$.
We are particularly interested in a question of information transport: Given input data on system C (a "core"), can it be recovered from system A (as for example from a boundary region A)? This question can be answered positive if an associated map to the one just described is isometric. Our main objective is to investigate which random tensor network states induce isometric mappings. These states will be called transporting or holographic in the following.\\
First, begin from input data $ \theta $ on the system C, described as a state $ \ket{\theta} \in \hbb_C$. The available data on the remainder of the system after transport is $ \Phi\ket{\theta}= \bra{\theta}\phi\rangle  \in\hbb_A \otimes \hbb_{B} $. As we only consider information recoverable from A, we work with the reduced density matrix 
\begin{equation}\label{reducedstate1}
	\rho_A(\theta) = \Tr_B[\Phi\ket{\theta}\bra{\theta}\Phi^\dagger] 
\end{equation}
More generally, then, for data described by a mixed state $ \rho_c \in \text{BL}(\hbb_C) $, the transported state on A is 
\begin{equation}\label{QChannel1}
	\mathcal{T}_{C,A}(\rho_c) = \Tr_B[\Phi\rho_c \Phi^\dagger]
\end{equation}
So, the transport through the system determines a superoperator $ \mathcal{T}_{C,A} = \Tr_B[\Phi(-) \Phi^\dagger] : BL(\hbb_C)\rightarrow BL(\hbb_A)$. We can write its result in components with respect to a basis $ \{\ket{a}\} $ as $ \bra{a}\mathcal{T}_{C,A}(X)\ket{a'}  = \Tr_C[X \Phi^\dagger (\ket{a'}\bra{a}\otimes \mathbb{I}_B)\Phi]$\\
Assume now that the dimension of $ \hbb_C $ does not exceed that of $ \hbb_A $ so that isometry between spaces of operators is possible in principle. Equip the spaces with the Hilbert-Schmidt norm; If the map is isometric, we have a transporting state. This is the case if for all operators $ X,Y \in BL(\hbb_C) $
\begin{equation}\label{QChannel2}
	\Tr_A[\mathcal{T}_{C,A}(X)^\dagger\mathcal{T}_{C,A}(Y)] = \Tr_C[ X^\dagger Y].
\end{equation}
We can write this condition as a trace over two copies of subsystem C:
\begin{align}\label{QChannel3}
	\Tr_{C^2}[(X^\dagger \otimes Y) 
	\sum_{a,a'} (\Phi^\dagger\otimes\Phi^\dagger) (\ket{a'}\bra{a}\otimes 1_B \otimes\ket{a}\bra{a'}\otimes 1_B)(\Phi\otimes\Phi)] 
	=\Tr_{C^2}[(X^\dagger \otimes Y) \mathcal{S}_C] \\
	\Leftrightarrow\Tr_{C^2}[(X^\dagger \otimes Y) 
	(\Phi^\dagger\otimes\Phi^\dagger) (\mathcal{S}_A\otimes 1_B^{\otimes 2})(\Phi\otimes\Phi)] 
	=\Tr_{C^2}[(X^\dagger \otimes Y) \mathcal{S}_C] 
\end{align}
where we have introduced a \textit{swap operator} $ \mathcal{S}_C $ which exchanges the two factors in the tensor product $ \hbb^{\otimes 2}_C $
So, we need to require
\begin{equation}\label{IsometryCond1}
	(\Phi^\dagger\otimes\Phi^\dagger) (\mathcal{S}_A\otimes 1_B^{\otimes 2})(\Phi\otimes\Phi) = \mathcal{S}_C
\end{equation}
Reverting to the state picture makes this
\begin{equation}\label{IsometryCond2}
	\Tr_{(AB)^2}[(\ket{\phi}\bra{\phi})^{\otimes 2}(\mathcal{S}_A\otimes 1_B^{\otimes 2} ) ] = \mathcal{S}_C
\end{equation}
Which is the general condition to get an isometry between operator spaces.\\
However, we can instead start by demanding a bit more from the map $ \mathcal{T}_{C,A} $ from the beginning. In particular, we can require it to be a quantum channel\footnote{A completely positive, trace preserving map.}: First notice that the conjugation by $ \Phi $ is completely positive, and the partial trace is CPTP\cite{wildeClassicalQuantumShannon2017}. Therefore, the transport is a quantum channel if we require conjugation by $ \Phi $ to be trace preserving. This is the case precisely iff $ \Phi^\dagger\Phi = \mathbb{I}_C $, so if the map $ \Phi $ is an isometry from $\hbb_C$ to $\hbb_{AB}$. If we assume T to be a quantum channel, the condition for T to be an isometry between B(H) simplifies to the following:
\begin{align}\label{IsometryCond3}
	\langle X,Y\rangle_C = \Tr_C[X^\dagger Y] = \Tr_C[X^\dagger \Phi^\dagger\Phi Y \Phi^\dagger\Phi]  
	= \Tr_{AB}[\Phi X^\dagger \Phi^\dagger  \Phi Y \Phi^\dagger]   = \Tr_{(AB)^2}[(\Phi X^\dagger \Phi^\dagger \otimes \Phi Y \Phi^\dagger)(\mathcal{S}_A\otimes\mathcal{S}_B)] \\
	\langle \mathcal{T}_{C,A}(X),\mathcal{T}_{C,A}(Y)\rangle_A = \Tr_{A}[\Tr_B[\Phi X^\dagger \Phi^\dagger]  \Tr_B[\Phi Y \Phi^\dagger]] 
	= \Tr_{(AB)^2}[(\Phi X^\dagger \Phi^\dagger\otimes \Phi Y \Phi^\dagger) (\mathcal{S}_A\otimes\mathbb{I}_{B^2})]
\end{align} 
So the isometry condition for quantum channels is 
\begin{equation}\label{IsometryCond4}
	(\Phi^\dagger\otimes\Phi^\dagger)(\mathcal{S}_A\otimes\mathbb{I}_{B^2}) (\Phi\otimes\Phi) = (\Phi^\dagger\otimes\Phi^\dagger)(\mathcal{S}_A\otimes\mathcal{S}_B)(\Phi\otimes\Phi)
\end{equation}
In particular, notice that for $ B=\emptyset $ the condition is automatically fulfilled. \textit{In this work, we will content ourselves with establishing when the transport superoperator is a quantum channel - so, equivalently, when $ \Phi $ is an isometry.}
The method we use is entirely analogous to the one used in previous works\cite{colafranceschiHolographicMapsQuantum2021,colafranceschiHolographicEntanglementSpin2022,haydenHolographicDualityRandom2016,qiHolographicCoherentStates2017}.
Assume that the dimension of C is lower or equal to that of AB. We rewrite
\begin{equation}\label{IsometryCond5}
	\Phi^\dagger \Phi = \mathbb{I}_C \;\text{     as     } \;\Tr_{AB}[\ket{\phi}\bra{\phi}] = \mathbb{I}_C.
\end{equation}
More explicitly, the map has components $ \Phi_{O,I} $ (O labeling a basis in AB, I in C). Then
\begin{equation}\label{IsoReform1}
	(\Phi^\dagger \Phi)_{I,I'} = \sum_{O}  (\Phi^\dagger)_{I,O} \Phi_{O,I'} = \sum_{O}  \bra{I}(\Phi^\dagger)\ket{O} \bra{O}\Phi\ket{I'} 
	= \sum_{O}  \bra{O}\Phi\ket{I'} \overline{\bra{O}\Phi\ket{I}}  
\end{equation}
And by using the defining relation $ \bra{O}\Phi\ket{I} \coloneqq \bra{O}\bra{I}\phi\rangle$:
\begin{equation}\label{IsoReform2}
	(\Phi^\dagger \Phi)_{I,I'} = \sum_{O}  \bra{O}\bra{I'}\ket{\phi}\bra{\phi}\ket{O}\ket{I} 
	= \bra{I'}\left(\sum_{O}  \bra{O}\ket{\phi}\bra{\phi}\ket{O}\right)\ket{I} = \bra{I'}\Tr_{AB}[\ket{\phi}\bra{\phi}]\ket{I}
\end{equation}
We can then answer the question of isometry by calculating the purity of the reduced state $ \rho_C $, for example as the Rényi 2-entropy\cite{tomamichelFrameworkNonAsymptoticQuantum2013,baezRenyiEntropyFree2011}:
\begin{equation}\label{Renyi1}
	e^{-S_2(\rho_C)} = \frac{\Tr[\rho_C^2]}{\Tr[\rho_C]^2}
\end{equation}
If this expression reaches its minimum of $ \dim(\hbb_C)^{-1} $, the map is a quantum channel.
Via the replica trick\footnote{Letting $ \mathcal{S}$ be the operator swapping two copies of a Hilbert space $\hbb$, we have $ \Tr[(A\otimes B)\mathcal{S} ] = \Tr[AB] $, while$ \Tr[(A\otimes B) ] = \Tr[A]\Tr[B] $.}, we can then rewrite this as traces over two copies of the system
\begin{equation}\label{Renyi2}
	\frac{\Tr_{\hbb_C^2}[\rho_C^{\otimes 2} \mathcal{S}_C]}{\Tr_{\hbb_C^2}[\rho_C^{\otimes 2}]}
	=
	\frac{\Tr_{\hbb^2}[(\ket{\phi}\bra{\phi})^{\otimes 2} \mathcal{S}_C]}{\Tr_{\hbb^2}[(\ket{\phi}\bra{\phi})^{\otimes 2}]}
\end{equation}
We can now apply the technique of averaging over the initial states. This yields the average purity of the reduced state, which will allow for a general statement about typical holographic properties. A crucial point to consider is the variance of the quantities we wish to compute: If the variance is small, we can not only approximate averages of functions by functions of averages, but can also be sure that the typicity statement holds value as most states will be close to that average. As was demonstrated by Hayden, Qi et al\cite{haydenHolographicDualityRandom2016}, the variance can be bounded from above to be arbitrarily small in the limit that the auxiliary leg dimensions (also known as bond dimensions) of the tensor network become large. This means, in the context of spin networks, that all involved area spins must be fairly large.\footnote{What fairly large means is debateable and could lie between spin values of 10 and 100. } When this assumption holds,
\begin{equation}\label{key}
	\avg{e^{-S_2(\rho_C)}}_U = \avg{\frac{\Tr_{\hbb^2}[(\ket{\phi}\bra{\phi})^{\otimes 2} \mathcal{S}_C]}{\Tr_{\hbb^2}[(\ket{\phi}\bra{\phi})^{\otimes 2}]}}_U
	\approx
	\frac{\avg{\Tr_{\hbb^2}[(\ket{\phi}\bra{\phi})^{\otimes 2} \mathcal{S}_C]}_U}{\avg{\Tr_{\hbb^2}[(\ket{\phi}\bra{\phi})^{\otimes 2}]}_U}
\end{equation}
from where one can use linearity of the trace to replace $ (\ket{\Psi_x}\bra{\Psi_x})^{\otimes 2} $ by the average-square state
\begin{equation}\label{key}
	R = \int_{SU(\hbb)}d\mu(U) \rho(U) (U \ket{\Psi_{ref}}\bra{\Psi_{ref}} U^\dagger)^{\otimes 2}
\end{equation}
This operator acting on two copies of the single-particle Hilbert space has the property that it is invariant under unitary conjugation:
\begin{equation}\label{key}
	V^{\otimes 2} R (V^\dagger)^{\otimes 2} = R
\end{equation}
by left-invariance of the Haar measure. Crucially, this requires the group to be a finite dimensional Lie group - this integral does not exist on the infinite unitary group, so our Hilbert spaces must stay finite dimensional. \\
With this property, we can easily find what $ R $ is - the only two operators invariant under this action are the identity and the swap operator, in the form
\begin{equation}\label{key}
	R = \frac{1}{\dim(\hbb)(\dim(\hbb)+1)} ( \mathbb{I} + \mathcal{S}).
\end{equation}
Since we average over each site seperately, we really replace the initial random vertex states by
\begin{equation}\label{key}
	\bigotimes_{x} \frac{1}{\dim(\hbb_x)(\dim(\hbb_x)+1)} ( \mathbb{I}_x + \mathcal{S}_x).
\end{equation}
The real trick happens now: To make the tensor product above tractable, we recognise that, when expanded as a sum, each term will have a number of swap operators, and identity operators do not matter. Each term can then be labeled by the set of sites with swap operators on it, a $ -1 $ indicating a swap. \\
The method by Hayden et al is to introduce on each site a $ \pm 1 $-valued \textit{Ising spin} $ \sigma_x $, which indicates whether a swap is on that site or not. This means the product turns into the sum over Ising configurations
\begin{equation}\label{key}
	\prod_x \frac{1}{\dim(\hbb_x)(\dim(\hbb_x)+1)} \sum_{\vec{\sigma}} \bigotimes_x \mathcal{S}^{\frac{1-\sigma_x}{2}}_x
\end{equation}
To explain, each term in the original sum is mapped to a unique Ising configuration such that the region of swap operators is the region of Ising spin-downs. Then, every configuration must be summed over.
This turns the numerator and denominator of the average purity into \textit{Ising partition functions}:
\begin{equation}\label{key}
	Z_{1|0} = \sum_{\vec{\sigma}} \Tr[\Pi_{\Gamma}^{\otimes 2}\bigotimes_x (\mathcal{S}^{\frac{1-\sigma_x}{2}}_x) \mathcal{S}^{1|0}_C ]
	= 
	\sum_{\vec{\sigma}} e^{-H_{1|0}(\vec{\sigma})}
\end{equation}
and evaluation of the average purity is turned into a calculation of Ising-like partition sums. In the case of large bond dimensions, we can approximate the sums by their ground state values as the lowest bond dimension functions as a notion of inverse temperature. The result is that an isometry, so minimal purity of the reduced state, is attainable depending on the size of the local input and output legs, as well as the graph structure.

\section{Bulk-to-boundary holography for single spin sectors}
It is the goal of this work to study a tensor network-like class of states, derived from spin network states, with respect to their holographic properties. The Hilbert space of $ N $ tetrahedra is, due to the Peter-Weyl decomposition,
\begin{equation}\label{HSdecomp1}
	\hbb^N \cong (L^2(G^D/G))^N\cong \bigotimes^N_{ x=1}(\bigoplus_{\mathbf{j}^x} V^{\mathbf{j}^x}\otimes \mathcal{I}_{\mathbf{j}^x}) \cong \bigoplus_{{\mathbf{j}^x:x}} \bigotimes_{ x}V^{\mathbf{j}^x}\otimes \mathcal{I}_{\mathbf{j}^x},
\end{equation}
which we can abbreviate by collecting all the independent representation spins into a single label $ \ind{j} $:
\begin{equation}\label{HSDecomp2}
	\hbb^N \cong \bigoplus_{\ind{j}} V^{\ind{j}}\otimes \mathcal{I}_{\ind{j}}.
\end{equation}
Each term in this decomposition will be called a \textit{spin sector} and we will seperate the study of states here by the amount of sectors involved: \textit{First, a single sector will be considered. The original work of this thesis is to extend the results there to the case where spin sectors are superposed}.\\
In this part, we will review previous studies on random spin tensor networks with a single spin sector - so, where all edge spins have been fixed once and for all. These states, when not randomised, are labeled by a twisted simplicial geometry and have definite values for the areas of all faces of the simplices. Still, they have volume information encoded in the intertwiner degrees of freedom at each vertex of the dual graph, which may be put into quantum superposition without changing the spins on the edges.
We wish to construct a geometry from $ N $ individual simplices whose facial areas have been fixed. The Hilbert space of these is given by
\begin{equation}\label{FixedSpinProductHS}
	\hbb_{\ind{j}} = \bigotimes_{x} \hbb_{\mathbf{j}^x} = \bigotimes_{x} \mathcal{I}_{\mathbf{j}^x} \otimes V^{\mathbf{j}^x}.
\end{equation}
We pick a product state $ \ket{\Psi } = \bigotimes_{ x}\ket{\Psi_x} $ and wish to glue the tetrahedra according to a simplicial complex dual to a given 4-valent, connected 4-coloured graph $ \gamma $ with open edges. To do this, we project the product state onto a maximally entangled link state for each pair of faces to be glued. Name the total edge set of the graph $ E $ and its subset of internal links $ \Gamma $, while the open boundary links will be called $ \bnd $. Alternatively, the valency of the graph may be either 4 or 1, where the open edges are then capped off with a 1-valent vertex. This point of view allows us to see our open edges as true edges of a graph.
We now project the product state onto 
\begin{equation}\label{FixedSpinInternalLinkState}
	\ket{\Gamma} = \bigotimes_{e \in \Gamma} \ket{e} \qquad \ket{e} = \frac{1}{\sqrt{d_{j_e}}} \sum_{m_e} (-1)^{j_e+m_e}\ket{j_e m_e}_{s(e)}\ket{j_e ,-m_e}_{t(e)}
\end{equation}
where $ s(e),t(e) $ name the source and target vertices of the edge $ e $, respectively. Each individual state is normalised to $ 1 $. 
$ \ket{\Gamma}\bra{\Gamma} $ is the projector which performs the glueing. If, instead, we project \textit{out} the internal links, we have a bulk/boundary state 
\begin{equation}\label{key}
	\ket{\phi} = \bra{\Gamma} \Psi\rangle \in \hbb_{\bnd,\ind{j}|_{\bnd}}\otimes \mathcal{I}_{\ind{j}}
\end{equation}
containing only relevant information on the intertwiners and boundary edges. We will call this type of state, once randomised using unitaries, a \textit{Random spin tensor network} (RSTN) state. One can now see this state as defining a map from bulk data, given by intertwiners, to the boundary space.

\subsection{Bulk-to-boundary isometry}
Taking as $ \hbb_C = \mathcal{I}_{\ind{j}} $, $ \hbb_B = \emptyset $ and $ \hbb_A = \hbb_{\bnd,\ind{j}|_{\bnd}} $, one can immediately apply the technique of random tensor networks.\cite{colafranceschiHolographicEntanglementSpin2022} Assume, that all involved spins are large enough to suppress fluctuations in the unitary average over individual vertex states. This amounts to making the geometry \textit{more semiclassical} in the same sense as large angular momentum values behaving approximately semiclassically and thus making variances of the involved quantities smaller.
Then,
\begin{equation}\label{key}
	\avg{\frac{\Tr_{\hbb^2}[(\ket{\phi}\bra{\phi})^{\otimes 2} \mathcal{S}_C]}{\Tr_{\hbb^2}[(\ket{\phi}\bra{\phi})^{\otimes 2}]}}_U
	\approx \frac{\avg{\Tr_{\hbb^2}[(\ket{\phi}\bra{\phi})^{\otimes 2} \mathcal{S}_C]}_U}
	{\avg{\Tr_{\hbb^2}[(\ket{\phi}\bra{\phi})^{\otimes 2}]}_U}
\end{equation} Writing each vertex state 
\begin{equation}\label{key}
	\ket{\Psi_x} = U_x \ket{Ref},
\end{equation}
a unitary average over each vertex gives, through Schur's lemma,
\begin{equation}\label{FixedSchurAvg}
	\avg{\ket{\Psi_x}\bra{\Psi_x}^{\otimes2}}_{U_x} = \frac{\mathbb{I}_{\hbb_{j^x}^{\otimes 2}} + \mathcal{S}_{j^x}}{\dim(\hbb_{j^x}) (\dim(\hbb_{j^x})+1)} 
\end{equation}
which passes through the traces due to linearity. Then, we can turn the tensor product over these into a sum by introducing bookkeeping variables $ \sigma_x \in \{\pm 1\}$ on each vertex of the graph. When collecting them over all vertices into the vector $ \vec{\sigma} $, they label a term in the sum by giving vertices with a $ \sw_x $ operator a $ \sigma_x=-1 $. Then the sum is
\begin{equation}\label{BigSchurAvg}
	\avg{\bigotimes_{ x} \ket{\Psi_x}\bra{\Psi_x}}_{U} = 
	\bigotimes_{ x} \avg{\ket{\Psi_x}\bra{\Psi_x}}_{U_x} =
	\frac{1}{\prod_x \dim(\hbb_x) (\dim(\hbb_x)_x+1)}\sum_{\vec{\sigma}} \bigotimes_x \sw_x^{\frac{1-\sigma_x}{2}}
\end{equation}
Luckily, the prefactor is independent of the configuration and thus drops out in the quotient \ref{AvgRenyi1} and we will ignore it from now on. The average purity is then the quotient of
\begin{equation}\label{key}
	Z_{1|0} = \sum_{\vec{\sigma}} 
	\Tr[ (\rho_\Gamma)^{\otimes 2} \bigotimes_x \sw_x^{\frac{1-\sigma_x}{2}} (\mathcal{S}_I)^{1|0}].
\end{equation}
Each term in the sum can now be understood as an amplitude in a statistical partition function $ Z_{1|0} $, simply by rewriting it as an exponential:
\begin{equation}\label{key}
	Z_{1|0} = \sum_{\vec{\sigma}} e^{-\mathcal{H}_{1|0}(\vec{\sigma})}
\end{equation}
and where the Hamiltonians are
\begin{equation}\label{key}
	\mathcal{H}_{1|0}(\vec{\sigma}) = \sum_{e \in \bnd} \frac{1-\sigma_{s(e)}}{2} \log(d_{j_e}) + \sum_{e \in \Gamma} \frac{1-\sigma_{s(e)}\sigma_{t(e)}}{2} \log(d_{j_e}) + \sum_x \frac{1-\sigma_x b_x}{2}\log(\mathcal{D}_{\mathbf{j}^x})
\end{equation}
where we have introduced the \textit{bulk pinning field $ b_x $ given by 1 for $ Z_0 $ and by $ -1 $ on bulk vertices for $ Z_1 $.}
Given this setup, we can compute the purity in the regime of high spins. The reason is that we may see the lowest spin in the sector as a notion of temperature for the partition sum - it also specifies the energy gap between the ground and first excited states of the Hamiltonians. 
If this lowest spin is large enough, excited states of the Ising system will not contribute much to the partition sum. In this regime, we may work with just the ground states to approximate the purity.
The ground state of $ \mathcal{H}_{0} $ is easily found to be the all-up Ising configuration. For $ \mathcal{H}_{1} $, it is less trivial: One rather typically has two regions of the graph seperated by a domain wall where one side is Ising-up and the other down.  However, to achieve an isometry, the all-up configuration must still be the ground state. Flipping a spin somewhere down should thus increase the energy:
\begin{equation}\label{key}
	\Delta_z \mathcal{H}_{1} = 
	\sum_{e \in E: s(e)=z}  \log(d_{j_e}) - \log(\mathcal{D}_{\mathbf{j}^z}) \stackrel{!}{>} 0
\end{equation}
which is easily satisfied - but once on flips larger regions, this may turn negative. In fact, if one flips a region $ X $ down, then the condition of positivity is
\begin{equation}\label{key}
	\Delta_X \mathcal{H}_{1} = 
	\sum_{e \in E: s(e)\in X, t(e)\notin X}  \log(d_{j_e}) - \sum_{x\in X} \log(\mathcal{D}_{\mathbf{j}^x}) \stackrel{!}{>} 0	
	\qquad\text{ or }\qquad 	\prod_{e \in E: s(e)\in X, t(e)\notin X}  d_{j_e} > \prod_{x \in X} \mathcal{D}_{\mathbf{j}^x}
\end{equation}
which is easily violated for some graphs, for example in the homogeneous case where all spins are equal. Thus, while the all-up configuration is always a \textit{local} minimum of the energy, one needs it to be a \textit{global} one. In particular, there is an interplay between the graph structure and geometric quantities on it that determines which states admit such a holography. It was found that the larger the overall inhomogeneity of spins on the graph, the easier it is to achieve isometric behaviour. This directly corresponds to having small intertwiner spaces, which is easy to understand from the perspective that information from the bulk needs to be mapped to the boundary, no matter how small the region. More succinctly, the holography condition is that $ \dim(\hbb_{\partial X}) > \dim(\mathcal{I}_X)$, which is the minimal condition for having an isometry from $ \mathcal{I}_X $ to $ \hbb_{\partial X} $. So, to have isometry for the whole graph, we need precisely that every subset of the graph fulfils the minimal condition.
A similar calculation can be done for the boundary-to boundary isometry case, where first a bulk state must be fixed.\\
By the same methods, one can find a type of lattice RT formula for boundary areas:
\begin{equation}\label{key}
	S_2(A) \sim \sum_{e \in \Sigma_A} \log(d_{j_e}) 
\end{equation}
which depends on the domain wall $ \Sigma_A $ associated to a Hamiltonian with distinguished boundary region $ A $. Said domain wall plays the role of the minimal surface of an RT formula. In this way, the mapping to the Ising model provides a precise implementation of the geometry of holographic behaviour.\\
It is noteworthy that whether or not a bulk-to-boundary isometry exists cannot be calculated in the case of superpositions. This is only possible on the fixed-spin level due to factorisation of the Hilbert space.\footnote{There may be a possibility of doing this factorisation through the addition of edge modes. However, this is beyond the scope of this work and reserved for future investigations.} As such, we will instead investigate boundary-to-boundary mappings in the following, where a projection onto a fixed bulk state factorises the remainder of the Hilbert space.
\newpage
\chapter{RSTN with multiple spin sectors}
\section{Superposition of spins} 
We now consider the general case of states which feature a superposition of spins, but within the class of RSTNs.
The total Hilbert space of N quanta, $ \mathbb{H} := L^2(G^D/G_{\text{Diag}})^N = \bigotimes_{ x}\hbb_x \cong \bigoplus_{\vec{\textbf{j}}} \mathbb{H}_{\vec{\textbf{j}}} $ splits into \textit{spin sectors}.\\
Each sector factorises as $ \mathbb{H}_{\vec{\textbf{j}}} \cong \mathcal{I}_{\ind{j}} \otimes V_{\ind{j}} = \bigotimes_x \mathcal{I}^{\textbf{j}^x} \otimes \bigotimes_x V^{\textbf{j}^x}  $ with $ \mathcal{I}^{\textbf{j}^x} $ the intertwiner space and $ V^{\textbf{j}^v} \cong \bigotimes_{\alpha=1}^{D} V^{j^x_{\alpha}} $ the space of open semilinks associated with the tetrahedron $ x $, of dimension $ d_{j^x_\alpha} = 2j^x_\alpha +1 $. Here, we have labeled the links with $ \alpha \in \{1,2,\dots,D\} $ as colours.\\
We once again want to take a product N-particle state and project it on some data, then reduce the resulting state to a boundary region $ C $ and calculate its Rényi entropy. Through this, we can test whether that region is mapped isometrically onto its boundary complement.\\

We will assume to work with a fixed graph and use the following definitions:
\begin{itemize}
	\item Fix the connectivity pattern $ \Gamma $ of links between the $ N $ vertices, for example as an adjacency matrix. We require the graphs to be colored, meaning in particular that vertices can not be connected by more than one link. Some links are left open.
	\item Let $ (L)E $ be the set of (semi)links of the graph. Each of these is then either an \textit{internal} link of the form $ (x,y;\alpha) $ or a \textit{boundary} ($ \bnd $) link $ (x;\alpha) $ belonging either to the \textit{outer boundary} $ \bnd_{out} $ or the \textit{inner boundary} $ \bnd_{in} $, which form a partition of the boundary links. When calculating the entropy of a region $ A \subset \bnd_{out}$, we could choose it to be in the outer boundary, while the inner one provides external information, for example from low area regions of the spin network that we traced out. For internal links $ e\in \Gamma $, we make the distinction between a single link and its two constituent semilinks by calling the set of internal semilinks $ \Gamma_L $. For boundary links, this distinction is vacuous. 
	
	\item Let $ \ket{e_j} := \frac{1}{\sqrt{d_j}}\sum_m (-1)^{j+m}\ket{j m}_1\ket{j -m}_2  $ be a maximally entangled state of 2 semilinks, in the spin-j sector. We superpose them to $ \ket{e} = \sum_{j_e \in \frac{\mathbb{N}}{2}} g_{j_e} \ket{e_{j_e}} $ with coefficients $ g_j $ of our choice. Then the full state to project on is $ \ket{\Gamma} = \bigotimes_e \ket{e} $.
	\item Let $ \ket{\zeta_{\ind{j}}} \in \mathcal{I}_{\ind{j}} $ be an intertwiner state in each spin sector, and $ \ket{\zeta} = \sum_{\ind{j}} \ket{\zeta_{\ind{j}}} $. Later on, we will replace this with a more general, mixed state $ \rho_I $. Do the same for a core state $ \ket{\theta} = \sum_{j_{in}}\ket{\theta_{j_{in}}} $ on which we project the inner boundary. For the most part, this last piece of data can be considered optional, but in the appendix it proves useful.
	
	\item Finally, we will the notation $ ||A|| $ to denote the number of boundary links associated to a subset of the vertices of the graph, while $ |A| $ denotes the number of vertices.
\end{itemize}
Now start with the product state
\begin{equation}\label{newstatesofpreparia}
	\ket{\Psi} = \bigotimes_x \ket{\Psi_x} 
\end{equation}
and project it onto the data we have given:\footnote{Technically, this scalar product is not defined, so this presents an abuse of notation. However, the projection on a subspace of states with the given intertwiner and boundary data is a well-defined operator and behaves in exactly the same way as this naive implementation of projection.} 
\begin{equation}\label{boundarystatenew}
	\ket{\Phi} = \bra{\theta}\bra{\zeta}\bra{\Gamma} \Psi \rangle \in  \mathbb{H}_{\partial\gamma,out} 
\end{equation}
Importantly, the projection forces the contributions of the intertwiners and links to be from the same spin sectors in the following sense:
\begin{equation}
	\begin{split}\label{EquivalentProjections}
		\bra{\theta}\bra{\zeta}\bra{\Gamma} \Psi \rangle =\sum_{\ind{j}}\bra{\theta}\bra{\zeta_{\ind{j}}}\bra{\Gamma} \Psi \rangle
		= \sum_{\ind{j}}\bra{\theta}\bra{\zeta_{\ind{j}}}\bra{\Gamma} \Psi_{\ind{j}} \rangle \\
		= \sum_{\ind{j}}\bra{\theta_{\ind{j}|_{\bnd_{in}}}}\bra{\zeta_{\ind{j}}}\bra{\Gamma_{\ind{j}|_{\Gamma}}} \Psi_{\ind{j}} \rangle 
		= (\sum_{\ind{j}}\bra{\theta_{\ind{j}|_{\bnd_{in}}}}\bra{\zeta_{\ind{j}}}\bra{\Gamma_{\ind{j}|_{\Gamma}}}) \Psi_{} \rangle
		\coloneqqrev \bra{\theta,\zeta,\Gamma}\Psi_{} \rangle 
	\end{split}
\end{equation}
so that we make no restriction on the possible data we project on by instead using $ \ket{\theta,\zeta,\Gamma} $, which is diagonal in spin sectors.\\
The boundary Hilbert space the projected state lives in factorises as
\begin{equation}\label{BndSplits}
	\mathbb{H}_{\partial\gamma,out} \cong \bigoplus_{j_{\partial\gamma,out}} \mathbb{H}_{j_{\partial\gamma}} \cong \bigoplus_{j_A}\bigoplus_{j_{\bar{A}}}
	\mathbb{H}_{A,j_A} \otimes \mathbb{H}_{\bar{A},j_{\bar{A}}} \cong \bigoplus_{j_A}
	\mathbb{H}_{A,j_A} \otimes \bigoplus_{j_{\bar{A}}}\mathbb{H}_{\bar{A},j_{\bar{A}}}\cong
	\mathbb{H}_{A} \otimes \mathbb{H}_{\bar{A}}
\end{equation}
So we can reduce the state to the region $ A $ by tracing out the complement.\\

\subsection{Calculating average entropy}
Instead of calculating the entropy for a particular state, we will make a typicity statement about a class of states of interest with data as presented earlier. These states have a fixed graph structure and data on intertwiners, but are not limited to one spin sector. For them, we can write the average entropy as a partition function of a randomised Ising model. To start, we first average the exponential of the entropy over a distribution of starting states $ \ket{\Psi_x} = U_x \ket{\Psi_{ref}} $, where we choose some reference state as before. This distribution is chosen to be uniform over the unitary group relating different 1-particle-states. More explicitly, we perform an integral 
\begin{equation}\label{HaarMeasure}
	\avg{\ket{\Psi_x}\bra{\Psi_x}}_{U_x} := \int_{\mathcal{U}(\hbb_x)} d\mu_{Haar}(U_x) (U_x \ket{\Psi_{ref}}\bra{\Psi_{ref}} (U_x)^\dagger)^{\otimes 2}
\end{equation}
with the Haar measure on the unitary group of each vertex/tetrahedron's Hilbert space seperately. By linearity this average commutes with taking traces and we denote it by $ \avg{-}_U $ in the following.\\
Then, if we choose a large enough lower cutoff on participating spins $ \mathbb{j} $, we can suppress fluctuations in the quotient
\begin{equation}\label{AvgRenyi1}
	\avg{e^{-S_2(\rho_C)}}_U =
	\avg{\frac{\Tr_{\hbb^2}[(\ket{\phi}\bra{\phi})^{\otimes 2} \mathcal{S}_C]}{\Tr_{\hbb^2}[(\ket{\phi}\bra{\phi})^{\otimes 2}]}}_U 
	\approx \frac{\avg{\Tr_{\hbb^2}[(\ket{\phi}\bra{\phi})^{\otimes 2} \mathcal{S}_C]}_U}{\avg{\Tr_{\hbb^2}[(\ket{\phi}\bra{\phi})^{\otimes 2}]}_U}
	=: \frac{Z_1}{Z_0}
\end{equation}
as has been shown in random tensor networks. Then, we expand the boundary state in the numerator and denominator:
\begin{equation}\label{PF1}
	Z_{1|0} = \Tr[(\rho_I)^{\otimes 2} (\rho_\Gamma)^{\otimes 2} \avg{(\ket{\Psi}\bra{\Psi})^{\otimes 2}}_U (\mathcal{S}_C)^{1|0}]
\end{equation}
If we also introduce an upper cutoff on spins $ J $, we can then use Schur's lemma to evaluate \ref{HaarMeasure}, as the averaging makes the quantity $ Conj(U_x) $-invariant. The result is that
\begin{equation}\label{SchurAvg}
	\avg{\ket{\Psi_x}\bra{\Psi_x}}_{U_x} = \frac{\mathbb{I}_{\hbb_x^2} + \mathcal{S}_x}{\dim(\hbb_x) (\dim(\hbb_x)_x+1)} 
\end{equation}
Then, we can turn the tensor product over these into a sum by introducing bookkeeping variables $ \sigma_x \in \{\pm 1\}$ on each vertex of the graph. when collecting them over all vertices into the vector $ \vec{\sigma} $, they label a term in the sum by giving vertices with a $ \sw_x $ operator a $ \sigma_x=-1 $. Then the sum is
\begin{equation}\label{BigSchurAvg}
	\avg{\bigotimes_{ x} \ket{\Psi_x}\bra{\Psi_x}}_{U} = 
	\bigotimes_{ x} \avg{\ket{\Psi_x}\bra{\Psi_x}}_{U_x} =
	\frac{1}{\prod_x \dim(\hbb_x) (\dim(\hbb_x)_x+1)}\sum_{\vec{\sigma}} \bigotimes_x \sw_x^{\frac{1-\sigma_x}{2}}
\end{equation}
Luckily, the prefactor is independent of the configuration and thus drops out in the quotient \ref{AvgRenyi1} and we will ignore it from now on. \\
With this, the quantities \ref{PF1} can be expanded into
\begin{align}\label{PF2}
	Z_{1|0} &= \sum_{\vec{\sigma}} 
	\Tr[(\rho_I)^{\otimes 2} (\rho_\Gamma)^{\otimes 2} \bigotimes_x \sw_x^{\frac{1-\sigma_x}{2}} (\mathcal{S}_C)^{1|0}]\\
	&= \sum_{(\ind{j},\ind{k},\vec{\sigma})} \Tr_{\hbb_{\ind{j}}\otimes\hbb_{\ind{k}}}[(\rho_I)^{\otimes 2} (\rho_\Gamma)^{\otimes 2} \bigotimes_x \sw_x^{\frac{1-\sigma_x}{2}} (\mathcal{S}_C)^{1|0}]		
\end{align}
where we used that the Hilbert spaces split into orthogonal spin sectors. The traces in each term are now over spaces that factorise over vertices and links, and accordingly the single-site swap operators do so, too: $ \sw_x = \sw_{\mathcal{I},x}\prod_\alpha\sw_{\alpha,x} $. The swaps here are indiscriminate of spin sector.
The traces can then be evaluated over intertwiner, link and boundary parts seperately. \\
To illustrate the calculations, the intertwiner factor is 
\begin{align}\label{Intertw1}
	\Tr_{\mathcal{I}_{\ind{j}} \otimes \mathcal{I}_{\ind{k}}}[(\rho^I)^{\otimes 2} \bigotimes_x \sw_{I,x}^{\frac{1-\sigma_x}{2}}] 
	&=
	\Tr_{\bigotimes_{S_\downarrow}\mathcal{I}_{\mathbf{j}^x} \otimes \mathcal{I}_{\mathbf{k}^x}}[
	(\Tr_{\bigotimes_{S_\uparrow}\mathcal{I}_{\mathbf{j}^x}}[\rho^I]\otimes
	\Tr_{\bigotimes_{S_\uparrow}\mathcal{I}_{\mathbf{j}^x}}[\rho^I])
	\bigotimes_{S_\downarrow} \sw_{I,x}]\\
	&= \Tr_{}[
	\bigotimes_{S_\downarrow}\mathbb{P}_{\mathcal{I}_{\mathbf{j}^x}}
	(\rho^I \bigotimes_{S_\uparrow}\mathbb{P}_{\mathcal{I}_{\mathbf{j}^x}})_\downarrow
	\bigotimes_{S_\downarrow}\mathbb{P}_{\mathcal{I}_{\mathbf{k}^x}}
	(\rho^I \bigotimes_{S_\uparrow}\mathbb{P}_{\mathcal{I}_{\mathbf{k}^x}})_\downarrow]
\end{align}
where we use the shorthand $ \mathbb{P}_{\mathcal{I}_{\mathbf{k}^x}} $ for the projector onto the intertwiner space $ \mathcal{I}_{\mathbf{k}^x}$ and $ (-)_\downarrow $ denotes reduction to the subspace $ \bigotimes_{x: \sigma_x = -1} \mathcal{I}_x $ associated to $ S_\downarrow $.\\
So, a copy of $ \rho^I $ is first reduced to the region $ S_\downarrow = \{x:\sigma_x = -1\} $, but with the spins set by $ \ind{j}(\ind{k}) $ on $ S_\uparrow $. Then, the two copies are multiplied and traced over the remainder of vertices, but with spins fixed again to be $ j $ or $ k $ \textit{between} them. In particular, for certain values of $ j,k $ on $ S_\downarrow $, this factor will vanish depending on the intertwiner state $ \rho^I $ chosen. For example, if it is diagonal in spin sectors, meaning $ (\rho^I)_{\ind{j},\ind{k}} \sim \delta_{\ind{j},\ind{k}} $, this gives a constraint on which pairs of spin sectors have a nonvanishing intertwiner factor, of the form
\begin{equation}\label{key}
	\Delta_I(\ind{j},\ind{k},\vec{\sigma})= \prod_{x\in S_\downarrow} \delta_{\mathbf{j}^x,\mathbf{k}^x}
\end{equation}
We will write this contribution from intertwiners as a general boolean-valued factor $\Delta$, a $ \sigma $-independent term and an exponential as follows:
\begin{equation}\label{Intertw2}
	\Tr_{\mathcal{I}_{\ind{j}} \otimes \mathcal{I}_{\ind{k}}}[(\rho^I)^{\otimes 2} \bigotimes_x \sw_{I,x}^{\frac{1-\sigma_x}{2}}] 
	=
	\Delta_I(\vec{\sigma},\ind{j},\ind{k}) 
	\Tr_{\mathcal{I}_{\ind{j}}}[\rho^I] 
	\Tr_{\mathcal{I}_{\ind{k}}}[\rho^I] 
	e^{-\Sigma_I(\vec{\sigma},\ind{j},\ind{k})}
\end{equation}
where we include an entropy-like quantity
\begin{equation}\label{IntertwEntropy}
	\Sigma_I(\vec{\sigma},\ind{j},\ind{k})
	=
	-
	\log[
	\frac
	{\Tr_{\bigotimes_{S_\downarrow}\mathcal{I}_{\mathbf{j}^x} \otimes \mathcal{I}_{\mathbf{k}^x}}[
		(\Tr_{\bigotimes_{S_\uparrow}\mathcal{I}_{\mathbf{j}^x}}[\rho^I]\otimes
		\Tr_{\bigotimes_{S_\uparrow}\mathcal{I}_{\mathbf{j}^x}}[\rho^I])
		\bigotimes_{S_\downarrow} \sw_{I,x}]}
	{\Tr_{\mathcal{I}_{\ind{j}}}[\rho^I] 
		\Tr_{\mathcal{I}_{\ind{k}}}[\rho^I] }
	]
\end{equation}
and it is understood that the $ \Delta$-factor forces the term to be 0 if $ \Sigma_I $ reaches $ \infty $. Its presence serves the purpose of setting the right hand side to be 0 whenever the left hand side is, in a transparent manner. In sums over spin sectors later on, these types of factors serve as constraints or indicators on which spin sectors contribute. In particular, we note three special cases:
\begin{itemize}
	\item $ \ind{j} = \ind{k} $: $ \Sigma_I(\vec{\sigma},\ind{j},\ind{k}) = S_2((\rho^I_{\ind{j}})_\downarrow) $ the Rényi 2-entropy of the reduced intertwiner state.
	\item $ \vec{\sigma}= \vec{1} $: $ \Sigma_I = 0 $.
	\item $ \vec{\sigma}= \vec{-1} $: $ \Sigma_I = -
	\log[
	\frac
	{\Tr_{}[
		(\rho^I)_{\ind{j},\ind{k}}
		(\rho^I)_{\ind{k},\ind{j}}
		]}
	{\Tr_{\mathcal{I}_{\ind{j}}}[\rho^I] 
		\Tr_{\mathcal{I}_{\ind{k}}}[\rho^I] }
	] $.
\end{itemize}
This pattern continues: We will split the contributions from the other parts of the graph in the same way into an indicator, a contribution independent of the Ising spins and an exponential.\\
For the internal links, we find that the space to be traced over factorises, as does the state of interest. For each internal link of the graph, there are two semilinks, and so  
\begin{equation}\label{key}
	\hbb_{\Gamma} \cong \bigotimes_{l\in \Gamma_L} \hbb_{l} \;\qquad \hbb_l =  \bigoplus_{j_l} V^{j_l}
\end{equation}
which means that the restriction to the $ \ind{j} $-spin sector just leaves each (semi)link with a single spin value:
\begin{equation}\label{InternalSpaceFactorises}
	\hbb_{\Gamma,\ind{j}} \cong \bigotimes_{e\in \Gamma} \hbb_{e,j_e} \;\qquad \hbb_{e,j_e} =  V^{j_e}_{s(e)}\otimes V^{j_e}_{t(e)}
\end{equation}
so there is a degree of freedom of dimension $ 2j_e +1 $ at each end of the link $ e $. The projection operator also factorises over internal links: 
\begin{equation}\label{InternalStateFactorises}
	\rho^\Gamma = \bigotimes_{e\in \Gamma} \rho^\Gamma_e \; \qquad \rho^\Gamma_e = \ket{e}\bra{e} \; \qquad \ket{e} = \sum_{j_e} \frac{g_{j_e}}{\sqrt{d_{j_e}}}\sum_{m_e}(-1)^{j_e + m_e} \ket{j_e m_e}_{s(e)}\ket{j_e m_e}_{t(e)}
\end{equation}
Internal links will either have 2 swap operators or 1 (depending on whether they are fully contained in $ S_{\uparrow,\downarrow} $ or not, respectively). If we name the subset of internal links on the boundary of the two regions $ \partial S $, then
\begin{align}\label{IntLinks}
	\Tr_{\hbb_{\Gamma,\ind{j}} \otimes \hbb_{\Gamma,\ind{k}}}
	[(\rho^\Gamma)^{\otimes 2} \bigotimes_{l \in \Gamma_L} \sw_{l}^{\frac{1-\sigma_x}{2}}]
	&=
	\prod_{ e\in \Gamma}\Tr_{\hbb_{e,j_e}\otimes\hbb_{e,k_e}}[(\ket{e}\bra{e})^{\otimes 2} \sw_{e}^{\frac{1-\sigma_s(e)\sigma_t(e)}{2}}]\\	
	&=
	\prod_{e\in \partial S} \delta_{j_e,k_e} \prod_{e\in \Gamma} d_{j_e}^{-\frac{1-\sigma_s(e)\sigma_t(e)}{2}} 
	\prod_{ e\in \Gamma} |g_{j_e}|^2|g_{k_e}|^2.
\end{align}
For this, we state the result for both the single-and no-swap case, where for simplicity $ s(e)=x,y=t(e) $:
\begin{equation}\label{key}
	\begin{split}
		\Tr_{\hbb_{e,j_e}\otimes\hbb_{e,k_e}}[(\ket{e}\bra{e})^{\otimes 2} \sw_{e}] = \\
		\sum_{m^x,m^y,n^x,n^y}\bra{j m^x}_x\bra{j m^y}_y  
		\bra{k n^x}_x\bra{k n^y}_y
		(\ket{e}\bra{e})^{\otimes 2} \sw_{e}
		\ket{j m^x}_x\ket{j m^y}_y  
		\ket{k n^x}_x\ket{k n^y}_y\\
		=
		\sum_{m^x,m^y,n^x,n^y}\bra{j m^x}_x\bra{j m^y}_y  \ket{e}
		\bra{k n^x}_x\bra{k n^y}_y\ket{e}
		\bra{e}\bra{e} \sw_{e}
		\ket{j m^x}_x\ket{j m^y}_y  
		\ket{k n^x}_x\ket{k n^y}_y\\
		=
		\sum_{m^x,m^y,n^x,n^y}\bra{j m^x}_x\bra{j m^y}_y  \ket{e}
		\bra{k n^x}_x\bra{k n^y}_y\ket{e}
		\bra{e}\bra{e} 
		\ket{j m^x}_x\ket{k n^y}_y  
		\ket{k n^x}_x\ket{j m^y}_y\\
		=
		\sum_{m^x,m^y,n^x,n^y}(-1)^{j+m^x+k+n^x}
		\frac{g_{j_e}}{\sqrt{d_{j_e}}} \delta_{m^x+m^y,0}
		\frac{g_{k_e}}{\sqrt{d_{k_e}}}\delta_{n^x+n^y,0}
		\bra{e}\ket{j m^x}_x\ket{k n^y}_y  
		\bra{e}\ket{k n^x}_x\ket{j m^y}_y\\
		=	
		\sum_{m^x,m^y,n^x,n^y}
		\frac{g_{j_e}}{\sqrt{d_{j_e}}} \delta_{m^x+m^y,0}
		\frac{g_{k_e}}{\sqrt{d_{k_e}}} \delta_{n^x+n^y,0}
		\frac{\overline{g_{j_e}}}{\sqrt{d_{j_e}}} 
		\frac{\overline{g_{k_e}}}{\sqrt{d_{k_e}}}  \delta_{n^x+m^y,0}^2\delta_{j_e,k_e}^2\\
		=
		\delta_{j_e,k_e}\frac{|g_{j_e}|^4}{d_{j_e}^2}\sum_{m^x} 1 = \delta_{j_e,k_e}\frac{|g_{j_e}|^2|g_{k_e}|^2}{d_{j_e}}
	\end{split}
\end{equation}
\begin{equation}\label{key}
	\begin{split}
		\Tr_{\hbb_{e,j_e}\otimes\hbb_{e,k_e}}[(\ket{e}\bra{e})^{\otimes 2} ]  \\
		=
		\sum_{m^x,m^y,n^x,n^y}\bra{j m^x}_x\bra{j m^y}_y  \ket{e}
		\bra{k n^x}_x\bra{k n^y}_y\ket{e}
		\bra{e} 
		\ket{j m^x}_x\ket{j m^y}_y  
		\bra{e}\ket{k n^x}_x\ket{k n^y}_y\\
		=
		\sum_{m^x,m^y,n^x,n^y}
		\frac{g_{j_e}}{\sqrt{d_{j_e}}} \delta_{m^x,m^y}^2
		\frac{g_{k_e}}{\sqrt{d_{k_e}}} \delta_{n^x,n^y}^2
		\frac{\overline{g_{j_e}}}{\sqrt{d_{j_e}}} 
		\frac{\overline{g_{k_e}}}{\sqrt{d_{k_e}}}
		=
		|g_{j_e}|^2|g_{k_e}|^2
	\end{split}
\end{equation}
Similarly, the boundary has its own factor: By introducing a boundary pinning field on the boundary vertices, with $ h_x = -1  $ for $ Z_1 $ if $ x $ is a boundary vertex at $ C $ and $ 0 $ else, we find by the same kind of calculation
\begin{equation}\label{Boundary1}
	\Tr_{\hbb_{\bnd,\ind{j}}\otimes\hbb_{\bnd,\ind{k}}}[\bigotimes_{e \in \bnd}\mathcal{S}_{e}^{\frac{1-\sigma_{s(e)}}{2}} \mathcal{S}_C
	] = \Tr_{\hbb_{\bnd,\ind{j}}\otimes\hbb_{\bnd,\ind{k}}}[\bigotimes_{e \in \bnd} \mathcal{S}_e^{\frac{1-\sigma_{s(e)} h_{t(e)}}{2}}] = \prod_{e: \sigma_x h_e = -1} \delta_{j_e,k_e} d_{j_e}^{-\frac{1-\sigma_s(e)} h_{t(e)}{2}}
	\prod_{ e\in \bnd} d_{j_e}d_{k_e}.
\end{equation}
From this we can see that the result would be uniform across all links if we chose the normalisation factors to be $ |g_j| = \sqrt{d_j} $. As should be clear from a quick glance, this is not normalisable if we take the cutoff $ J $ to infinity. For this calculation as before, the cutoff becomes a crucial assumption.\\

\subsection{The random Ising model}
Now, we can write \ref{PF1} as
\begin{equation}\label{PF3}
	Z_{1|0} = \sum_{(\ind{j},\ind{k},\vec{\sigma})} \Delta_{1|0}(\ind{j},\ind{k},\vec{\sigma}) e^{B(\ind{j})+B(\ind{k})} e^{-\mathcal{H}_{1|0}(\ind{j},\ind{k},\vec{\sigma})}
\end{equation}
where
\begin{equation}\label{IsingHammy}
	\mathcal{H}_{1|0}(\ind{j},\ind{k},\vec{\sigma}) 
	=
	\sum_{e \in \bnd} (\frac{1-\sigma_{s(e)} h_{t(e)}}{2})\log (d_{j_e})
	+
	\sum_{e \in \Gamma} (\frac{1-\sigma_{s(e)} \sigma_{t(e)}}{2})\log (d_{j_e})
	+
	\Sigma_I(\vec{\sigma},\ind{j},\ind{k})
\end{equation}
\begin{equation}\label{IsingWeight}
	K_{\ind{j}}= e^{B(\ind{j})} = \prod_{ e\in \bnd} d_{j_e} \prod_{ e\in \Gamma} |g_{j_e}|^2 \Tr_{\mathcal{I}_{\ind{j}}}[\rho^I] 
\end{equation}
and the (partial) constraint factor is given by
\begin{equation}\label{IsingConstraint}
	\Delta_{1|0}(\ind{j},\ind{k},\vec{\sigma}) = \prod_{e: \sigma_x h_e = -1} \delta_{j_e,k_e} \prod_{e\in \partial S} \delta_{j_e,k_e} \, \Delta^I(\ind{j},\ind{k},\vec{\sigma})
\end{equation}
and constraints from intertwiners are to be incorporated in the final factor. For example, if one chooses a product state over vertices for intertwiners, $\rho^I = \otimes_x \rho^I_x$, we get 
\begin{equation}
	\Delta^I(\ind{j},\ind{k},\vec{\sigma}) = \prod_{x:\sigma_x=-1} \delta_{\mathbf{j}^x,\mathbf{k}^x}
\end{equation}Let us also define, for reference, the quantities 
\begin{equation}\label{PFfixedSpins}
	Z^{(\ind{j},\ind{k})}_{1|0} = \sum_{\vec{\sigma}} \Delta_{1|0}(\ind{j},\ind{k},\vec{\sigma}) e^{-\mathcal{H}_{1|0}(\ind{j},\ind{k},\vec{\sigma})}
\end{equation}
which enable us to phrase the discussion of \ref{PF3} nicely. 
By defining the normalised distribution over spin sectors
\begin{equation}\label{IsingDistrib}
	P(\ind{j},\ind{k}) = \frac{K_{\ind{j}}K_{\ind{k}}}{Z_{0}} Z^{(\ind{j},\ind{k})}_{0}
\end{equation}
we see our quantity of interest as a probability average
\begin{equation}\label{ProbAvg1}
	\frac{Z_1}{Z_0} = \sum_{(\ind{j},\ind{k})}  P(\ind{j},\ind{k}) 
	\sum_{\vec{\sigma}}
	\frac{e^{-\mathcal{H}_{1}(\ind{j},\ind{k},\vec{\sigma})}}{Z^{(\ind{j},\ind{k})}_{0}}  \Delta_{1}(\ind{j},\ind{k},\vec{\sigma})
\end{equation} 
In particular, we can write 
\begin{align}\label{LocalEntropy}
	\mathcal{H}_{1}(\ind{j},\ind{k},\vec{\sigma}) &= \mathcal{H}_{0}(\ind{j},\ind{k},\vec{\sigma}) 
	+
	\sum_{e \in \bnd} \frac{(1-h_e) }{2}\sigma_x \log (d_{j_e})\\
	&= \mathcal{H}_{0}(\ind{j},\ind{k},\vec{\sigma}) 
	+
	\sum_{e \in C}\sigma_x \log (d_{j_e}) 
	= \mathcal{H}_{0}(\ind{j},\ind{k},\vec{\sigma}) 
	+
	Q(\ind{j},\ind{k},\vec{\sigma}).
\end{align}
So the difference is a quantity supported on the boundary region $ C $.\\
Then, the quantity at the right end of \ref{ProbAvg1} takes the form of a type of statistical average:
\begin{equation}
	\sum_{\vec{\sigma}}
	\frac{e^{-\mathcal{H}_{1}(\ind{j},\ind{k},\vec{\sigma})}}{Z^{(\ind{j},\ind{k})}_{0}}  \Delta_{1}(\ind{j},\ind{k},\vec{\sigma})
	=
	\frac{\sum_{\vec{\sigma}} e^{-\mathcal{H}_{0}(\ind{j},\ind{k},\vec{\sigma})} e^{-Q(\ind{j},\ind{k},\vec{\sigma})}\Delta_{1}(\ind{j},\ind{k},\vec{\sigma})}
	{\sum_{\vec{\sigma}}  e^{-\mathcal{H}_{0}(\ind{j},\ind{k},\vec{\sigma})}
		\Delta_{0}(\ind{j},\ind{k},\vec{\sigma})}
\end{equation}
If one disregards the constrained sum due to the $\Delta$-factors, this quotient is an average in a Gibbs-type canonical ensemble for the Hamiltonian $\mathcal{H}_0$.\\
We adopt the following interpretation: The average exponentiated Rényi entropy is given as a quantity associated to a \textit{random Ising model} on our fixed graph whose couplings $ \log(d_j) $ are distributed according to the function $ P $. In fact, disregarding the difference between the $ \Delta $-factors, the quantity is actually $ \avg{\avg{e^{-Q}}_{Z_0}}_P $, so if $ \mathcal{H}_0 $ is taken as the Hamiltonian, we just take the average expectation value of $ e^{-Q} = \prod_{e \in C} d_{j_e}^{-\sigma_x} $. The individual Ising partition functions for fixed spins are given by \ref{PFfixedSpins}.\\
Inspiration for this comes from SYK-type models, where a similar, but Gaussian, average is performed over couplings and induces holographic behaviour as well\cite{anousQuantumSpinGlass2021,christosSpinLiquidSpin2022}.
Two things must be highlighted before proceeding further. First, the role of the constraint factors. Second, the shape of the distribution of couplings and what controls it.\\
The constraints remove certain terms from the sum if the area spins on a given set of links is not the same in the tuple $ (\ind{j},\ind{k}) $. If we take the real-space viewpoint on Rényi k-entropy\cite{nilssonRoadEmergentSpacetime2019}, we can interpret them in a geometric way. Say we wish to compute the entropy of a local QFT's state in a certain region $ A $. Thus, we fix a state $ \ket{\Psi} $ in the bipartite Hilbert space $ \hbb_A\otimes\hbb_{A^c} $. A geometric way to compute said entropy is then to consider k copies of the domains of the fields, but glued together at the region A in a certain way by attaching cells and arranging the copies in a 'chain' with independent fields in each copy, but matched using boundary conditions along each glueing. Given this, the entropy is given as a partition function of the field theory living on this extended domain. \\
For our case, we have a similar situation: 2 copies of the graph are needed and constraints connect the values of area (spins) between the two copies on subregions determined by the Ising spin configuration. As such, we can give the following interpretation: In computing the entropy, we glue, for a fixed Ising configuration, the two copies with given area spins together in a subregion (see figure \ref{fig:glueing}). However, we only glue them if the areas agree on the given region - analogous to how we only consider tetrahedra glued when their areas are maximally entangled. In a sense, one may only glue what fits together geometrically. In the end, we sum over all Ising configurations, so different regions will be glued together.
The intuition to be taken from this is that if graphs with different area values are too far apart \textit{geometrically}, the pair will not contribute to the entropy. \\
\begin{figure}
	\centering
	\includegraphics[width=0.6\linewidth]{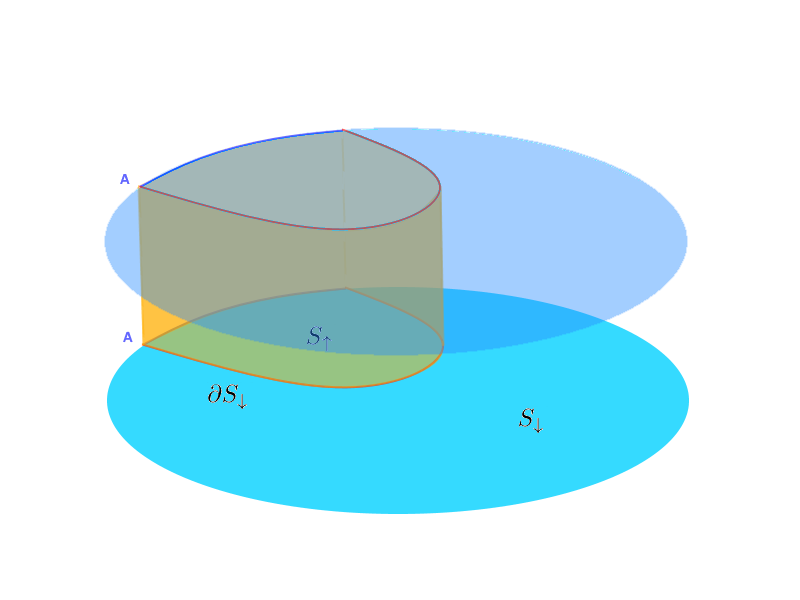}
	\caption{A schematic of the glueing imposed by the constraints. Along the minimal surface $ \partial S_\downarrow $, there is a fictitious glueing of the two spin networks where areas are required to be equal. Conversely, any surface where a glueing occurs needs to agree in spin labels in the two geometries, similar to the situation in a spin foam.}
	\label{fig:glueing}
\end{figure}
Next, it is clear from \ref{IsingWeight},\ref{IsingDistrib} that only the normalisations of each spin sector contributing to the states matter for the distribution of areas. If one chooses the intertwiner state to contain only a single spin sector (so only one fixed discrete geometry is in the state), the distribution will become sharp on that geometry (in the sense of areas). Additionally, the unbounded growth of the boundary factor shows that only a class of intertwiner states obeying a 'boundary condition' can have a sensible entropy in our setting once the upper cutoff on spins is removed. For each boundary link, $ (d_{j_e}\cdot \Tr_{\ind{j}}[\rho^I](j_e))_{j_e} $ needs to be a summable sequence.
Similarly we need also that for each internal link,$ (g_{j_e})_{{j_e}} \in \mathfrak{l}^2(\frac{\mathbb{N}_0}{2})$ and $ (\Tr_{\ind{j}}[\rho^I])_{\ind{j}} \in \mathfrak{l}^1((\frac{\mathbb{N}_0}{2})^|E|) $.\\

\subsection*{Alternative interpretation: A sum over geometries}
As an alternative interpretation, one considers the set of spins $(\ind{j},\ind{k})$ on the two copies as a background geometry for the Ising models \ref{PFfixedSpins}. Then, our partition function is gained by promoting said background to a dynamical variable and summing over its configurations with the weight $P$. In the cases where the glueing is imperfect, not all Ising configurations are permissible - the Ising system can only be spin-down on the subset of links and sites where the geometry is well-defined. In this interpretation, one may see our result as a discrete path integral over geometries for the Rényi 2-entropy. However, the interesting point there is that a \textit{single} state of a quantum geometry already gives rise to such a sum over geometries. In a sense, the superposition provides already a sufficient setting for summing over spatial geometries.

To generalise this idea, consider a Hilbert space $\hbb = \bigoplus_{\ind{j}} \hbb^{\gamma}_{\ind{j}}\otimes \hbb^M_{\ind{j}}$ which splits over sectors of geometric data $ \ind{j} $. Each sector then has geometric and additional matter degrees of freedom. If we consider
an expectation value of some operator,
\begin{equation}
	\avg{X}_\rho = \sum_{\ind{j}}
	(\frac
	{\Tr_{\ind{j}}[\rho ]}
	{\Tr_{}[\rho ]})
	(\frac
	{\Tr_{\ind{j}}[\rho X]}
	{\Tr_{\ind{j}}[\rho ]})
	= \sum_{\ind{j}}
	P(\ind{j})
	\avg{X}_j
\end{equation}
it can be expressed as a probability average in the same way as we had before. Specifically, if the state has no interference between sectors, so only $\rho_{\ind{j},\ind{j}}$ exist, then $\avg{X}_j = \avg{X}_{\rho_{\ind{j},\ind{j}}}$. So, if the sectors correspond to geometries, this formulation means that in states which are classical superpositions of geometries averages are given by weighted averages in the individual geometries. This gives a very straightforward explanation to how our weighted Ising model comes to be, and puts it into a larger picture of quantum gravity models with superposed geometries.\\
\subsection*{Another alternative: spin foam models}\label{SFMPF}
We may even see the sums as something akin to a spin foam model. To see this, forget about the intertwiners on a spin network on $ \gamma $, which is then just labeled by its graph and the representations $ \ind{j} $. Introduce a 2-complex $ W((\gamma,\ind{j})\mapsto (\gamma,\ind{k});\sigma) $ trivially connecting the two graphs in the obvious way. Then, assign to the vertical 1-faces the Ising spins and to the 2-faces the representations of the spin networks they end on. This may not be a unique assignment, but by suitably restricting the labels, the weights assigned to this 2-complex will still be unique. In fact, the role of the $ \Delta $-constraints in this formulation is precisely that of restricting the sum over all such spin foams with fewer data to that subset for which the weight is unambiguously calculable. Said weight is simply given by the same trace as in \ref{PF2}. The calculation may therefore be understood as follows:\\
Fixing the graph $ \gamma $, state $ \rho_I $ and region $ C $, we define two spin foam model partition sums 
\begin{equation}\label{SFMIsing}
	Z_{1|0} = \sum_{W_{\text{comp}}} w_2(\ind{j}\cup\ind{k})  w_1(\ind{j}\cup\ind{k};\vec{\sigma})
\end{equation}
where the class of spin foam models is analogous to the one we introduced just now - they trivially connect two copies of $ \gamma $ and have $ \mathbb{Z}_2 $-labels on their vertical 1-faces representing the Ising spins, which replace the intertwiner labels. The weights of the model depend on the intertwiner state and the region under consideration and are the same as before up to multiplicity factors coming from counting $ (\ind{j},\ind{k}) $ configurations which lead to the same amplitude. They factorise into a part depending only on the 2-faces and one with dependence on the 1-faces. Some of them will vanish on certain spin foams, in which case these are excluded from the sum. So, the $ \Delta $-constraints that are always present in the partition sums are implemented in the spin foam models by restricting to a set of \textit{compatible} foams.

\subsection{Characterising holographic states}
We turn to the question of holography again. Necessary conditions on states that minimise the purity can be straightforwardly derived. First, see from \ref{PF3} that the terms summed over are all nonnegative:
\begin{equation}\label{key}
	\Tr_{\hbb_{\ind{j}}\otimes\hbb_{\ind{k}}}[(\rho_I)^{\otimes 2} (\rho_\Gamma)^{\otimes 2} \bigotimes_x \sw_x^{\frac{1-\sigma_x}{2}} (\mathcal{S}_C)^{1|0}]
\end{equation} 
First trace out all the degrees of freedom without swap operators; This turns the nonnegative $ \rho_I\rho_\Gamma $ into another nonnegative operator. Then, the swappers make the remaining trace into the trace of its square, which is nonnegative. All projectors onto fixed-spin subspaces involved are nonnegative operators. We thus see that every term is nonnegative in principle.
So, the only ways to reduce the sum in \ref{ProbAvg1} is to make terms smaller or leave them out entirely. For this, we first study the $ \Delta $-constraints. Note that $ \Delta_{1|0}(\ind{j},\ind{j},\vec{\sigma}) = 1 $ regardless of the configuration, as by our previous considerations. However, we can choose our intertwiner and link data such that the remaining cases vanish as often as possible. Let 
\begin{equation}\label{key}
	A_{\ind{j},\ind{k}} :=\{e\in L : j_e = k_e\}
\end{equation}
be the set of links where the areas of two spin sectors agree. Valid glueings between the two geometries can only happen on this set. If now all involved spin sectors in our chosen data have $ A_{\ind{j},\ind{k}} = \emptyset $ for $ \ind{j}\neq\ind{k} $, as for example in the unit matrix $ \mathbb{I} $, then there is no constraint as only glueings between copies of the same geometry can happen. These states are classical superpositions of definite geometries (again, only in the sense of definite areas). In this case, the sum collapses to the diagonal pairs:
\begin{equation}\label{key}
	\frac{Z_1}{Z_0} 
	= 
	\sum_{\ind{j}}  P(\ind{j},\ind{j}) 
	\sum_{\vec{\sigma}}
	\frac{e^{-\mathcal{H}_{1}(\ind{j},\ind{j},\vec{\sigma})}}{Z^{(\ind{j},\ind{j})}_{0}}
	=
	\sum_{\ind{j}}  P(\ind{j},\ind{j}) 
	\avg{e^{-Q}}_{Z^{\ind{j}_0}}
\end{equation}
The geometries in these superpositions are clearly distinguished by their areas. In our optimisation, we can restrict to this subset of states. Next, we need to find, for a fixed spin sector $ \ind{j} $, the value of $ \avg{e^{-Q}}_{Z^{\ind{j}_0}} $. This is simply a result from previous research from the fixed-spin case\cite{colafranceschiHolographicMapsQuantum2021}.\\
If all spins in a given sector are large enough, we can perform a crucial approximation to the partition sums.
In the Ising model, we may approximate the partition function by its ground state contribution if the excited states have very low weight. This is the case if the couplings of the model are very large, as any spin flip will increase the energy by an amount proportional to that coupling constant. When the spins are all large, we may approximate, in particular, 
\begin{equation}\label{key}
	Z^{(\ind{j},\ind{k})}_{0} \approx 1 , \qquad Z^{(\ind{j},\ind{k})}_{1} \approx \exp(-\mathcal{H}_{1|0}(\ind{j},\ind{k},\vec{\sigma}_{GS}) )
\end{equation}
which massively simplifies the distribution $ P $, as well:
\begin{equation}\label{key}
	Z_0 = \sum_{\ind{j},\ind{k}}K_{\ind{j}}K_{\ind{k}}Z^{(\ind{j},\ind{k})}_{0} \approx
	(\sum_{\ind{j}}K_{\ind{j}})^2
	\qquad P(\ind{j},\ind{k}) \approx p_{\ind{j}}p_{\ind{k}} \qquad p_{\ind{j}} = \frac{K_{\ind{j}}}{\sum_{\ind{k}}K_{\ind{j}}}
\end{equation}
In particular, given that $ Z_0 = \avg{Tr[\rho]^2}_U $, we can interpret the factorisation of the partition function as the statement $ \avg{Tr[\rho]^2}_U = \avg{Tr[\rho]}^2_U $ in the high-spin regime.
The given spins serve as an external scale for the fixed-spin Ising models defined by \ref{PFfixedSpins} and provide a notion of temperature to it. The diagonal Hamiltonians take the simplified form
\begin{equation}\label{key}
	\mathcal{H}_{1|0}(\ind{j},\ind{j},\vec{\sigma}) 
	=
	\sum_{e \in \bnd} (\frac{1-\sigma_{s(e)} h_{t(e)}}{2})\log (d_{j_e})
	+
	\sum_{e \in \Gamma} (\frac{1-\sigma_{s(e)} \sigma_{t(e)}}{2})\log (d_{j_e})
	+
	S_2((\rho^I_{\ind{j}})_\downarrow)
\end{equation}
Then let us restrict our study to those states where the minimal area spin is large enough to suppress contributions to partition sums coming from the non-minimal configurations. Assuming that the energy gap in the two Hamiltonians $\mathcal{H}_{1|0}$ is equal, we can achieve a straightforward bound on the relative error
\begin{equation}
	|\frac{\frac{Z^j_1}{Z^j_0}-e^{-E_1}}{e^{-E_1}} | \leq \frac{2^N -1}{e^{\Delta E}}.
\end{equation}
In particular, if we assume the energy gap grows with the minimal area present in the geometry, as it does when $S_2(\rho^I) = 0$, where $\Delta E = \log(d_{j_{min}})$, then raising the minimal area will suppress all higher configurations to a certain degree. The minimal configuration, in turn, is built from a minimal surface $\Sigma$ in the geometry. The result from the fixed-spin case is schematically 
\begin{equation}
	e^{-E_1} \approx (\prod_{e \in \Sigma} d_{j_e})^{-1} e^{-S_2((\rho^I)_\downarrow)}
\end{equation}
In seeking a smaller $\frac{Z_1}{Z_0}$, we then see that including only large geometries is beneficial to making the reduced spin network state more mixed. We may for example adjust our probability weights $P$ to do this in order to get a smaller value. The optimal case is then one where all areas are equal to the upper cutoff $ J $ we imposed earlier, and thus only one single spin sector is present in the state. We emphasize that this is only the states which have a higher mixedness than the others. For them, we can estimate 
\begin{equation}
	\frac{Z^j_1}{Z^j_0} \approx  d_{J}^{-||\Sigma||} e^{-S_2((\rho^I_J)_\downarrow)}.
\end{equation}
And then, as the spin sector is homogeneous, we know that holographic behaviour is not possible. So, the most mixed state over the set of states for arbitrary spin sectors is not holographic.\\
However, as the following example shows, this is an ill-posed optimisation. We should instead fix the spin sectors under consideration and only adjust the remaining parameters. 
\subsection{Example calculation}
\begin{figure}
	\centering
	\includegraphics[width=0.5\linewidth]{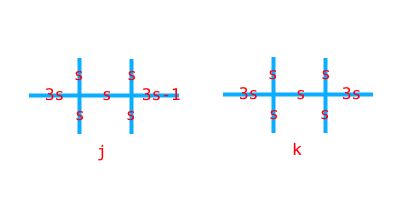}
	\caption{ We consider a superposition of two spin sectors which are only different on a single link. The values are chosen as to give a controllably low intertwiner space dimension of 1 or 2.}
	\label{fig:glueing}
\end{figure}
To illustrate the complexity of the calculations involved, we shall give an innocuous example. The graph in question is the simplest nontrivial one, and we only consider two distinct spin sectors, which agree on all but one link. This means that there are only four Ising configurations. We choose the area spins such that only one intertwiner space is nontrivial. We will also refer to the vertices just by L(eft) or R(ight) for convenience. In spin sector $\ind{k}$, all intertwiner spaces are 1-dimensional, while in $\ind{j}$, the right one is 2-dimensional. We can then give the intertwiner state 
\begin{equation}
	\rho^I= \begin{bmatrix}
		\rho^I_{\ind{j},\ind{j}} & \rho^I_{\ind{j},\ind{k}} \\
		\rho^I_{\ind{k},\ind{j}} & \rho^I_{\ind{k},\ind{k}}
	\end{bmatrix}\qquad
	\rho^I_{\ind{j},\ind{j}} = \begin{bmatrix}
		a & b \\
		\bar{b} & d
	\end{bmatrix}
	\qquad   \rho^I_{\ind{j},\ind{k}} = \begin{bmatrix}
		u \\
		v
	\end{bmatrix} \qquad \rho^I_{\ind{k},\ind{k}} = w = 1-(a+d)
\end{equation}
From this, we can derive the reduced entropies 
\begin{equation}
	S_2((\rho^I_{\ind{j},\ind{j}})_R) =S_2((\rho^I_{\ind{j},\ind{j}})) = -\log\left[ \frac{a^2+d^2+|b|^2}{(a+d)^2} \right] \qquad S_2((\rho^I_{\ind{j},\ind{j}})_L) = 0
\end{equation}
while the ones for $\ind{k}$ vanish.\\
We choose the area $ C $ to be the rightmost link, where the two sectors disagree. The individual sectors induce, at least for large spins, an isometry, making the only question what parameters need to be chosen to make their superposition induce an isometry.\\
Explicit calculation as in Appendix C gives us an expression for the purity, in the large-spin limit: 
\begin{equation}
	\frac{Z_1}{Z_0} 
	=
	\frac{2 w^2-2 w+1}{6 s}
	+
	\frac{(1-2 w)^3}{36 s^2}
	+
	O\left(\left(\frac{1}{s}\right)^3\right)
\end{equation}
Particularly, the lower bound of $ \frac{1}{12s} $ is achieved in the regime of large $ s $ for $(a,d,w)\approx(\frac{1}{4},\frac{1}{4},\frac{1}{2})$. The dimension of the boundary link space is precisely $ (6s+1)+(6s-1) = 12s $, so an isometry can be achieved with these parameters.\\
From this simple example, we already gain valuable information. First, the set of spin sectors under consideration must be fixed to have a more tractable, as well as interesting, minimisation problem. Additionally, it is clear that the intertwiner state has to be adjusted depending on the boundary region chosen and that different spin sectors may contribute with different weights. Therefore, the question of characterising holographic states becomes a fairly involved question, as the number of parameters for the intertwiner states grows very fast with number of vertices, spin sectors and spin values. To illustrate this, consider only $K$ spin sectors which are all homogeneous. The intertwiner spaces for each sector have dimensions $\{D_k=(2j_k+1)^N\}^K_{k=1}$, such that the intertwiner state has $(\sum_k D_k)^2-1$ real parameters before imposing positivity constraints. For large enough $N$ or large spins, this scales as $4^N (\sum_k s^N_k)^2$. The curse of dimensionality binds us to the regime of very low $N$ if we wish to do numerical investigations without heavy approximations. \\
\begin{figure}
	\centering
	\includegraphics[width=0.5\linewidth]{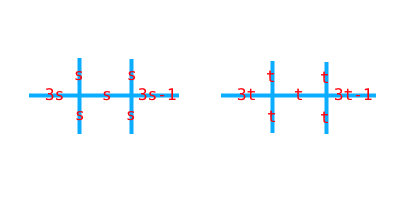}
	\caption{ For the second example, consider two spin sectors of the same form, differing only in spin. Once again, intertwiner space dimensions are kept at 1 or 2.}
	\label{fig:glueing}
\end{figure}
We will consider another instructive example. Consider the same graph structure and boundary region as before, but make the two spin sectors of the same form, differing only in their average area. Then, as no two edge spins agree anywhere, the constraints make all off-diagonal partition sums vanish. The sum simplifies to 
\begin{equation}
	\frac{Z_1}{Z_0} = P(j,j) \frac{Z_1^j}{Z_0^j} + P(k,k) \frac{Z_1^k}{Z_0^k}
\end{equation}
Take both of these spins to be fairly large, with fraction $\nu \coloneqq \frac{d_t}{d_s}=\frac{2t+1}{2s+1} \approx \frac{t}{s}$. Then, we may approximate the quotients as before by their ground state value - e.g., for choosing $C$ to be the right link as before,
\begin{equation}
	\frac{Z_1^j}{Z_0^j} = d_{3s-1}^{-1}
\end{equation} and same for the other sector, replacing s with t. We can also easily write the probability weights for the two sectors as 
\begin{equation}
	P(j,j)= \left[1+(\frac{|g_t|^2}{|g_s|^2}\frac{d_t^4d_{3s}d_{3s-1}}{d_s^4d_{3t}d_{3t-1}}\frac{c_k}{c_j})\right]^{-2}, \qquad 
	P(k,k)= \left[1+(\frac{|g_t|^2}{|g_s|^2}\frac{d_t^4d_{3s}d_{3s-1}}{d_s^4d_{3t}d_{3t-1}}\frac{c_k}{c_j})^{-1}\right]^{-2}
\end{equation}
In the large spin limit and taking $ \frac{|g_t|^2}{|g_s|^2} \approx 1 $, this is approximately
\begin{equation}
	P(j,j)= \left[1+(\nu^6\frac{c_k}{c_j})\right]^{-2}, \qquad 
	P(k,k)= \left[1+(\nu^6\frac{c_k}{c_j})^{-1}\right]^{-2}.
\end{equation}
Note that this is the same as approximating the geometries by homogeneous ones.
After factorising, this yields
\begin{equation}
	\frac{Z_1}{Z_0} = d_{3s-1}^{-1}(\frac{1}{(1+\nu^{6}a)^2}
	+\frac{\nu^{-1}}{(1+\nu^{-6}a^{-1})^2})
\end{equation} where $a= c_k/c_j$. We can now characterise which states are holographic: 
We wish the previous expression to be 
\begin{equation}
	(\prod_e d_{j_e}+d_{k_e})^{-1} \approx (\prod_e d_{j_e}(1+\nu))^{-1} =  (1+\nu)^{-1} (d_{3s-1})^{-1}.
\end{equation}
So the isometry constraint is
\begin{equation}\label{key}
	(1+\nu)(\frac{1}{(1+\nu^{6}a)^2}
	+\frac{\nu^{-1}}{(1+\nu^{-6}a^{-1})^2}) \approx 1
\end{equation}
which has solution $ a \approx \nu^{-5}$, or equivalently $ (c_j,c_k) = (\frac{1}{1+\nu^{-5}},\frac{1}{1+\nu^5}) $. This is independent of the intertwiner state chosen, as it was for the individual sectors.
It is thus easy to see that if one geometry (j) is large in terms of areas and the other (k) small, we must give a lot of weight to the smaller geometry to make their combination induce an isometry. As they approach each other (corresponding to $ \nu \rightarrow 1 $), the two sectors require equal weight, as one might expect from intuition.\\
Let us generalise this setting: Assume that we work with $ K $ spin sectors sharing no two spins, labeled $ s_k $, and fix a boundary region such that for each sector, the fixed-spin purity is the minimal value, which makes the sector induce an isometric map. By our arguments in the appendix, this should generically leave a lot of parameters free. So, $ \frac{Z_1^k}{Z_0^k} = \dim(\hbb_{C,n})^{-1} = (\prod_{e\in C}d_{s_k,e})^{-1} $. For this, we assume intertwiner data has been adjusted as necessary. Then, we may still adjust the $ c_n $. The condition for isometry becomes 
\begin{equation}\label{key}
	\begin{split}
		\frac{Z_1}{Z_0}\dim(\hbb_C) = 
		\sum^K_{n=1} P(n,n) \prod_{e \in C} \sum_{j \in \{s_{k,e}: k \in [1:K]\}} \frac{d_j}{d_{s_{n,e}}} = \\
		\sum^K_{n=1} P(n,n) \prod_{e \in C}  \left(\frac{d_{s_{n,e}}}{{\sum^K_{m=1}}d_{s_{m,e}}}\right)^{-1}
		=
		\sum^K_{n=1} P(n,n) \prod_{e \in C}  \rho_e^{-1}
		\stackrel{!}{=} 1.
	\end{split}
\end{equation}
In the end, this condition is a polynomial one: Writing, with $ \tilde{K}_n = \prod_{ e\in \bnd} d_{j_{n,e}} \prod_{ e\in \Gamma} |g_{j_{n,e}}|^2 $,  $ P(n,n) = \frac{c_n^2 \tilde{K}_n^2}{\sum_m c_m^2 \tilde{K}_m^2} $ and $ u_n = \prod_{e \in C} \rho^{-1}_e $, this is equivalent to
\begin{equation}\label{key}
	\sum_n \tilde{K}_n^2 c_n^2 (u_n-1) = 0,
\end{equation}
which is just a multinomial of degree 2 in the variables $ \{c_n\}_n $. As long as some $ u_n < 1 $, this should admit solutions in the region of interest $ 0<c_n <1 , \sum_n c_n = 1$. For these $ u_n $, the value of $ c_n $ can then be large, while the other sectors will need small $ c_n $ weights to make the sum 0. Additionally, large spin sectors will have their weights be suppressed by $ \tilde{K}_n^{-2} $. The existence of holographic states is thus not much of an issue - as long as there are enough parameters and the spin sectors are not too uncooperative, we should be able to reach the level set of holographic states by adjusting a few parameters. \\
The next question becomes the dimensionality of this set. As shown in the appendix and used analogously here, we expect the set of holographic state parameters to be of codimension one in the full parameter space. So, even though we fixed some parameters per spin sector already, we are technically free to change some more once we consider the superposition - there is not even a requirement for each individual sector to be holographic, per se. \\
Let us attempt the above in yet slightly larger generality. To achieve holography, we require $ \frac{Z_1}{Z_0}\dim(\hbb_C) =1 $, or, in the regime of large spins
\begin{equation}\label{key}
	0 = \sum_{m,n} \tilde{K}_n c_n \tilde{K}_m c_m (\frac{Z^{m,n}_1}{Z^{m,n}_0} - \frac{1}{\dim(\hbb_C)}) = 	\sum_{n} \tilde{K}^2_n c^2_n  (\frac{Z^{n,n}_1}{Z^{n,n}_0} - \frac{1}{\dim(\hbb_C)}) + 2\sum_{m\neq n} \tilde{K}_n c_n \tilde{K}_m c_m (\frac{Z^{m,n}_1}{Z^{m,n}_0} - \frac{1}{\dim(\hbb_C)})
\end{equation}
which we can easily analyse: In the diagonal part of the sum, $ \frac{Z^{n,n}_1}{Z^{n,n}_0} $ is the purity of the part of the full state contained in sector $ n $. This means that the usual entropy bound applies: $ 1 \geq \frac{Z^{n,n}_1}{Z^{n,n}_0} \geq \frac{1}{\dim(\hbb_{C,n})} \geq \frac{1}{\dim(\hbb_{C})} $, meaning all terms in the diagonal part are nonnegative. So, to achieve holography, the off-diagonal terms are crucial. In the best case, where $ Z^{m,n}_1 = 0 $ off of the diagonal, the entire off-diagonal sum is negative and holography might be achieved by adjusting the $ c_n $. \\
\, \\
As our final generalisation, take $ \beta_{m,n} = \frac{Z^{m,n}_1}{Z^{m,n}_0}\dim(\hbb_C) - 1  $ and $ \alpha_n = \tilde{K}_n c_n $, then the condition can be written as $ \avg{\alpha,\beta \alpha} = 0 $.  We then can find solutions iff the $ K\times K$ matrix $ \beta $ has negative or $ 0- $eigenspaces. In the simplest case, $ \beta_{m,n} = \delta_{m,n} \frac{1}{\rho_n} -1 $, where $ \rho_n = \frac{\dim(\hbb_{C,n})}{\dim(\hbb_{C})} $ sums to 1. \\
This matrix has the general determinant $ \det(\frac{1}{\rho})(1 - \Tr(\rho)) = 0 $, so there is a 0-eigenspace which admits a solution. This space is in fact spanned by the solution $ \alpha_n = \dim(\hbb_{C,n}) $. Take an $ \alpha $ to be from this eigenspace and recover the corresponding $ c_n = \frac{\dim(\hbb_{C,n})}{\tilde{K}_n} $. Their sum is not 0, so we may rescale $ \alpha $ within this eigenspace to normalise the sum to $ \sum_n c_n = 1 $, yielding holography for the state. The final result for the weights of the individual sectors for superpositions where one can take $ Z_1^{m,n} \sim \delta_{m,n} $ and the individual spin sectors holographic, is then
\begin{equation}\label{key}
	c_n = \left(\sum^K_{m=1}\frac{\prod_{e\in C^c}d_{j_{n,e}} \prod_{e\in \Gamma}|g_{j_{n,e}}|^2}{\prod_{e\in C^c}d_{j_{m,e}} \prod_{e\in \Gamma}|g_{j_{m,e}}|^2}\right)^{-1}
\end{equation}
in the high-spin regime.
In particular, this result extends to cases where the off-diagonal partition sums may not be zero, but numerically negligible. Also, the general condition for a 0-eigenspace to exist is, in the same vein as earlier, that 
\begin{equation}\label{key}
	\sum_{m,n} (Q^{-1})_{m,n} = 1
\end{equation}
(where $ Q_{m,n} = \frac{Z^{m,n}_1}{Z^{m,n}_0}\dim(\hbb_C)$). For diagonal $ Q $, this is precisely the statement that $ \sum_{n} (\frac{Z^{n,n}_1}{Z^{n,n}_0})^{-1}  = \dim(\hbb_{C})  $ - in other words, that each sector is by itself holographic. In this way, we have found a general sufficient and necessary criterion for holography for the case where the off-diagonal sums are negligible.\\
A particularly simple case is when each internal link has the same weight for the possible values of spins, $ g_{j_{m,e}} = \frac{1}{\sqrt{K}} $:
Then the weight is given by 
\begin{equation}\label{key}
	\tilde{K}_n c_n = (\frac{\dim(\hbb_{C^c,n})}{\dim(\hbb_{C^c})})^{-1}
\end{equation}
Heuristically, this confirms what we saw before: For spin sectors with larger geometry, the weight must be smaller than for those with small geometries. In fact, our earlier example gets precisely the same values of $ c_1 = \frac{1}{1+\nu^5} $ as before.

\section{Averages of observables}
We can also study boundary observables in this setting of typical spin network states. For this, we use the following analogue of the replica trick:
\begin{equation}\label{key}
	\avg{X}_\rho\avg{Y}_\rho = \frac{\Tr(\rho X)\Tr(\rho Y)}{\Tr(\rho )^2} = \frac{\Tr(\rho^{\otimes 2} (X\otimes Y))}{\Tr(\rho\otimes\rho )} =
\end{equation}
which allows us to use the same averaging method as before. The reason we introduce two observables $ X,Y $ at this point is the following fact:
\begin{equation}\label{key}
	\avg{\avg{XY}_\rho\avg{\mathbb{I}}_\rho}_U = \avg{\avg{XY}_\rho}_U \neq \avg{\avg{X}_\rho\avg{Y}_\rho}_U
\end{equation}
which hints that the measure of correlation of the two observables is not entirely erased by the average over the vertex wavefunctions. We can thus not only calculate average expectation values, but also average correlations. Importantly, we can derive the same type of random Ising model as before, with the boundary part replaced by a quantity related to $ X,Y $:
The $ (\ind{j},\ind{k}) $-boundary contribution in the numerator is now 
\begin{equation}\label{Boundary1}
	\Tr_{\hbb_{\bnd,\ind{j}}\otimes\hbb_{\bnd,\ind{k}}}[\bigotimes_{e \in \bnd}\mathcal{S}_{e}^{\frac{1-\sigma_{s(e)}}{2}} (X\otimes Y)
	] =
	\Tr_{\hbb_{\bnd,\ind{j}}}[\mathbb{I}]
	\Tr_{\hbb_{\bnd,\ind{k}}}[\mathbb{I}] e^{-c_{XY}(\ind{j},\ind{k},\vec{\sigma})}
\end{equation}
where 
\begin{equation}\label{key}
	c_{XY}(\ind{j},\ind{k},\vec{\sigma}) = -\log\left[\frac{
		\Tr[\Pi_{j_\downarrow}\Pi_{j_\uparrow} X \Pi_{j_\uparrow} \Pi_{k_\downarrow} \Pi_{k_\uparrow} Y \Pi_{k_\downarrow}\Pi_{j_\downarrow}]
	}{\Tr_{\hbb_{\bnd,\ind{j}}}[\mathbb{I}]
		\Tr_{\hbb_{\bnd,\ind{k}}}[\mathbb{I}]}\right].
\end{equation}
Here, we have labeled for simplicity $ \Pi_{j_\uparrow} $ the projector onto the boundary subspace of links in $ S_\uparrow $ with spins from sector $ \ind{j} $, others analogous. In particular, for the all-up configuration, this quantity turns into a sum, signalling again the glueing on the spin-down region. 
The Hamiltonian for the numerators $ Z^{\ind{j},\ind{k}}_1 $ is then given as
\begin{equation}\label{key}
	\mathcal{H}_1(\ind{j},\ind{k},\vec{\sigma}) = 
	\sum_{e\in \Gamma} \frac{1-\sigma_{s(e)}\sigma_{t(e)}}{2}\log(d_{j_e}) 
	+ c_{XY}(\ind{j},\ind{k},\vec{\sigma})
	+ \Sigma_I(\vec{\sigma},\ind{j},\ind{k})
\end{equation}
and new boundary constraints may arise from the form of the quantities $ X,Y $.
However, the distribution of couplings $ P(\ind{j},\ind{k}) $ in the randomised Ising model will remain the same by design. In this way, we may use the results of entropy calculations as a starting point for characterising other boundary quantities.\\
We will calculate the expectation value of the area\footnote{Here, we use a simplified approximation to usual models of the area operator where all eigenvalues are given by $ j $ instead of $ \sqrt{j(j+1)} $. As we deal with large spins, this should not be an issue.} of the boundary region $ C $ when the geometry is in a holographic state. For this, we calculate 
\begin{equation}\label{key}
	\avg{\avg{A}_\rho\avg{\mathbb{I}}_\rho}_U = \avg{\avg{A}_\rho}_U \qquad \text{for } A = \sum_{e\in C} A_e=\sum_{e\in C} \sum_{j_e} j_e \mathbb{I}_{j_e}.
\end{equation}
Observables of the form 
\begin{equation}\label{key}
	X = \bigotimes_{e \in L} X_e
\end{equation}
always factorise $ X = X_\uparrow \otimes X_\downarrow $. Here we consider operators of the form $ X_e = \sum_{j_e} \lambda_{e,j} \ket{j m}\bra{j m} $, which are diagonal in spins. For these, the Hamiltonian contribution is particularly simple. There is a constraint $ \Delta_{\bnd} = \prod_{e\in \bnd\cap S_\downarrow}\delta_{j_e,k_e} $, and the factor for boundaries turns out to be
\begin{equation}\label{key}
	e^{-c_{X\mathbb{I}(\ind{j},\ind{k},\vec{\sigma})}} = \frac{\prod_{e\in L} \lambda_{e,j}
		\prod_{e\in \bnd\cap S_\uparrow} d_{j_e}
		\prod_{e\in \bnd\cap S_\downarrow} d_{j_e}
		\prod_{e\in \bnd\cap S_\uparrow} d_{k_e}
	}{
		\prod_{e\in \bnd\cap S_\uparrow} d_{j_e}
		\prod_{e\in \bnd\cap S_\downarrow} d_{j_e}
		\prod_{e\in \bnd\cap S_\uparrow} d_{k_e}
		\prod_{e\in \bnd\cap S_\downarrow} d_{k_e}
	}
	= \frac{\prod_{e\in L} \lambda_{e,j}}{\prod_{e\in \bnd\cap S_\downarrow} d_{j_e}}.
\end{equation}
For the area operator, we put $ \lambda_{e,j} = 1 $ for all but one $e $, where $ \lambda_{e,j} = j_e $. Then, we sum this over all edges of $ C $. The numerator, even after sum, is independent of the Ising configuration - therefore, it can be factored out from the fixed-spin partition sums. Remarkably, then,
\begin{equation}\label{key}
	Z^{\ind{j},\ind{k}}_1 = A_{C,\ind{j}} \sum_{\vec{\sigma}} \Delta_1(\ind{j},\ind{k},\vec{\sigma}) e^{-\mathcal{H}_0(\ind{j},\ind{k},\vec{\sigma})}
\end{equation}
which indeed includes the same Hamiltonian as in $ Z_0 $ after all, $ (\prod_{e\in \bnd\cap S_\downarrow} d_{j_e})^{-1} = e^{-\sum_{e \in \bnd}\frac{1-\sigma_{s(e)} h_{t(e)}}{2}\log(d_{j_e})} $ if we take $ h_x = 1 $ everywhere. This means that the only difference between numerator and denominator is in the $ \Delta $-factors. In particular, for $ \ind{k}=\ind{j} $ the quotient is always exactly 1. In the regime of high spins, we will thus approximate both numerator and denominator by the ground state value $ 1 $. 
Then, say for a superposition of $ K $ spin sectors,
\begin{equation}\label{key}
	\avg{A} = \sum^K_{n=1}P(m,n) \frac{Z^{m,n}_1}{Z^{m,n}_0}
	=\sum^K_{m,n=1}p_m p_n A_{C,n} = \sum^K_{n=1} p_n A_{C,n},
\end{equation}
which means that we may use our earlier result for the distribution. In the same assumptions as before:
\begin{equation}\label{key}
	p_n = \frac{\dim(\hbb_{C,n})}{\dim(\hbb_{C})} = \frac{\prod_{e\in C}\frac{d_{j_{n,e}}}{2}}{\sum_m \prod_{e\in C}\frac{d_{j_{m,e}}}{2}}
\end{equation}
So that for large enough spins we may approximate
\begin{equation}\label{key}
	\avg{A} = \sum_n p_n A_{C,n} \approx \frac{\sum_nA_{C,n}\prod_{e\in C}j_{n,e} }{\sum_m \prod_{e\in C}j_{m,e}} = 
	\frac{\sum_n A_{C,n}^2 }{(\sum_m A_{C,m})^2} \sum_m A_{C,m} = e^{-S_2((A_{C,n}))} \sum_m A_{C,m}
\end{equation}
which expresses that the expected area is proportional to the sum of the individuals, up to a factor between $ 1 $ and $ \frac{1}{K} $. The more spin sectors are superposed, the smaller the actual area may become. However, none of the areas can be equal by our earlier assumption - at best there is a difference of $ ||C|| $ between the values of $ A_{C,n} $. If, however, we let $ A_{C,n} = A_1 + n||C|| $, the prefactor still scales like $ \frac{4}{3K} $ as K grows larger. So in practice, the effective area of a boundary segment that gets holographically mapped onto the boundary complement may be between the \textit{mean} or the \textit{sum} of the individual areas in the superposition.\\
We can even make the same calculation for the variance $ \avg{A^2}-\avg{A}^2 $ of this same area in a similar way. Since the Hamiltonian factors for each of the two involved averages are
\begin{equation}\label{key}
	e^{-c_{A^2\mathbb{I}}(\ind{j},\ind{k},\vec{\sigma})} = \frac{\prod_{e\in L} \lambda_{e,j}^2}{\prod_{e\in \bnd\cap S_\downarrow} d_{j_e}} \text{ and }\qquad
	e^{-c_{AA}(\ind{j},\ind{k},\vec{\sigma})} = \frac{\prod_{e\in L} \lambda_{e,j}\lambda_{e,k}}{\prod_{e\in \bnd\cap S_\downarrow} d_{j_e}},
\end{equation}
the fixed-spin partition sums are given by 
\begin{equation}\label{varianceContributions}
	\frac{Z^{m,n}_1}{Z^{m,n}_0}_{A^2,1} = A_{C,n}^2\qquad \frac{Z^{m,n}_1}{Z^{m,n}_0}_{A,A} = A_{C,m}A_{C,n}
\end{equation}
so that the unitary average of the expectations turns out to be precisely the variance in the distribution $ p $:
\begin{align}\label{key}
	\avg{\avg{A^2}_\rho-\avg{A}_\rho^2}_U
	= \sum_n p_n A_{C,n}^2 - (\sum_n p_n A_{C,n})^2 
	= \avg{A^2}_p - \avg{A}_p^2& \\
	= \left(\frac{\sum_n A_{C,n}^3}{(\sum_n A_{C,n})^3} 
	- \left(\frac{\sum_n A_{C,n}^2}{(\sum_n A_{C,n})^2}\right)^2\right) (\sum_m A_{C,m})^2 
	= (e^{-S_3((A_{C,n}))} - &e^{-2S_2((A_{C,n}))}) (\sum_m A_{C,m})^2
\end{align}
For the large-K limit and the sequence of areas as before, the prefactor behaves approximately as $ \frac{2}{9K^2} $. 
On the other hand, if one takes only 2 geometries, one of which is much larger than the other, we find the opposite result: the prefactor of the area gets arbitrarily close to $ 1 $ ($ \sim 1-\frac{2A_2}{A_1} $), while the variance's still becomes very small ($ \sim \frac{4 A_2}{A_1} $).\\
We can finally see that not only do geometric properties of even simple types of quantum gravity states like the ones we studied depend sensitively on the way they are superposed, but also the relative sizes of geometries in the superposition. However, in holographic states, these dependencies are easily controlled and explicitly studied.

\chapter{Extensions and Limitations}
We will now highlight a few ways our analysis can be extended to other interesting questions, while highlighting which conceptual or practical problems arise.
\section{Alternative distributions}\label{AltDist}
When picking tetrahedral states at random, we must specify a distribution. This choice crucially influences the typicity result - for example, the support of the distribution may be different and produce a more 'coarse-grained' result, say by averaging over all tetrahedra simultaneously, as is presented in the appendix.\\
So, the choice of distribution matters for the statement we wish to make. The previously used constant distribution (with respect to the Haar measure) is characterised by having maximal entropy among \textit{all} distributions: The Lagrangian
\begin{equation}\label{key}
	L[\rho] = -\int_{SU(\hbb)}d\mu(U) \rho(U) \log(\rho(U)) - \alpha (\int_{SU(\hbb)}d\mu(U) \rho(U)  - 1 )
\end{equation}
has equation of motion $ \log(\rho(U)) + 1+\alpha = 0 $, so the distribution is constant.\\
We take this as the 'most conservative' guess for a distribution, as maximal entropy distributions are typically understood as having the least amount of useful information 'put in'. For the constant distribution, it implies that we choose our tetrahedra completely indiscriminately.\\
We can now characterise interesting distributions by the same principle: Say, perhaps, we wish to only pick tetrahedra with a certain \textit{mean area}
\begin{equation}\label{key}
	\bar{A} = \sum_{\mathbf{j}} (\frac{1}{D} \sum_{\alpha} j_\alpha) \; \mathbb{I}_\mathbf{j} ,
\end{equation}
which means we impose a further constraint on the distribution:
\begin{equation}\label{key}
	L[\rho] = -\int_{SU(\hbb)}d\mu(U) \rho(U) \log(\rho(U)) - \alpha (\int_{SU(\hbb)}d\mu(U) \rho(U)  - 1 ) 
	-\beta (\int_{SU(\hbb)}d\mu(U) \rho(U) \Tr[
	\bar{A} U \rho_0 U^\dagger
	]  - \mathcal{A})
\end{equation}
This Lagrangian has the similarly simple equation of motion with corresponding solution
\begin{equation}\label{key}
	\log(\rho(U)) + 1+\alpha +\beta \Tr[\bar{A} U \rho_0 U^\dagger] = 0 \qquad\implies\qquad \rho(U) = \frac{e^{-\beta\Tr[\bar{A} U \rho_0 U^\dagger]}}{Z(\beta)}
\end{equation}
which is just an analogue to the standard Boltzmann weight. This is the 'least biased' distribution still only giving only tetrahedra with the sought-for average surface area. In principle, we could have chosen other quantities such as the volume. More importantly, though, we have to begin from some reference state $ \rho_0 $. To consider its influence, notice that while the distribution depends on it, the only thing really playing a part is its conjugacy class. In other words, the only meaningfully distinct $ \rho_0 $ are not connected by conjugation with a unitary. So, our choices are reduced to $ \mathcal{D}(\hbb)/SU(\hbb) $, or, in lieu of a true quotient, the set of path components of the orbit space. Luckily, in finite dimensions, this is easy to describe:\\
If the matrix $ \rho_0 $ is diagonalisable, it is so by an $ SU(\hbb) $-matrix conjugation. So any diagonalisable matrix, modulo such conjugations,is specified by its eigenvalues, which, for density matrices on $ \mathbb{C}^m $, form a standard $ (m-1) $-simplex. In this simplex, we have to remove the corner vertices, since these correspond to pure states, which require a seperate treatment. In fact, the set of pure states is transitively acted upon by $ SU(\hbb) $, so any two pure states are connected through it. Collapsing the orbits then yields a single point. So, the full set of choices $ \rho_0 $ is given by an $ (m-1) $-simplex whose corner vertices are identified.
This means that choosing $ \rho_0 $ pure, the distribution does not have further parameters coming from it.\\
However, the distribution constructed this way has a reduced symmetry, as well. This can be seen especially well in the \textit{average state}
\begin{equation}\label{key}
	\hat{K}_A = \int_{SU(\hbb)}d\mu(U) \rho(U) \hat{U}\hat{\rho_0} \hat{U}^\dagger
\end{equation}
which satisfies
\begin{equation}\label{e}
	VK_AV^\dagger = K_{VAV\dagger} \qquad\forall V \in 
	SU(\hbb)
\end{equation}
so, is only invariant under the stabiliser, or commutant, of $ A $ in the unitary group. We can now see the success of the random tensor technique in a new light: It produces an average state with a particularly high symmetry in which evaluation of quantities is fairly easy to do. By introducing further constraints, we reduce this effective symmetry to the largest one that respects them. In the case above, the symmetry $ SU(\bigoplus_{\mathbf{j}} \hbb_\mathbf{j}) $ is broken down to $ \bigtimes_{\mathbf{j}}SU(\hbb_\mathbf{j}) $. It seems fitting to then instead consider an averaging only over this smaller group, in which the distribution is again invariant under all transformations. In practice, such finer averages make the analysis more involved. To see this, an average over all tetrahedra simultaneously has been performed in the appendix for demonstration, where the result turns out to be much easier to calculate.

\section{Superposition of graph structures}
We can leave the class of classical spin network states further by replacing the link states by more arbitrary ones. Instead of a maximally entangled state, consider for each open semilink of a vertex the $ (N-1) $ other possible vertices it can be connected to with the same colour $ \alpha $. The most general coloured state of these semilinks is a mixed state $ \rho^\Gamma_\alpha \in \bigoplus_{\{j^x_\alpha\}} \bigotimes_x V^{j^x_\alpha}$, and then the most general coloured link state is simply $ \rho^\Gamma = \bigotimes_\alpha \rho^\Gamma_\alpha $. However, this simple looking generalisation opens up a new, more pressing question. How is the boundary space identified? We can think of two solutions to this. 

The first is to identify 'sufficiently non-entangled' semilinks in the state, which are then taken as a boundary. This might be done by considering arbitrary bipartitions of the set of links and computing their entanglement. A maximal set of semilinks under the proposed entanglement threshold will be then a good boundary of the state. This, however, is only algorithmically determined. To actually work with this selected boundary, and ask questions, this choice must be characterised in a closed form expression.
The second solution is to reverse the problem: Instead of starting from a given link state, we first fix a boundary and then select only link states in the complement of the boundary semilinks.

An interesting point to make is that the operator on the links in $Z_{1|0}$ used so far is 
\begin{equation}
	\rho_L = \bigotimes_{e_{xy} \in \Gamma} \mathbb{I}_{e_{xy}} \otimes \bigotimes_{e_{x}\in \bnd }\mathbb{I}_{e_{x}}.
\end{equation}
So in words, it is a state where every link is uncorrelated with any other link and each individual one is maximally mixed. From this, we can think of a few generalisations. The first is to consider convex superpositions of different geometries, each with their own boundary. The second is to replace the maximally mixed states on each link by something more general. The third is to allow correlations between different links. While the first two options are fairly easy to implement, the third makes the analysis considerably less explicit by not letting us use local per-link terms in the Ising model. All of these are valid generalisations of the states considered so far, so a choice of study must be motivated by some application. 
Consider a generic coloured state of $ N $ tetrahedra with fixed spins and intertwiner data. One way to apply our results so far is to ask whether it is possible to produce a spin tensor network from the existing state. Then, our problem is the following: Given a generic mixed state, approximate it by a convex linear combination of spin tensor network states with different combinatorial patterns.\\
Additionally, we might consider a superposition of different node numbers. This confronts us with the nontrivial choice of how to embed the fixed-node-number spaces in a larger one. The simplest choice is to make these spaces orthogonal -  this choice is made in free GFTs. Another choice is to impose a kind of consistency condition that makes "similar looking" graphs close in the scalar product, even if they have a different number of nodes - this would be similar to the cylindrical consistency conditions of LQG. The GFT choice, of course, allows one to apply a different set of tools to study the state space.
Notably, the same issue with identifying boundaries arises in this case.
\section{GFT Dynamics in holographic calculations}
Let us consider the influence of dynamics of a GFT model on the previous calculations. The type of dynamics we have in mind is that defined by a coherent state path integral, implementing, in a weak sense, constraints that we wish to place on the degrees of freedom of the microscopic tetrahedra. 
The first thing we wish to explore is how we can recover the previous results from a 'free theory' type, quadratic action. For this, it is instructive to first consider an analogy with free scalar field theories on $  \mathbb{R}^n $.

\subsection{The free field theory as weak imposition of constraints}
Let us consider the unconstrained, 'off-shell' Hilbert space $ \hbb= L^2(\mathbb{R}^n) $of a free particle, here for simplicity in Euclidean spacetime. The on-shell constraint of the particle is given by $ P^2=m^2 $ for the momentum of the particle. This defines a projection $ \Pi $ as well as a constraint operator $ C = \frac{P^2}{m^2} - 1 $ and makes the on-shell space of states its kernel. In terms of the off-shell scalar product $ <,> $ on $ \hbb $, we can write a scalar product that respects this constraint as $ <a,b>_D = <a,\Pi b>  $. We can also achieve a projection onto the constrained subspace in different ways: First, we can take $ <a,b>_D = \lim_{\beta\rightarrow \infty} <a,e^{-\beta C} b>$, as states for which $ C\neq 0 $ will be exponentially suppressed. A different way of achieving the same thing is through a highly oscillatory integral: If again the constraint is not satisfied, then $ e^{isC} $ will integrate to 0 over $ s $. So, an integral $ \int_{0}^{\infty} e^{is(C+i\epsilon)} ds$ with a regulator $ \epsilon >0 $ will 'wash away' the parts that have $ C\neq 0 $. Then, a sensible scalar product is
\begin{equation}\label{key}
	<a,b>_D = \lim_{\epsilon \rightarrow 0} <a,b>_{D,\epsilon}  =\lim_{\epsilon \rightarrow 0} \int_{0}^{\infty} <a,e^{is(C+i\epsilon)} b> ds 
	= \lim_{\epsilon \rightarrow 0}<a,\mathcal{G}_\epsilon b>
\end{equation}
where we define the regularised \textit{Green 2-point operator} 
\begin{equation}\label{key}
	\mathcal{G}_\epsilon = \int_{0}^{\infty}e^{is(C+i\epsilon)} = i(C+i\epsilon)^{-1}
\end{equation}
where the second equality is by Schwinger's identity. Writing out the constraint yields
\begin{equation}\label{key}
	\mathcal{G}_\epsilon = \frac{i}{P^2 - m^2 + i \epsilon}
\end{equation}and inserting a resolution of the identity by momentum states to get the scalar product of coordinate eigenstates
\begin{equation}\label{key}
	<a,b>_{D,\epsilon} = \mathcal{G}_\epsilon(x,y) = \bra{x}\mathcal{G}_\epsilon\ket{y} 
	= \int_{\mathbb{R}^n} d\mu(p) \frac{i \bra{x}p\rangle \bra{p}y\rangle}{p^2 - m^2 + i \epsilon}
	= \int_{\mathbb{R}^n} d\mu(p) e^{-ip(x-y)} \frac{i }{p^2 - m^2 + i \epsilon}	
\end{equation}
which is just the usual 2-point function of free scalar field theory. From the unconstrained Hilbert space point of view, we can thus see the on-shell field theory either as the result of implementing the on-shell constraint or as imposed dynamics on an unconstrained field.
The partition function
\begin{equation}\label{key}
	Z_\epsilon[J] = \exp\left( \int J(x)\mathcal{G}_\epsilon(x,y)J(y)dxdy\right)
\end{equation} produces this scalar product and can be obtained from a coherent state path integral
\begin{equation}\label{key}
	\det(\frac{\Delta}{\pi})^{\frac{1}{2}}\int \mathcal{D}\phi e^{-(\phi,\Delta \phi) + (\phi,J)}
\end{equation}where $ \Delta $ is the formal inverse of the Green's function. However, as we have seen, that formal inverse is just the constraint itself. Thus, flipping the definition around, we can take the dynamics of the free theory to be given through the path integral
\begin{equation}\label{key}
	Z_C[J] = \det(\frac{C}{\pi})^{\frac{1}{2}}\int \mathcal{D}\phi e^{-(\phi,C \phi) + (\phi,J)}
\end{equation}
whose 2-point function will reproduce the scalar product in the constrained space. 1-particle states will be created through functional differentials with respect to J or equivalently through insertions of the coherent field. Arbitrary 1-particle states can then be put into the scalar product by superposition. This procedure extends to n-particle states straightforwardly and allows for a direct generalisation to arbitrary dynamics. In the above integral, for $ J=0 $ the extremising set of coherent field configurations is given by those that satisfy the constraint. For more general actions, then, the interpretation is that one implements a constraint given by the equations of motion, but weakly so.\\
We might use this approach of implementing dynamics in various ways to incorporate GFT dynamics. The way which we choose here is to rephrase the imposition of the gauge invariance constraint as a Schwinger-Dyson equation arising from the GFT path integral. The action
\begin{equation}\label{key}
	S(\phi) = \int [dg] \phi(g) [\vec{J}_{Diag}^2 - S(S+1)] \phi^{\dagger }(g)
\end{equation}
will enforce every vertex's intertwiner space to be a representation of spin $ S $ under the action of the diagonal subgroup of $ SU(2)^D $. The choice $ S=0 $ then retrieves the standard case of gauge invariant intertwiner spaces. The Schwinger-Dyson equation then asserts
\begin{equation}\label{key}
	[\vec{J}_{Diag}^2(g) - S(S+1)]\avg{\phi(g)\phi^{\dagger}(h)} = -i \hbar \delta(g,h)
\end{equation}
and many others. If we rephrase the entropy calculation in terms of GFT correlation functions, we can then impose the dynamics 'straightforwardly'. A test of this would be to recover the gauge invariant calculation starting from the unconstrained one.

\subsection{Expressing the trace as a GFT integral}
We will actually take a slightly different approach to proceed, which is however inspired by the above argument. We draw inspiration from condensed matter physics and write the trace 
\begin{equation}\label{key}
	\Tr [e^{-\beta H} X]
\end{equation}
as a coherent state path integral in periodic, Euclidean time. Here we use $ H = C $ both as a constraint and a Hamiltonian operator. This allows us to see the GFT constraint as the ground state space of the Hamiltonian $ H $ - so the GFT becomes the study of ground state properties of $ H $. The system on which the Hamiltonian will live is the same as before - A many-particle system of tetrahedra. If, in addition to the constraint, we have additional data on the tetrahedra, say from a scalar field $ \phi $, then the Hamiltonian should not vanish completely and provides actual time evolution. In that case, we may see the system as those additional data with its own Hamiltonian $ H_{Matter} $, but coupled to a background system in its ground state.
One way to analyse this trace is, as said, to go to a path integral representation. Then, by the usual way of coherent state bosonic path integrals,
\begin{equation}\label{key}
	\Tr [e^{-\beta H} X] = \int_{\phi(0)=\phi(\tau)} \mathcal{D}[\phi(\tau)\bar{\phi}(\tau)] e^{-S(\bar{\phi},\phi)} \bra{\phi(\beta)} X \ket{\phi(0)}
\end{equation}
\begin{equation}\label{key}
	S(\bar{\phi},\phi) = \int_0^\beta d\tau [ \bar{\phi}\partial_\tau\phi + H(\bar{\phi},\phi) ]
\end{equation}
So we integrate over fields which have an additional periodic time variable. In the limit of $ \beta \rightarrow \infty $, the action becomes an integral over $ \mathbb{R} $, and we lose the periodicity requirement. \\
Our purpose here is to use this method, roughly, to analyse traces of the form $ \Tr_{\mathbb{H}^{\otimes 2}}[(e^{-\beta H})^{\otimes 2} X] $, so we will have 2 independent field theories of $ \phi,\chi $ which mix only through matrix elements of $ X $. The Hilbert spaces in question are once again the (unconstrained/constrained) Fock ones. Let $ \mathbf{m},\mathbf{n} $ label a basis of the one-particle Hilbert space. Then, if X is given by some first-quantised 1-particle operator, we can write it as 
\begin{equation}\label{key}
	X = \sum_{\mathbf{m_1},\mathbf{n_1},\mathbf{m_2},\mathbf{n_2}} X_{\mathbf{m_1},\mathbf{n_1};\mathbf{m_2},\mathbf{n_2}} \ket{\mathbf{m_1}}_1\bra{\mathbf{n_1}}_1\otimes \ket{\mathbf{m_2}}_2\bra{\mathbf{n_2}}_2
\end{equation}
And its second quantised form will simply be 
\begin{equation}\label{key}
	X = \sum_{\mathbf{m_1},\mathbf{n_1},\mathbf{m_2},\mathbf{n_2}} X_{\mathbf{m_1},\mathbf{n_1};\mathbf{m_2},\mathbf{n_2}}
	\hat{\phi}_{\mathbf{m_1}}^\dagger\hat{\phi}_{\mathbf{n_1}} \otimes
	\hat{\chi}_{\mathbf{m_2}}^\dagger\hat{\chi}_{\mathbf{n_2}}	
\end{equation}
Similar extensions from 1st to 2nd quantisation apply for higher-body operators. In our case, we are interested in $ X $ being a density matrix for $ N\otimes N $ particles times a swap operator. As such, the expression for $ X $ will involve $ N $ fields of $ \phi,\phi^\dagger $ and $ \chi,\chi^\dagger  $ each. The density matrix part will factorise over the two independent field operators while the swap operator mixes them. In our 1-body example, the trace is then a combination of thermal correlation functions:
\begin{equation}\label{key}
	\Tr [(e^{-\beta H})^{\otimes 2} X] = \sum_{\mathbf{m_1},\mathbf{n_1},\mathbf{m_2},\mathbf{n_2}} X_{\mathbf{m_1},\mathbf{n_1};\mathbf{m_2},\mathbf{n_2}}
	\Tr [e^{-\beta H}\hat{\phi}_{\mathbf{m_1}}^\dagger\hat{\phi}_{\mathbf{n_1}}] 
	\Tr [e^{-\beta H}\hat{\chi}_{\mathbf{m_2}}^\dagger\hat{\chi}_{\mathbf{n_2}}]
\end{equation}
\begin{equation}\label{key}
	\avg{X}_{\beta} = \sum_{\mathbf{m_1},\mathbf{n_1},\mathbf{m_2},\mathbf{n_2}} X_{\mathbf{m_1},\mathbf{n_1};\mathbf{m_2},\mathbf{n_2}} 
	\avg{\hat{\phi}_{\mathbf{m_1}}^\dagger\hat{\phi}_{\mathbf{n_1}}}_{\beta}
	\avg{\hat{\chi}_{\mathbf{m_2}}^\dagger\hat{\chi}_{\mathbf{n_2}}}_{\beta}
\end{equation}
We should then get the expectation value we want by calculating the correlation functions in the zero temperature limit and contracting the tensor $ X_{\mathbf{m_1},\mathbf{n_1};\mathbf{m_2},\mathbf{n_2}}  $ with them. The image is that the tensor network given by $ X $ is capped off by the Feynman diagrams of the GFT, which are spin foams connecting the two ends of a cobordism between N particles/tetrahedra. In this sense, the spin foams are glued together into a closed discrete spacetime by putting the tensor network on the ends.\\
\begin{figure}
	\centering
	\includegraphics[width=0.7\linewidth]{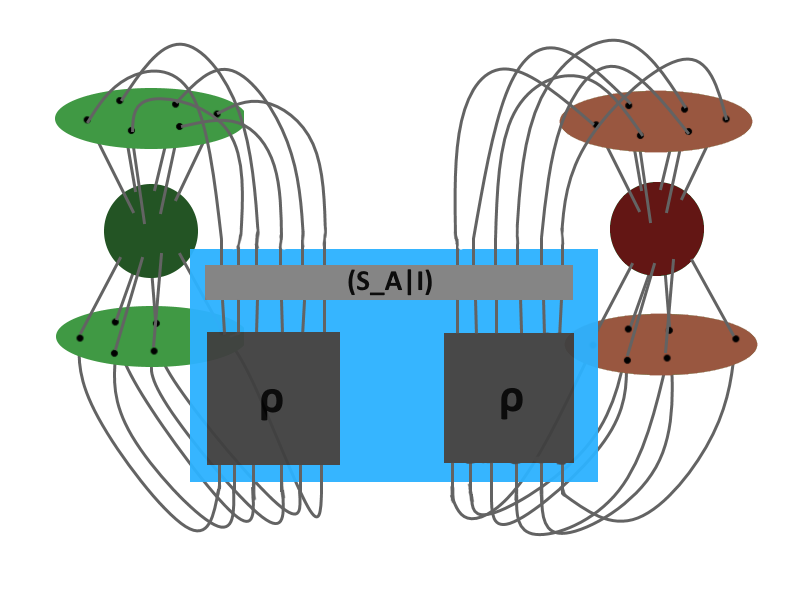}
	\caption{Schematic of how contractions happen in the setting of the traced two copies of a GFT. Two copies of the 2N-point functions of the GFT are contracted with density matrices. Preliminary image.}
	\label{fig:dynamics}
\end{figure}

Importantly, one need not use the condensed matter path integral for this. Any definition of correlation functions may function as a way to specify the GFT dynamics that influences the entropy calculation. In particular, one might ask how the above is related to the timeless GFT path integral: One gets the usual GFT path integral by integrating over fields which have trivial time dependence $ \phi(\tau ) = \phi $. They may still depend on group parameters or modes, but all nonzero Matsubara frequency modes of the periodic time path integral are dropped. Then, the integral is simply
\begin{equation}\label{key}
	\Tr [e^{-\beta H} X] = \int \mathcal{D}\phi\mathcal{D}\bar{\phi} e^{-\beta H(\bar{\phi},\phi) } \bra{\phi} X \ket{\phi}
\end{equation}
so one recovers the typical form of GFT path integrals. This reduction can be equivalently seen as replacing $ e^{-\beta H} $ by $ :e^{-\beta H}: $ in the initial trace. Of course, this means one is in principle calculating a very different object and sufficient care may be necessary to see which definition is more appropriate.\\
The simplest example of this is in the free GFT case, where we can choose our basis of field modes such that the correlation function $ \avg{\hat{\phi}_{\mathbf{m}}^\dagger\hat{\phi}_{\mathbf{n}}}_{\infty} = \delta_{\mathbf{m},\mathbf{n}} $. For example, we can reduce go from the unconstrained to the Gauge constrained Hilbert space by switching the modes from $ (\mathbf{j},\mathbf{m},\mathbf{n}) $ to $ (\mathbf{j},\mathbf{m}; S, \iota_S) $, where the label S specifies the representation under the diagonal subgroup of $ SU(2)^4 $ and $ \iota_S $ a basis of that space. One receives the constrained Hilbert space as the subspace for which $ S=0 $. However, in a theory which places a nontrivial spin on each tetrahedron, say through torsion in the classical spacetime induced by fermions, it might be necessary to dynamically set this constraint to something weaker. Then, the unconstrained Fock space might be necessary again. Either way, in the limit of $ \beta\rightarrow \infty $ the trace reduces to the already known contraction of tensor networks, with an added symmetry in the indices due to the permutation symmetry of the bosonic fields. Thus the calculation is in principle the same as before, but with a \textit{bosonic} or \textit{unlabeled} tensor network state.\\
Next we can consider typical simplicial GFT interactions. In 4D, these are of the form $ \hat{\phi}^5 + h.c. $ Regardless of colouring, we can see that due to our operator consisting of $ 2N $ fields $ \hat{\phi} $, interactions will change the relevant correlation function at order $ \mathcal{O}(\lambda^2) $ in the GFT coupling.\\
For a tensorial/bubble interaction with bipartite field set, this is different and will contribute at lower order as well. 
However, from this point on the conceptual issues are cleared up and the calculation involves once again 'just' contractions of tensors. Perturbative corrections to the entropy calculation will have to be renormalised and an analysis of the renormalisation group flow might be necessary. At each order in perturbation theory, then, we will have a sum over contracted spin foams.\\
Instead of calculating the correlation functions for a given GFT, we can instead ask once again which conditions we might impose on them to achieve holography. So, we see the CFs as another element in the tensor contraction which, along with intertwiner and core states, may be tweaked to achieve maximal entropy. In this way, we might find conditions for more generic spin tensor network states to admit holography, or see how a different scalar product, closer to that of LQG, could change the effective traces and entropies. Alternatively, one might select a dynamics \textit{by requiring} as many states as possible to be holographic in some form. Such questions should be explored in future work to answer whether holography could be a guiding	principle for finding simple spin foam models or GFTs that produce the right behaviour for a theory of quantum gravity.
\subsection{Extension of the spin foam picture}
As we saw in section \ref{SFMPF}, one may see the calculation of the two Ising partition functions \ref{PF3} as a spin foam partition function. More specifically, the only spin foams being summed over have the form of being the trivial ones connecting two copies of the graph $ \gamma $. From here, many generalisations through the GFT approach are more obvious. For one, these trivial spin foams are the only ones available in a \textit{free} GFT's expansion. So, when calculating transition amplitudes between spin network states in such a GFT, only these will contribute to the sum, which is represented as a spin foam model. So, the partition functions used to calculate the entropy become once again numbers calculated from correlation functions in the GFT. By replacing the GFT with an interacting one, yet leaving the boundary as-is, we will obtain corrections though nontrivial spin foams connecting the two copies of the graph. The difference to the approach presented above is that we have a direct path to obtain the free GFT generating the spin foam model from which we calculate the entropy, and that there is only one copy of the GFT involved instead of two. Said GFT will have additional variables on each vertex given by the $ \mathbb{Z}_2 $ Ising spin.\\
Sums over graphs, even with varying numbers of vertices are also more straightforward in this setting, as this just involves a sum over larger class of Feynman diagrams of the GFT. However, the interpretation of these quantities is still as conceptually unclear as in the approach outlined before.
\subsection{Implementing dynamically distinguishable quanta}
As another problem to consider, we need to write a tensor network state of indistinguishable tetrahedra in a 2nd quantised setting somehow to apply GFT dynamics. We propose a simple way to implement this on a technical level, along with some criteria for when this works.\\
The issue is that, while GFT dynamics are specified in a second-quantised form, we wish to study a tensor network state with fixed connectivity under the new scalar product. While we can technically implement this by starting from a first-quantised tensor network, then producing the equivalent second-quantised one, this produces different results from the ones we had before, even for the free theory, by virtue of symmetrising the network. This means that somehow, a labeling of the 2nd quantised state must be introduced to compare the results properly. From an operational point of view, an effective labeling of vertices can arise through additional degrees of freedom. Let us consider an $ N $-particle state in the symmetric Fock space $ \mathbb{F}_s(L^2(G^D\times \mathbb{R}^D)) $ with the special form
\begin{equation}\label{key}
	\ket{\Psi} = \int d\mu(g_1)\dots d\mu(g_N) f(g_1,\dots ,g_N) \ket{(g_1,x_1),\dots ,(g_N,x_N)},
\end{equation}
with some fully symmetrised function $ f $. We may understand the set of these states for fixed $ x_i \in \mathbb{R}^D $ as one living in the slightly different Fock space 
\begin{equation}\label{key}
	\mathbb{F}_s(L^2(G^D\times \mathbb{R}^D; d\mu_{Haar}\times d\mu_{Count})),
\end{equation}
in which only the measure on the second set of data has changed: Instead of integrating over all of $ \mathbb{R}^D $, only those fixed $ x_i $ are relevant now. Actually, we may see this $ L^2 $-space as 
\begin{equation}\label{key}
	L^2(G^D\times \mathbb{R}^D; d\mu_{Haar}\times d\mu_{Count}) \cong L^2(\sqcup_i G^D_i; \sum_i d\mu_{Haar,i}) \cong \bigoplus_i L^2( G^D_i; d\mu_{Haar,i})
\end{equation}
which is entirely analogous to a \textit{mode splitting} as on compact spaces: $ L^2(S_1) \cong \bigoplus_n \text{span}(e^{i2\pi nx}) $.
A key property of the Fock functor is that it has adjoints - it turns direct sums into tensor products, in particular.
\begin{equation}\label{key}
	\mathbb{F}_s(L^2(G^D\times \mathbb{R}^D; d\mu_{Haar}\times d\mu_{Count})) \cong  \bigotimes^N_{i=1} \mathbb{F}_s(L^2( G^D_i; d\mu_{Haar,i}))
\end{equation}
which is a simple factorisation that we can make use of: In the state above, none of the modes labeled by $ x_i $ have higher occupation numbers than 1. Therefore, when identifying tetrahedra by this mode label in some way, it makes sense to speak of \textit{the} tetrahedron with label $ x_i $. Moreover, the factorisation allows for a straightforward usage of reduced states and similar operations. In other words, we can see the starting state as one in the space of $ N $ distinguishable quanta, inducing a map
\begin{equation}\label{key}
	\bigotimes^N_{i=1} \mathbb{F}_s(L^2( G^D_i; d\mu_{Haar,i})) \rightarrow \bigotimes^N_{i=1} L^2( G^D_i; d\mu_{Haar,i})
\end{equation} mapping the 0-particle sectors to 0.\\
The remaining natural questions are when such states exist and how stable these identifications are under constraints or other dynamics. We think of the following criteria for this procedure of effective distinguishability to work:
\begin{enumerate}
	\item The system of indistinguishable must have a sufficiently large set of quantum numbers to identify them.
	\item Observables or similar quantities must exist that allow identifying values of these quantum numbers in a state.
	\item A state must be sharp in the values of these labels and said values must be distinguishable (on an operational level).
	\item The dynamics of the theory must make these values evolve continuously and without having them intersect too often. Possibly, the evolution of these labels should even be \textit{slow} compared to some typical timescales.
\end{enumerate}
A typical setting where these criteria are fulfilled is in low-energy solid state systems: A priori indistinguishable atoms arrange in a lattice, where their spatial positions are nearly conserved in time. This allows using these positions as labels to distinguish the atoms. From this example, one can already see that this effective distinguishability is made possible both by the states and the dynamics of the system only in a certain regime.
\chapter{Conclusion}
Notions of holography help in many different ways to understand and simplify quantum field theories with high degrees of complexity. In the case of gravity, there are several suggestions that degrees of freedom of QG might be equally well described by boundary theories.
However, most known cases of holography study dualities or recovery of information in a quasi-classical setting: There are always definite geometric quantities or even a semiclassical spacetime. Even in the tensor network case, one finds behaviour more reminiscent of a single metric or geometry. In this thesis, we made a few first steps in exploring what properties holographic quantum states must have, if they are true superpositions of distinct geometries. For this purpose, we applied techniques developed in the context of random tensor networks and combined them with ideas from systems with disorder to understand the specific case of RSTNs with multiple, but finitely many spin sectors. We uncovered a few key insights that likely have a more general analogue beyond our setting:
\begin{itemize}
	\item The Hilbert space of states does not factor into bulk and boundary parts, so one can only make typicity statements about classes of states with fixed bulk data.
	\item The purity of a reduced boundary state as well as averages of boundary observables can be mapped to a random Ising model with a distribution of couplings related to the relative weight the geometries have in the state.
	\item Such a random Ising model may also be understood as a kind of sum over background geometries for the Ising model to live on, thus providing a notion of quantum gravity path integral for it.
	\item For holographic states, for which the map from a boundary region $ C $ to its complement $ C^c $ is an isometry, said relative weight of geometries is required to be inversely proportional to the size of the spin sectors, in terms of their boundary area in $ C^c $.
	\item The expectation value of the area of $ C $ in such a state was found to be related in a very simple way to the sequence of areas of $ C $ in the involved spin sectors. In particular, the value is at least the $ mean $ and does not exceed the $ sum $ of the individual areas.
\end{itemize}
It is easy to envision many extensions of our work, particularly as we did not use most of our assumptions on the graph structure, like valency, colouring, etc. In principle, also the dimensions of the representation spaces are not essential and one might replace the $ SU(2) $-data by that of other Lie groups or other structures. We also highlighted possible extensions to consider superpositions of different graphs labeling the states, as well as the inclusion of dynamics in the form of a GFT path integral. We expect that, with more work in all of these directions, a more coherent picture of general properties of states in many QG approaches will emerge. In fact, as spin networks are used beyond discrete quantum gravity approaches, any context in which many-body quantum states and lattice gauge theory-like methods might make use of the extensions of RSTN holography that we have started to explore here. We expect the results of this work to be of use in investigating holographic properties of a large number of superposed spin sectors and entanglement patterns, which will play a key role in the description of semiclassical geometries in quantum gravity.\\
\chapter{Appendix}
\section{Average over all tetrahedra}
We can perform, to contrast our main approach, an average over all tetrahedral states as a whole, as opposed to averaging over individual tetrahedra's states. While this simplifies the calculations significantly, we lose the correspondence to a random Ising model and the bulk and boundary decouple as far as the boundary entropy is concerned. Still, we find that there is a leftover transition of a minimal surface that is not seen explicitly.
We choose as the state to be projected a general  
\begin{equation}\begin{split}\label{generalf}
		\ket{\Psi}= U \ket{\Psi_0}=\bigoplus_{\ind{j}}\ket{\Psi_{\ind{j}}}  = \bigoplus_{\ind{j}}\sum_{\ind{k}}U_{\ind{j},\ind{k}}\ket{\Psi_{0,\ind{k}}}
	\end{split}
\end{equation}
where $ U $ is a general unitary $ \mathbb{H} \mapsto \mathbb{H} $ and $ \ket{\Psi_0} $ is a reference state.	
Now, we can define the boundary state 
\begin{equation}\label{boundarystatenew}
	\ket{\Phi} = \bra{\theta}\bra{\zeta}\bra{\Gamma} \Psi \rangle \in  \mathbb{H}_{\bnd_{out}} = \bigoplus_{j_{\bnd_{out}}} \mathbb{H}^{\bnd}_{j_{\bnd_{out}}}
\end{equation}as before.

We are interested in the second Rényi entropy of this reduced state in dependence on the connectivity $ \Gamma $ and the presence of an interior $ \core $ and intertwiner data $ \rho_I $, from now on assumed to be a mixed state. For this, we need to use the replica trick again. 
\begin{equation}\label{replicatricknew}
	e^{-S_2(A)} 
	= \frac{\text{Tr}_{A} [ \rho_A^2 ]}{\text{Tr}_{A} [ \rho_A ]^2} 
	= \frac{\text{Tr}_{\mathbb{H}_{\partial\gamma}^{\otimes 2}} [ (\rho\otimes \rho) \mathcal{S}_A ]}{\text{Tr}_{\mathbb{H}_{\partial\gamma}^{\otimes 2}} [ \rho\otimes \rho ]} 
	\coloneqqrev \frac{Z_1}{Z_0}
\end{equation}
where the swap operator mixes all spin sectors.
By using the definition \ref{boundarystatenew}, we can write the objects $ Z_{1|0} $ as 
\begin{equation}\label{psumsnew}
	\begin{split}
		Z_{1|0} &=\Tr_{(\mathbb{H}_{\partial\gamma,in}\otimes\mathbb{H}_{\partial\gamma,out}\otimes\mathbb{H}_{b})^{\otimes 2}} \left[
		\core^{\otimes 2}
		\rho_I^{\otimes 2}
		\left(\ket{\Gamma}\bra{\Gamma}\right)^{\otimes 2} \left(\ket{\Psi}\bra{\Psi}\right)^{\otimes 2}( \mathcal{S}_A | \mathbb{I}) \right] 
	\end{split}		
\end{equation}
where we note that the trace is now performed not over the Hilbert space $ \mathbb{H} $, but $ \mathbb{H}^{\bnd} \otimes \hbb^I \otimes  \hbb^{\Gamma} $. 
However, because the state $ \Psi $ is from $ \mathbb{H} $ and thus 'diagonal' in $ \mathbb{H}_{\partial\gamma}\otimes \mathbb{H}_{b} $, we can switch to $ \mathbb{H} $ instead.
\begin{equation}\label{psumsnew2}
	Z_{1|0} = \Tr_{ \mathbb{H}^{\otimes 2} } \left[
	\core^{\otimes 2}
	\rho_I^{\otimes 2}
	\left(\ket{\Gamma}\bra{\Gamma}\right)^{\otimes 2} \left(\ket{\Psi}\bra{\Psi}\right)^{\otimes 2}( \mathcal{S}_A | \mathbb{I}_{\mathbb{H}} ) \right].
\end{equation}
We now need to perform an average over some distribution of unitaries $ U $ to proceed. 
The simplest way to do this is for a uniform choice of states $ \Psi $, corresponding to the Haar measure on the unitary group $ \mathcal{U}(\mathcal{H}_J) $, where
\begin{equation}\label{cutoffspace}	
	\mathcal{H}_J \coloneqq \bigoplus_{\ind{j},\mathbb{j} \leq j^x_\alpha \leq J} \mathbb{H}_{\ind{j}} \quad \text{with dimension } \mathcal{D}_J
\end{equation}
For this case, we may use the same application of Schur's lemma as before. If $ \avg{f}_U $ denotes the average of a function $ f $ on $ U(\mathcal{H}_J) $ with respect to the Haar measure, we have that 
\begin{equation}\label{cutoffSchur}
	\avg{\left(\ket{\Psi}\bra{\Psi}\right)^{\otimes 2}}_U = \frac{\mathbb{I}_{\mathcal{H}_J\otimes\mathcal{H}_J} + \mathcal{S}_{\mathcal{H}_J\otimes\mathcal{H}_J}}{\mathcal{D}_J(\mathcal{D}_J+1)}
\end{equation}
where S swaps the two copies of $ \mathcal{H}_J $.
We note two things. First, it is not possible to take the limit $ J \rightarrow \infty $ here, and do this argument in the full, untruncated Hilbert space. One can see this from the normalisation making the RHS vanish, while the LHS stays of trace 1 in that limit.
Second, contrast this to the previously considered case of a seperate average over each spin sector seperately. In that case, we had 
$ \avg{\ket{f_{\ind{j}}}\bra{f_{\ind{j}}}\otimes\ket{f_{\ind{j}}}\bra{f_{\ind{j}}}}_U = \frac{\mathbb{I}_{\ind{j}} + \mathcal{S}_{\ind{j}}}{\mathcal{D}_{\ind{j}}(\mathcal{D}_{\ind{j}}+1)}$, so that the current $ \avg{\left(\ket{\Psi}\bra{\Psi}\right)^{\otimes 2}}_U $ would be equivalent to $ \bigotimes_x \bigoplus_{\textbf{j}^x} \frac{\mathbb{I}_{\ind{j}} + \mathcal{S}_{\ind{j}}}{\mathcal{D}_{\ind{j}}(\mathcal{D}_{\ind{j}}+1)} $. This, inserted into the trace \ref{psumsnew2} would lead to a much more fine grained, vertex-dependent expression, which we do not have here.
Now, we insert this into \ref{psumsnew2} and proceed as usual.

\begin{equation}\label{psumsnew3}
	\begin{split}
		\avg{Z_{1|0}}_U = \Tr_{ \mathbb{H}^{\otimes 2} }
		\left[ 
		\core^{\otimes 2}
		\left(\ket{\zeta}\bra{\zeta}\right)^{\otimes 2}
		\left(\ket{\Gamma}\bra{\Gamma}\right)^{\otimes 2} \overline{\left(\ket{\Psi}\bra{\Psi}\right)^{\otimes 2}}( \mathcal{S}_A | \mathbb{I} ) \right]\\
		=\Tr_{ \mathbb{H}^{\otimes 2} } \left[ 
		\core^{\otimes 2}
		\left(\ket{\zeta}\bra{\zeta}\right)^{\otimes 2}
		\left(\ket{\Gamma}\bra{\Gamma}\right)^{\otimes 2}
		\frac{\mathbb{I}_{\mathcal{H}_J\otimes\mathcal{H}_J} + \mathcal{S}_{\mathcal{H}_J\otimes\mathcal{H}_J}}
		{\mathcal{D}_J(\mathcal{D}_J+1)}( \mathcal{S}_A | \mathbb{I} ) \right]\\
		=\frac{1}{\mathcal{D}_J(\mathcal{D}_J+1)}
		\Tr_{ \mathbb{H}^{\otimes 2} } \left[ 
		\core^{\otimes 2}
		\left(\ket{\zeta}\bra{\zeta}\right)^{\otimes 2}
		\left(\ket{\Gamma}\bra{\Gamma}\right)^{\otimes 2}
		\left(\mathbb{I}_{\mathcal{H}_J\otimes\mathcal{H}_J} + \mathcal{S}_{\mathcal{H}_J\otimes\mathcal{H}_J}\right)( \mathcal{S}_A | \mathbb{I} ) \right]\\
	\end{split}
\end{equation}
If we also neglect fluctuations in the low spin regime\footnote{This requires that the lower cutoff must scale polynomially in the number of vertices of the graph, as in $ \mathbb{j} >> N^k $ for some $ k > \frac{2}{\Delta E} $ with the spectral gap of the later Ising model.}, we may discard the $ J $-dependent prefactor in the quotient and write 
\begin{equation}\label{entropyfinalform1}
	\avg{e^{-S_2(A)}}_U = \avg{\frac{Z_1}{Z_0}}_U \approx \frac{\avg{Z_1}_U}{\avg{Z_0}_U} 
\end{equation}
However, we can no longer perform the conversion to an Ising model as before. The reason is that for that conversion, we need a tensor product $ \bigotimes_x\left(\mathbb{I}_{\mathcal{H}_x\otimes\mathcal{H}_x} + \mathcal{S}_{\mathcal{H}_x\otimes\mathcal{H}_x}\right) $ of operators acting on the tetrahedra individually. Working with completely generic classes of states $ \Psi $ has removed the local structure from the problem entirely, and thus averaging over it will not allow us to recover that local data.\\
Note that this is to be expected: An average \textit{removes} information from a distribution or random variable. The larger, or coarser, the average we perform, the more data we remove in the process. On the other hand, removing said data can pinpoint typical behaviour and allow for simpler calculations. In our case, the removal of local data clearly makes the calculation simpler - so simple, in fact, that we will not be able to talk about holographic surfaces or entanglement wedges or similar concepts, as those objects are not needed for the entropy calculation.\\
In fact, we can perform the calculation as-is. For this, it is actually convenient to work in the form of \ref{psumsnew}, where the traces over bulk and boundary can be performed seperately.
\begin{equation}\label{key}
	Y_{1|0} =\Tr_{(\mathbb{H}_{\partial\gamma}\otimes\mathbb{H}_{b})^{\otimes 2}} \left[
	\core^{\otimes 2}
	\rho_I^{\otimes 2}
	\left(\ket{\Gamma}\bra{\Gamma}\right)^{\otimes 2}
	\left(\mathbb{I}_{\mathbb{H}_{\partial\gamma}\otimes\mathbb{H}_{\partial\gamma}}\mathbb{I}_{\mathbb{H}_{b}\otimes\mathbb{H}_{b}} 
	+
	\mathcal{S}_{\mathbb{H}_{\partial\gamma}\otimes\mathbb{H}_{\partial\gamma}}\mathcal{S}_{\mathbb{H}_{b}\otimes\mathbb{H}_{b}}\right)
	( \mathcal{S}_A | \mathbb{I}) 
	\right]
\end{equation}
The key fact is that the bulk state has no support on the boundary, so it does not matter for the trace over the boundary space at all. We study both parts seperately and find:
\begin{align}\label{key}
		Y_{0} &=\Tr_{(\mathbb{H}_{\partial\gamma}\otimes\mathbb{H}_{b})^{\otimes 2}} \left[
		\core^{\otimes 2}
		\rho_I^{\otimes 2}
		\left(\ket{\Gamma}\bra{\Gamma}\right)^{\otimes 2}
		\left(\mathbb{I}_{\mathbb{H}_{\partial\gamma}\otimes\mathbb{H}_{\partial\gamma}}\mathbb{I}_{\mathbb{H}_{b}\otimes\mathbb{H}_{b}} +
		\mathcal{S}_{\mathbb{H}_{\partial\gamma}\otimes\mathbb{H}_{\partial\gamma}}\mathcal{S}_{\mathbb{H}_{b}\otimes\mathbb{H}_{b}}\right) \right]\\
		&=\Tr_{(\mathbb{H}_{\partial\gamma})^{\otimes 2}} \left[
		\core^{\otimes 2} \mathbb{I}_{\mathbb{H}_{\partial\gamma}\otimes\mathbb{H}_{\partial\gamma}}  \right]
		\Tr_{(\mathbb{H}_{b})^{\otimes 2}} \left[ 
		\rho_I^{\otimes 2}
		\left(\ket{\Gamma}\bra{\Gamma}\right)^{\otimes 2}  \right] \\
		&+\Tr_{(\mathbb{H}_{\partial\gamma})^{\otimes 2}} \left[   \core^{\otimes 2}
		\mathcal{S}_{\mathbb{H}_{\partial\gamma}\otimes\mathbb{H}_{\partial\gamma}} 
		\right]
		\Tr_{(\mathbb{H}_{b})^{\otimes 2}} \left[ 
		\rho_I^{\otimes 2}
		\left(\ket{\Gamma}\bra{\Gamma}\right)^{\otimes 2}  \mathcal{S}_{\mathbb{H}_{b}\otimes\mathbb{H}_{b}} \right]\\
		&=\Tr_{(\mathbb{H}_{\partial\gamma})^{\otimes 2}} \left[
		\core^{\otimes 2} \right]
		\Tr_{(\mathbb{H}_{b})^{\otimes 2}} \left[ 
		\rho_I^{\otimes 2}
		\left(\ket{\Gamma}\bra{\Gamma}\right)^{\otimes 2}  \right] 
		+\Tr_{(\mathbb{H}_{\partial\gamma})^{\otimes 2}} \left[   \core^{\otimes 2}
		\mathcal{S}_{\mathbb{H}_{\partial\gamma}\otimes\mathbb{H}_{\partial\gamma}} 
		\right]
		\Tr_{(\mathbb{H}_{b})^{\otimes 2}} \left[ 
		\rho_I^{\otimes 2}
		\left(\ket{\Gamma}\bra{\Gamma}\right)^{\otimes 2}  \right]\\
		&=\left(
		\Tr_{(\mathbb{H}_{\partial\gamma})^{\otimes 2}} \left[
		\core^{\otimes 2} \right]
		+
		\Tr_{(\mathbb{H}_{\partial\gamma})^{\otimes 2}} \left[   \core^{\otimes 2}
		\mathcal{S}_{\mathbb{H}_{\partial\gamma}\otimes\mathbb{H}_{\partial\gamma}} 
		\right]
		\right)
		\Tr_{\mathbb{H}_{b}} \left[\rho_I
		\ket{\Gamma}\bra{\Gamma}\right]^2 
\end{align}
While
\begin{equation}\label{key}
	\begin{split}
		Y_{1} =\Tr_{(\mathbb{H}_{\partial\gamma}\otimes\mathbb{H}_{b})^{\otimes 2}} \left[
		\core^{\otimes 2}
		\rho_I^{\otimes 2}
		\left(\ket{\Gamma}\bra{\Gamma}\right)^{\otimes 2}
		\left(\mathbb{I}_{\mathbb{H}_{\partial\gamma}\otimes\mathbb{H}_{\partial\gamma}}\mathbb{I}_{\mathbb{H}_{b}\otimes\mathbb{H}_{b}} +
		\mathcal{S}_{\mathbb{H}_{\partial\gamma}\otimes\mathbb{H}_{\partial\gamma}}\mathcal{S}_{\mathbb{H}_{b}\otimes\mathbb{H}_{b}}\right) 
		\mathcal{S}_A\right]\\
		=\left(
		\Tr_{(\mathbb{H}_{\partial\gamma})^{\otimes 2}} \left[
		\core^{\otimes 2} \mathcal{S}_A \right] 
		+
		\Tr_{(\mathbb{H}_{\partial\gamma})^{\otimes 2}} \left[   
		\core^{\otimes 2}
		\mathcal{S}_{\mathbb{H}_{\partial\gamma}\otimes\mathbb{H}_{\partial\gamma}} 
		\mathcal{S}_A
		\right]
		\right)
		\Tr_{\mathbb{H}_{b}} \left[\rho_I
		\ket{\Gamma}\bra{\Gamma}\right]^2 
	\end{split}
\end{equation}

Which leads to the interesting result that the entropy does not depend on the bulk at all - in fact, the global average completely erased the information about the bulk combinatorics:
\begin{equation}\label{key}
	\avg{e^{-S_2(A)}}_U =  \frac{Y_1}{Y_0} 
	=
	\frac
	{
		\Tr_{(\mathbb{H}_{\partial\gamma})^{\otimes 2}} \left[
		\core^{\otimes 2} \mathcal{S}_A \right] 
		+
		\Tr_{(\mathbb{H}_{\partial\gamma})^{\otimes 2}} \left[   
		\core^{\otimes 2}
		\mathcal{S}_{\mathbb{H}_{\partial\gamma}\otimes\mathbb{H}_{\partial\gamma}} 
		\mathcal{S}_A
		\right]
	}	
	{
		\Tr_{(\mathbb{H}_{\partial\gamma})^{\otimes 2}} \left[
		\core^{\otimes 2} \right]
		+
		\Tr_{(\mathbb{H}_{\partial\gamma})^{\otimes 2}} \left[   \core^{\otimes 2}
		\mathcal{S}_{\mathbb{H}_{\partial\gamma}\otimes\mathbb{H}_{\partial\gamma}} 
		\right]
	}
\end{equation}
We can simplify this further using the fact that $ \core $ only has support on the inner boundary, and $ \mathcal{S}_A $ only on the outer one. For example,
\begin{equation}\label{key}
	\Tr_{(\mathbb{H}_{\partial\gamma})^{\otimes 2}} \left[   
	\core^{\otimes 2}
	\mathcal{S}_{\mathbb{H}_{\partial\gamma}\otimes\mathbb{H}_{\partial\gamma}} 
	\mathcal{S}_A\right]
	=
	\Tr_{(\mathbb{H}_{\partial\gamma,out})^{\otimes 2}} \left[   
	\mathcal{S}_{\mathbb{H}_{\partial\gamma,out}} 
	\mathcal{S}_A\right]
	\Tr_{(\mathbb{H}_{\partial\gamma,in})^{\otimes 2}} \left[   
	\core^{\otimes 2}
	\mathcal{S}_{\mathbb{H}_{\partial\gamma,in}} \right]		
\end{equation}
which, when used in the above, yields
\begin{equation}\label{key}
	\begin{split}
		\avg{e^{-S_2(A)}}_U =  \frac{Y_1}{Y_0} =
		\frac
		{
			\Tr_{(\mathbb{H}_{\partial\gamma,out})^{\otimes 2}} \left[
			\mathcal{S}_A \right] 
			+
			e^{-S_2(\core)}
			\Tr_{(\mathbb{H}_{\partial\gamma},out)^{\otimes 2}} \left[   
			\mathcal{S}_{\mathbb{H}_{\partial\gamma,out}} 
			\mathcal{S}_A
			\right]
		}	
		{
			\Tr_{(\mathbb{H}_{\partial\gamma,out})^{\otimes 2}} \left[
			\mathbb{I} \right]
			+
			e^{-S_2(\core)}
			\Tr_{(\mathbb{H}_{\partial\gamma,out})^{\otimes 2}} \left[  
			\mathcal{S}_{\mathbb{H}_{\partial\gamma,out}} 
			\right]
		}\\
		=
		\frac
		{
			\dim(\mathbb{H}_{A}) \dim(\mathbb{H}_{\bar{A}})^2
			+
			e^{-S_2(\core)}
			\dim(\mathbb{H}_{A})^2 \dim(\mathbb{H}_{\bar{A}})
		}	
		{
			\dim(\mathbb{H}_{\partial\gamma,out})^2
			+
			e^{-S_2(\core)}
			\dim(\mathbb{H}_{\partial\gamma,out})
		} 
		=
		\frac
		{
			h^{||\bar{A}||} 
			+
			e^{-S_2(\core)}
			h^{||A||}
		}	
		{
			h^{||\bnd_{out}||}
			+
			e^{-S_2(\core)}
		}
	\end{split}
\end{equation}
where we defined $ h = \dim(\bigoplus_{\mathbb{j} \leq j^x_\alpha \leq J} V^{j^x_\alpha}) = \frac{d_J(d_J+1)-d_{\mathbb{j}}(d_{\mathbb{j}}+1)}{2} $.
So,
\begin{equation}\label{key}
	-\log(\avg{e^{-S_2(A)}}_U ) =
	-\log(\frac
	{
		h^{||\bar{A}||} 
		+
		e^{-S_2(\core)}
		h^{||A||}
	}	
	{
		h^{||\bnd_{out}||}
		+
		e^{-S_2(\core)}
	})
\end{equation}
which, for large enough $ h $, has limiting behaviour:
\begin{equation}\label{key}
	\avg{S_2(A)}_U\approx -\log(\avg{e^{-S_2(A)}}_U ) \approx \min\{S_2(\core) + ||\bar{A}||\ln(h),||A||\ln(h)\}
\end{equation}
and in particular $ \avg{S_2(\bnd)}_U = \min\{||\bnd_{out}||\ln(h),S_2(\core)\}$.
(This approximation is very good for even low values of $ h $.) 
It appears that the entropy will only depend on the size of the region A, the outer boundary and the entropy of the core. One might read this, in fact, as a form of a general Ryu-Takayanagi formula due to statistical considerations alone. Low spin fluctuations may change this result. The crucial fact is that after total randomisation, no connection between bulk and boundary exists anymore.\\

\section{Fixed-spin isometry}
We give a sufficient criterion for the fixed-spin case of geometries to induce isometries as well as an argument for the dimensionality of the set of holographic states.\\
For the single-spin sector case, we need to find the minimal configuration of $ \mathcal{H}_{1}(\ind{j},\ind{j}) $ and adjust it such that the ground state value becomes $ \sum_{e\in C} \log(d_{j_e}) $. The most obvious way to achieve this is when the all-up configuration is the ground state: Then, only boundary links in $ C $ contribute and we recover precisely the value we need. So, in a way similar to [BulkEntropyRef], we can tell by certain conditions on the spin sector or graph combinatorics whether the state is holographic.
Let us assume that we change the configuration around the all-up one. Then, the change in energy from flipping a single spin $ z $ with distance $ |z-C|>1 $ is always positive. If instead, we flip a spin with distance $ 1 $ from $ C $, so adjacent to that boundary region, we instead have the change 
\begin{equation}\label{FixedSpinEnergyChange}
	\sum_{\alpha: (z,\alpha) \notin C} \log(d_{j^z_\alpha}) - \sum_{\alpha: (z,\alpha) \in C} \log(d_{j^z_\alpha}) + S_2((\rho^I)_z).
\end{equation}
If this quantity is positive for all vertices with distance 1 from $ C $, the all-up configuration is a local minimum and the state is, if spins are large enough, approximately holographic. This condition translates to
\begin{equation}\label{FixedSpinEnergyChangeExp}
	\prod_{\alpha: (z,\alpha) \notin C} d_{j^z_\alpha}  \prod_{\alpha: (z,\alpha) \in C} d_{j^z_\alpha}^{-1}  > e^{-S_2((\rho^I)_z)}
\end{equation}
or more generally for spin-down regions $ X $:
\begin{equation}\label{FixedSpinEnergyChangeExp2}
	\prod_{e \in E: s(e)\in X, t(e)\notin X} d_{j_e}^{h_{t(e)}}
	> e^{-S_2((\rho^I)_X)}
\end{equation}
which should hold for all regions $ X $, with the understanding that h is 1 on all vertices except on the boundary region C. This gives a condition on the entropy of the intertwiner state; If the left hand side is smaller than $ 1 $, this constrains the intertwiners to be not too pure. As reducing a general random state to a subsystem like $ X $ typically produces high entropies, this should not be too much of a constraint in many cases. However, for this to work, there is still the necessity that the left hand side exceeds the lower bound on the purity of the reduced density matrix on the right:
\begin{equation}\label{key}
	\prod_{e \in E: s(e)\in X, t(e)\notin X} d_{j_e}^{h_{t(e)}} > \prod_{x\in X} \mathcal{D}_{\mathbf{j}^x}
\end{equation}
Otherwise, the holography condition cannot be achieved. Like the bulk-to-boundary case, this condition depends on the graph and spins. For homogeneous graphs, this translates to
\begin{equation}\label{Necond}
	||\partial X \setminus C||-||\partial X \cap C|| > |X|
\end{equation}
for all subsets of the graph, which can be violated on many graphs easily. One of the smallest type of region for homogeneous spins that can do this is the "once-fine-grained vertex",\\
\begin{figure}
	\centering
	\includegraphics[width=0.3\linewidth]{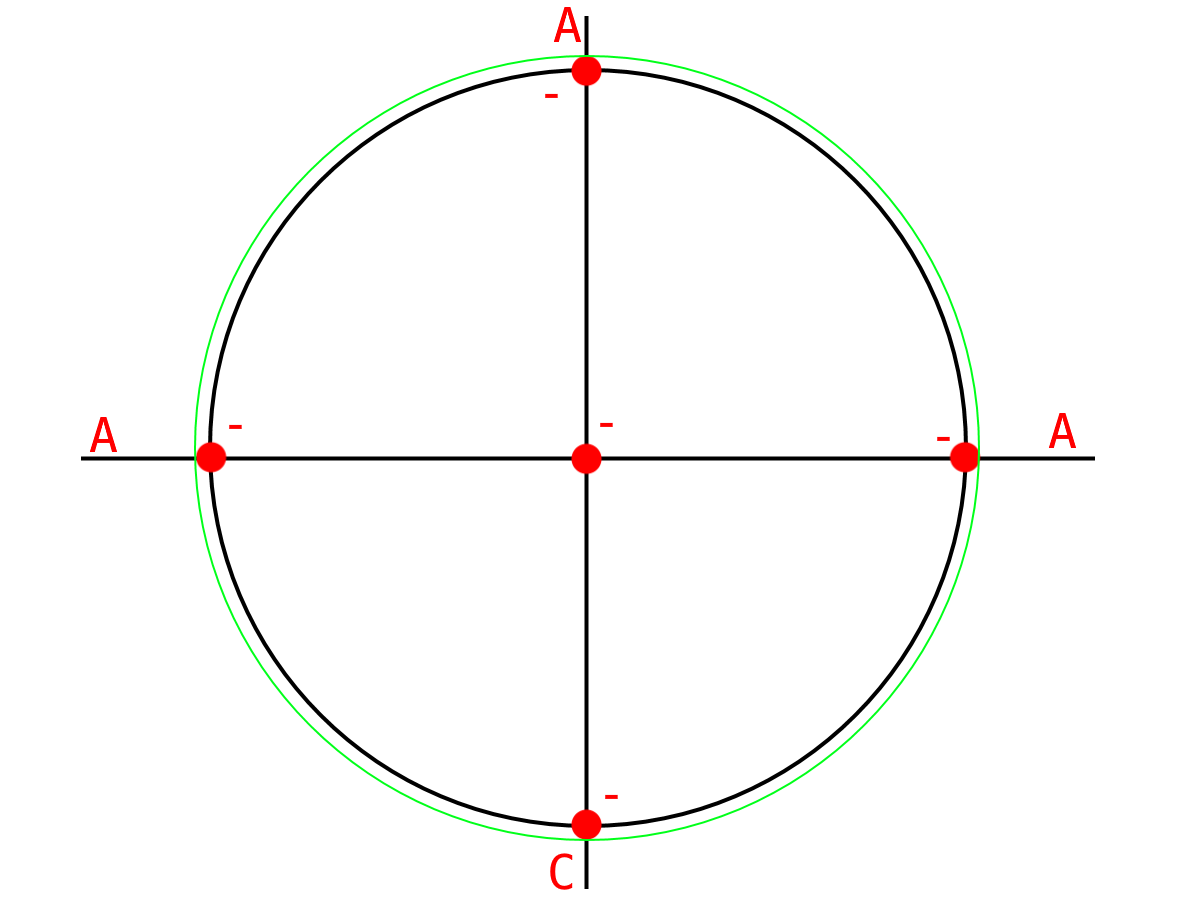}
	\caption{The once-fine-grained vertex graph. When choosing the regions A,C as outlined, flipping all spins down violates the condition \ref{Necond} for homogeneous spins, as $ 3-1 = 2 < 5 $. The spin-down region $ X $ has been enclosed in green here. }
	\label{fig:oncefinegrained}
\end{figure}
which has 5 vertices but 4 boundary edges. However, if the region $ X $ is close to $ C $, there will be even smaller geometries. It is easy to see that the condition for holography presented here is most easily satisfied for inhomogeneous geometries, where the intertwiner dimensions are as small as possible. Like in the bulk-to-boundary question, the interpretation here is that the minimal condition for isometry from $ \hbb_{\partial X \cap C} $ to $ \hbb_{\partial X \setminus C} $ is modified slightly by the presence of a bulk state $ \rho^I$ through which information can dissipate. Most succinctly,
\begin{equation}\label{key}
	\dim(\hbb_{\partial X \setminus C}) > e^{-S_2((\rho^I)_X)} \dim(\hbb_{\partial X \cap C})
\end{equation}
demonstrates that each region in the bulk must have an impure enough reduced bulk state to make the \textit{effective} input dimension on the right smaller than the output one. 
If equality is reached, though, one has a degeneracy in the ground state as flipping that spin $ z $ will not change the energy. Any degeneracies will persist into the limit of large spins and destroy holographic behaviour.

Given this simple criterion, which we may see as a region of parameters $ \rho^I $ to be tuned, we turn to the general question of characterising the space of holographic states. Our claim is the following: Fixing a graph structure and a spin sector, the set of intertwiner states yielding, on average, a holographic state, is either \textit{empty} or a subspace of $ \mathcal{D}(\mathcal{I}_{\ind{j}}) $, the space of density matrices, with \textit{codimension 1}. This means that, while the space is not dense in the full set of states, it is still of relatively high dimension - a lot of parameters may be tuned when starting from a holographic state. In particular, the dimension of the space of holographic states still scales exponentially with $ N $, the number of vertices. However, it is still of measure zero in the full space.\\
To illustrate this idea, consider our holography question as a differential geometric problem\footnote{We stealthily ignore the complication that the space of density matrices is not globally a smooth manifold of constant dimension - it is rather a stratified space made of smooth components. When restricting to the interiors of the strata, however, we can apply standard arguments.}: Let the map 
\begin{align}\label{FixedSpinMap}
	F: \mathcal{D}&(\mathcal{I}_{\ind{j}}) \longrightarrow \mathbb{R}\\
	&\rho_I \mapsto \sum_{\vec{\sigma}} e^{-\mathcal{H}_{1}(\ind{j},\ind{j}, \vec{\sigma})}
\end{align}
define the purity - our interest is the level set $ Hol(\gamma,\ind{j})\coloneqq F^{-1}(\dim(\hbb_{C,\ind{j}})^{-1}) $. By showing that the differential of this map is of rank 1 in the interior, we can use standard regular level set arguments to see that the level set is locally a smooth submanifold of codimension 1. 
For this purpose, first factor the map $ F $ into 
\begin{align}\label{FixedSpinMapFactored}
	\mathcal{D}&(\mathcal{I}_{\ind{j}}) &\stackrel{a}{\longrightarrow} &[0,1]^{2^N} &[0,1]^{2^N}\stackrel{b}{\longrightarrow} & &\mathbb{R}\\
	&\rho^I &\longmapsto &(e^{-S_2((\rho^I)_S)})_{S \subset \gamma}  &, (q_S)_{S \subset \gamma}&\longmapsto &\sum_{S \subset \gamma} \prod_{e \in \partial S \Delta C} d_{j_e}^{-1} \cdot q_S
\end{align}
So we can study the linearisation of $ a $ as 
\begin{equation}\label{LinearisationA}
	T_\rho a(\epsilon) = e^{-S_2((\rho+\epsilon)_\downarrow)} - e^{-S_2(\rho_\downarrow)} = \frac{2}{\Tr(\rho)^2 }\Tr\left[(\epsilon\otimes\rho )(\mathcal{S}_S-e^{-S_2(\rho)}\mathbb{I})\right] + \mathcal{O}(\epsilon^2) \coloneqqrev \avg{Q_S(\rho),\epsilon}_{HS}.
\end{equation}
Similarly, the linearisation of $ b $ is also a scalar product, as b itself is:
\begin{equation}\label{LinearisationB}
	T_q b (t) = \avg{\alpha,t} = \sum_{S \subset \gamma} \alpha_S \cdot t_S
\end{equation} where $ \alpha_S = \prod_{e \in \partial S \Delta C} d_{j_e}^{-1}  $. Then the composition of the two is simply
\begin{equation}\label{LinearisationF}
	T_\rho F (X) = \avg{\alpha,{Q(\rho),X}_{HS}} = \avg{
		\sum_{S \subset \gamma} \alpha_S Q_S(\rho)
		,X}_{HS}  =\avg{\tilde{Q}(\rho),X}_{HS}  
\end{equation}
which can be written as
\begin{equation}\label{LinearisationBRewrite}
	\frac{2}{\Tr(\rho)^2} \sum_{S \subset \gamma} \alpha_S \avg{\rho_S, (X-\frac{\Tr(X)}{\Tr(\rho)}\rho)_S}.
\end{equation}In particular, $ T_\rho F (\rho) = 0 $. However, we can perturb this around $ \rho $ to see that this map is nonzero, so of rank 1. Let $ \Tr(\epsilon) = 0 $, then
\begin{equation}\label{LinearisationPerturb1}
	T_\rho F (\rho+\epsilon) = \frac{2}{\Tr(\rho)^2} \sum_{S \subset \gamma} \alpha_S \avg{\rho_S, \epsilon_S} \approx \frac{1}{\Tr(\rho)^2} \sum_{S \subset \gamma} \alpha_S \left[ \avg{(\rho + \epsilon)_S,(\rho + \epsilon)_S} - \avg{\rho_S,\rho_S} \right].
\end{equation}
This can be, in turn, written as 
\begin{equation}\label{key}
	T_\rho F (\rho+\epsilon) = \sum_{S \subset \gamma} \alpha_S \left[ e^{-S_2((\rho + \epsilon)_S)} - e^{-S_2(\rho_S)} \right].
\end{equation}
So, starting from a given $ \rho $, we need to establish that the map is not identically 0. This is easily seen if $ e^{-S_2(\rho_S)} \neq 1 $, as then it is easy to perturb the state with $ \epsilon $ to increase the entropy. Then, each of the terms should become nonnegative and the sum will be, as well. As long as there is at least one positive term then, the map will be nonzero. The only complication arises from pure states, which incidentally have a lower dimensionality in the space of states and live in a different stratum from the rest. However, for nonpure states, this yields the desired result: as a map from nonpure intertwiner states, the map $ F $ has constant rank 1 and its regular level sets will have codimension 1. The intent of this argument is not to provide full rigour, but to highlight a crucial aspect of our endeavour to characterise holographic states: The space of parameters for which we achieve holography is relatively large in the sense of dimensionality.  
\section{Example calculations}
As we will see, even a simple example can become quite involved. \\
We first discuss the constraints on configurations. The case chosen here has intertwiner constraints which can be ignored, so we have the following:
\begin{enumerate}
	\item $\Delta_{0}(\ind{j},\ind{j},\vec{\sigma}) =\Delta_{0}(\ind{j},\ind{k},\vec{\sigma}) = 1$
	\item $\Delta_{0}(\ind{j},\ind{k},(+_L,+_R)) = 1 $
	\item $\Delta_{0}(\ind{j},\ind{k},(-_L,-_R)) = \prod_{e: \sigma_x = -1} \delta_{j_e,k_e} = 0$
	\item $\Delta_{0}(\ind{j},\ind{k},(+_L,-_R)) = \prod_{e: \sigma_x = -1} \delta_{j_e,k_e} = 0$
	\item $\Delta_{0}(\ind{j},\ind{k},(-_L,+_R))  = 1 $
\end{enumerate}
Essentially, as the spins on the right vertex do not agree between the two sectors, the Ising spin may not be down on it. The only difference in the numerator $\Delta$-factors is that the boundary pinning field flips around where spins have to agree. We take the pinning field to be $-1$ on the \textit{rightmost link} where the areas disagree:
\begin{enumerate}
	\item $\Delta_{1}(\ind{j},\ind{k},(+_L,+_R)) = 0$
	\item $\Delta_{1}(\ind{j},\ind{k},(-_L,-_R)) = 1$
	\item $\Delta_{1}(\ind{j},\ind{k},(+_L,-_R)) = 1$
	\item $\Delta_{1}(\ind{j},\ind{k},(-_L,+_R)) = 0$
\end{enumerate}
In other words, the right Ising spin must be down. This is a manifestation of the general rule that configurations where the value of the Hamiltonian would be ambiguous need to be excluded. Overall, even in this innocuous example we see that there can be a fair reduction of possible configurations contributing to the mixed partition sums. 
Then, we can calculate the Hamiltonian values individually.

\begin{tabular}{ |p{1.2cm}||p{1.9cm}|p{3.9cm}|p{3.9cm}|p{4cm}|  }
\hline
\multicolumn{5}{|c|}{Hamiltonian values for the 4 configurations} \\
\hline
& (+,+) &(+,-)&(-,+)&(-,-)\\
\hline
$\mathcal{H}_0(\ind{j},\ind{j})$   & $ 0 $    &$3\ln(2s+1)+\ln(6s-1)+ S_2 $&   $ 3\ln(2s+1)+\ln(6s-1) $ & $4\ln(2s+1)+\ln(6s-1)+\ln(6s+1)+ S_2 $\\
$\mathcal{H}_0(\ind{k},\ind{k})$&   0  & $3\ln(2s+1)+\ln(6s+1) $  &$3\ln(2s+1)+\ln(6s+1)$& $4\ln(2s+1)+2\ln(6s+1)$\\
$\mathcal{H}_0(\ind{j},\ind{k})$ &0 & \text{disallowed}&  $2\ln(2s+1)+\ln(6s+1)+ \Sigma_{(-,+)} $& \text{disallowed}\\
\hline
$\mathcal{H}_1(\ind{j},\ind{j})$    &$ \ln(6s-1) $ & $3\ln(2s+1)+ S_2 $&  $ 3\ln(2s+1)+\ln(6s+1)+\ln(6s-1) $& $4\ln(2s+1)+\ln(6s+1)+ S_2 $\\
$\mathcal{H}_1(\ind{k},\ind{k})$&   $\ln(6s+1)$  & $3\ln(2s+1)$&  $3\ln(2s+1)+2\ln(6s+1) $ & $ 4\ln(2s+1)+\ln(6s+1) $\\
$\mathcal{H}_1(\ind{j},\ind{k})$& disallowed  & $3\ln(2s+1)+\Sigma_{(+,-)}$   & \text{disallowed} & $ 3\ln(2s+1)+\ln(6s+1) $+$\Sigma_{(-,-)}$\\
\hline
\end{tabular}
We have used the shorthands 
\begin{align}
	\Sigma_{(\sigma_L,\sigma_R)} = \Sigma(\ind{j},\ind{k};(\sigma_L,\sigma_R))
	\qquad 
	S_2 = S_2(\rho^I_{\ind{j},\ind{j}}).
\end{align}
We thus find the six partition sums
\begin{enumerate}
	\item $Z^{\ind{j},\ind{j}}_0 = 1+ (2s+1)^{-3}(6s-1)^{-1}e^{-S_2}
	+(2s+1)^{-3}(6s-1)^{-1}
	+(2s+1)^{-4}(6s-1)^{-1}(6s+1)^{-1}e^{-S_2}$
	\item $Z^{\ind{k},\ind{k}}_0 = 1+(2s+1)^{-3}(6s+1)^{-1}
	+(2s+1)^{-3}(6s+1)^{-1}
	+(2s+1)^{-4}(6s+1)^{-2}$
	\item $Z^{\ind{j},\ind{k}}_0 = 1+(2s+1)^{-2}(6s+1)^{-1}e^{-\Sigma_{(-,+)}}$
	\item $Z^{\ind{j},\ind{j}}_1 = (6s-1)^{-1}+ (2s+1)^{-3}e^{-S_2}
	+(2s+1)^{-3}(6s-1)^{-1}(6s+1)^{-1}
	+(2s+1)^{-4}(6s+1)^{-1}e^{-S_2}$
	\item $Z^{\ind{k},\ind{k}}_1 = (6s+1)^{-1}+ (2s+1)^{-3}
	+(2s+1)^{-3}(6s+1)^{-2}
	+(2s+1)^{-4}(6s+1)^{-1}$
	\item $Z^{\ind{j},\ind{k}}_1 =  (2s+1)^{-3}e^{-\Sigma_{(+,-)}}
	+(2s+1)^{-3}(6s+1)^{-1}e^{-\Sigma_{(-,-)}}$
\end{enumerate}
and we note that as expected, the values of the $Z_0$ partition sums approach 1 as we increase the areas. Each term in the $Z_1$ sums also decays with some power of the area. Additionally, the decay in the mixed sums is stronger than the others due to suppression of certain configurations.\\
It is instructive to see the single-sector results at this point. For the first sector, the purity is
\begin{align}\label{key}
	\frac{Z^{\ind{j},\ind{j}}_1 }{	Z^{\ind{j},\ind{j}}_0 }
	&= 
	\frac{(6s-1)^{-1}+ (2s+1)^{-3}e^{-S_2}
		+(2s+1)^{-3}(6s-1)^{-1}(6s+1)^{-1}
		+(2s+1)^{-4}(6s+1)^{-1}e^{-S_2}}
	{1+ (2s+1)^{-3}(6s-1)^{-1}e^{-S_2}
		+(2s+1)^{-3}(6s-1)^{-1}
		+(2s+1)^{-4}(6s-1)^{-1}(6s+1)^{-1}e^{-S_2}}\\
	&\approx (6s-1)^{-1}
\end{align}
which is the reciprocal dimension of the boundary input space in the large-spin limit.
The same thing happens with the other sector, where $ \frac{Z^{\ind{k},\ind{k}}_1 }{Z^{\ind{k},\ind{k}}_0 } \approx (6s+1)^{-1} $. Thus both sectors individually yield an isometric map when spins are large enough.
Now, with the two weights 
\begin{equation}
	e^{B(\ind{j})} = (6s+1)(6s-1)(2s+1)^4(a+d) \qquad e^{B(\ind{k})} = (6s+1)^2(2s+1)^4 w
\end{equation}
and the entropies of the intertwiner state
\begin{equation}
	e^{-S_2} = t\coloneqq \frac{a^2+d^2+|b|^2}{(a+d)^2} 
	\qquad
	e^{-\Sigma_{(-,-)}} = e^{-\Sigma_{(+,-)}} = q \coloneqq
	\frac{|u^2|+|v|^2}{w(a+d)}
	\qquad
	e^{-\Sigma_{(+,+)}}
	=
	e^{-\Sigma_{(-,+)}}
	= 1
\end{equation}
this gives us $\frac{Z_1}{Z_0}$. The full expression is rather unilluminating, but we give a few special cases of interest. When the spin is taken to be asymptotically large:

\begin{align}\label{3rdOrderExpr}
		\frac{Z_1}{Z_0} 
		&=
		\frac{2 w^2-2 w+1}{6 s}
		+
		\frac{(1-2 w)^3}{36 s^2}\\
		&+
		\frac{27 (w-1)^2 e^{-S_2}+(61-54 q) w^2+(54 q-10) w+24 w^4-48 w^3+1}{216 s^3}
		+
		O\left(\left(\frac{1}{s}\right)^4\right).
\end{align}
This is particularly simple in that the $b$ and $u,v$ parameters do not contribute up to second order. Thus, in the high spin limit, these tend to not matter as much. Studying only the leading 2 orders, we can ask which parameters give us the most mixing.
These are easily found to be, for fixed s, $w\in \{\frac{1}{2}, \frac{2s+1}{2}\}$, which means that only $ w=\frac{1}{2} $ is a valid minimum. \\
If we take the third order into account, we instead have as minima
\begin{equation}
	(w,q) = (1,\frac{2}{27} \left(18 s^2-9 s+16\right)) \approx
	(1,\frac{4}{3} s^2)
\end{equation}
irrespective of the value of the entropy $e^{-S_2}$. Further orders preserve this minimum.
A numerical study with Mathematica of the minimum of the purity for given $ s $ reveals that the overall value decreases with spin inversely as expected:
\begin{equation}
	G(s):=\min_{a,d,w,t,q}(\frac{Z_1}{Z_0}(a,d,w,t,q,s)) \approx \frac{1}{12s}.     
\end{equation}
So, the maximal achievable entropy for fixed spin $s$ is then
\begin{equation}
	S_2 \approx  \ln(12s)
\end{equation}
We can then estimate the maximal dimension for a holographic subspace by $G^{-1} \approx 12s $, which is precisely the dimension of $\hbb_C$ in the studied case. Numerically, we find that this is achieved when $(a,d,w)\approx(\frac{1}{4},\frac{1}{4},\frac{1}{2})$ in the large-spin limit, with not much dependence on $q$ or $ t $. This, in particular, implies $ c_j = c_k = \frac{1}{2} $. The result agrees qualitatively with the second order result from \ref{3rdOrderExpr}, confirming that the large-spin regime is well approximated by it.\\
If we instead choose the region $C$ to be the upper right link, we get the same type of result: There is a state of maximal entropy which makes the induced map into an isometry, with $G(s)= \frac{1}{2s+1}$, and which is the minimum of $\frac{Z_1}{Z_0}$ for fixed spin sectors. The parameters of the minimum are, however, different: $(a,d,w)\approx(\frac{1}{3},\frac{1}{3},\frac{1}{3})$. Here, $ c_j = 2c_k $. This shows that whether an isometry exists or not can depend sensitively on the region under consideration.\\

  \bibliographystyle{plainnat}
  \bibliography{References}  

\cleardoublepage
\vspace*{2cm}
{\LARGE \sffamily{\textbf{Declaration of Academic Integrity}}}\\
\vspace*{0.2cm}\\
Hereby, I declare that I have written this thesis independently on my own and without any other sources than the ones mentioned.
\vspace*{3cm}\\
\noindent
\rule[0.5ex]{25em}{0.3pt}\\
Simon Langenscheidt \qquad \qquad Munich, \today

\end{document}